\documentclass[twocolumn,showpacs,preprintnumbers,amsmath,amssymb]{revtex4-1}

\usepackage{float}
\usepackage{graphicx}
\usepackage{dcolumn}
\usepackage{bm}
\usepackage{color}
\usepackage{xcolor}

\usepackage{amsmath}
\usepackage{amssymb}
\usepackage{amsfonts}
\usepackage{bbm}
\usepackage{hyperref}

\begin{document}

\title{Topological spin-orbit-coupled fermions beyond rotating wave approximation}

\author{Han Zhang$^{1,2} \dagger$}
\author{Wen-Wei Wang$^{1,2}\dagger$}
\author{Chang Qiao$^{1,2*}\dagger$}
\author{Long Zhang$^{3,4}\dagger$}
\author{Ming-Cheng Liang$^{1,2,5}$}
\author{Rui Wu$^{1,2}$}
\author{Xu-Jie Wang$^{1,2}$}
\author{Xiong-Jun Liu$^{1,2,4,6*}$}
\author{Xibo Zhang$^{1,2,4,5}$}
\email{Corresponding authors. Emails: qiaochang@pku.edu.cn, xiongjunliu@pku.edu.cn, and xibo@pku.edu.cn\\ $\dagger$ These authors contributed equally to this work.}
\affiliation{$^1$International Center for Quantum Materials, School of Physics, Peking University, Beijing 100871, China}
\affiliation{$^2$Collaborative Innovation Center of Quantum Matter, Beijing 100871, China}
\affiliation{$^3$School of Physics and Institute for Quantum Science and Engineering, Huazhong University of Science and Technology, Wuhan 430074, China}
\affiliation{$^4$Hefei National Laboratory, Hefei 230088, China}
\affiliation{$^5$Beijing Academy of Quantum Information Sciences, Beijing 100193, China}
\affiliation{$^6$International Quantum Academy, Shenzhen 518048, China}

\begin{abstract}
The realization of spin-orbit-coupled ultracold gases has driven a wide range of researches and is typically based on the rotating wave approximation (RWA). By neglecting the counter-rotating terms, RWA characterizes a single near-resonant spin-orbit (SO) coupling
in a two-level system.
Here, we propose and experimentally realize a new scheme for achieving a pair of two-dimensional (2D) SO couplings for ultracold fermions beyond RWA. This work not only realizes the first anomalous Floquet topological Fermi gas beyond RWA, but also significantly improves the lifetime of the 2D-SO-coupled Fermi gas. Based on pump-probe quench measurements, we observe a deterministic phase relation between two sets of SO couplings, which is characteristic for our beyond-RWA scheme and enables the two SO  couplings to be simultaneously tuned to the optimum 2D configurations. We observe intriguing band topology {\color{black}by measuring} two-ring band-inversion surfaces, quantitatively consistent with a Floquet topological Fermi gas in the regime of high Chern numbers. Our study can open an avenue to explore exotic SO physics and anomalous topological states based on long-lived SO-coupled ultracold fermions.

\end{abstract}


\maketitle

{\em Introduction.---}Spin-orbit (SO) coupling plays a crucial role in many prominent effects
in condensed matter physics. In particular, SO coupling is a pivotal ingredient of spin Hall and
anomalous Hall effects in spintronics~\cite{Awschalom09nature},
whose quantum versions belong to new classifications of fundamental quantum matter
dubbed topological phases, known as quantum spin Hall effect~\cite{bhz06science,qsh07science}
and quantum anomalous Hall effect~\cite{QiWuZhang06prb,yufang10science, Xue13science,qah14natphys,qah14prl,Xue14nsr}.
The topological phases, being a broad concept covering topological insulators~\cite{Kane10rmp,sczhang11rmp,asboth2016short}, topological semimetals~\cite{Yan17ARCMP,Armitage18rmp,Lv21rmp}, and topological superconductors~\cite{Kane10rmp,sczhang11rmp,Sato17rpp,Sharma22sst},
have now been extensively studied in solid state materials with SO couplings.

Ultracold atoms provide
versatile platforms in a controllable fashion to study SO physics~\cite{Dalibard11rmp,Goldman14rpp,Zhai15rpp, xjl18bookchapter}.
The realization of one-dimensional (1D) SO couplings in ultracold gases of {\color{black}bosons~\cite{Spielman11nature,Pan12prl,Spielman13natureSHE,Pan14np,Engels14nc,ChenYong14pra,Bloch14natphys,Ketterle16prl,Ketterle17nature, Spielman17njp,Lin18prlsoac,Jiang19prlsoac}}
and fermions~\cite{jing12prl,mz12prl,Spielman13prl,jingzhang14review,Lev16prx,livi16prl,Jo16pra,shimon17nature} offers
the opportunities to realize various neutral-atom topological states,
as well as unconventional quantum materials~\cite{Zoller03njp,Ohberg04prl,xiongjunliu06epjd,duan06prl, xjl07prl,xjl09prl,Spielman09pra,Spielman09nature,Spielman11pra,Spielman13nature,liu14prl,Mancini15science,Spielman15science,Gadway17sciadv, liu18pra,Peng22aappsb}.
Recently, the successes in realizing 2D and 3D SO couplings~\cite{JingZhang16natphys,shuai16science,JingZhang16prl,jo19natphys,xjl21science,LiLiu21scibull,Spielman21nc,barreiro22prl,Liang21arxiv} enabled a number of achievements including non-Abelian artificial gauge fields~\cite{JingZhang16natphys,shuai16science}, nodal-line semimetal~\cite{jo19natphys}, and Weyl semimetal~\cite{xjl21science,LiLiu21scibull}.
All these experiments 
are in a regime where the magnitude of SO interaction is much smaller than the atomic level splitting,
and thus built on a key approximation, i.e. the rotating wave approximation (RWA).
By neglecting counter-rotating terms that oscillate rapidly in the interacting picture,
the RWA characterizes one single near-resonant coupling in a two-level system~\cite{JaynesCummings63ProcIEEE, Knight93JMO,  AE75twolevelbook, Foot05atomicbook, HuiZhai21ultracoldbook}. Breaking the RWA allows for developing new methods such as time-optimal quantum control~\cite{Awschalom09science},
and can induce qualitatively new physical effects~\cite{BlochSiegert1940pr,Galbraith1983prl,Burnett08njp}.
Accordingly, new types of SO physics and exotic topological phases may be achieved in the regime without RWA. However,
till now, there is no report in theory or experiment on SO couplings beyond RWA for ultracold atoms.

In this article, we propose and experimentally realize 2D-SO-coupled topological fermions beyond RWA.
We prepare the spin-$1/2$ ultracold Fermi gas of strontium (${}^{87}$Sr) atoms
 in 2D optical Raman lattice, where two sets of Raman potentials of different resonance conditions are naturally generated without RWA, rendering an intrinsic Floquet SO-coupled system. By developing a highly nonlinear spin-discriminating method, we enhance the lifetime of 2D-SO-coupled fermions by an order of magnitude, achieving a long lifetime on the $100$-ms scale.
Using the recently developed pump-probe quench measurement method~\cite{Liang21arxiv}, we
explore a novel topological phase diagram by {\color{black}indirectly} measuring the band topology of 2D-SO-coupled fermions. In particular, various high-Chern-number Floquet topological bands are {\color{black}achieved.} This work may open a new avenue in engineering {\color{black}rich} long-lived topological systems with SO-coupled fermions.

\begin{figure}[t]
	\includegraphics[width = 8.3cm]{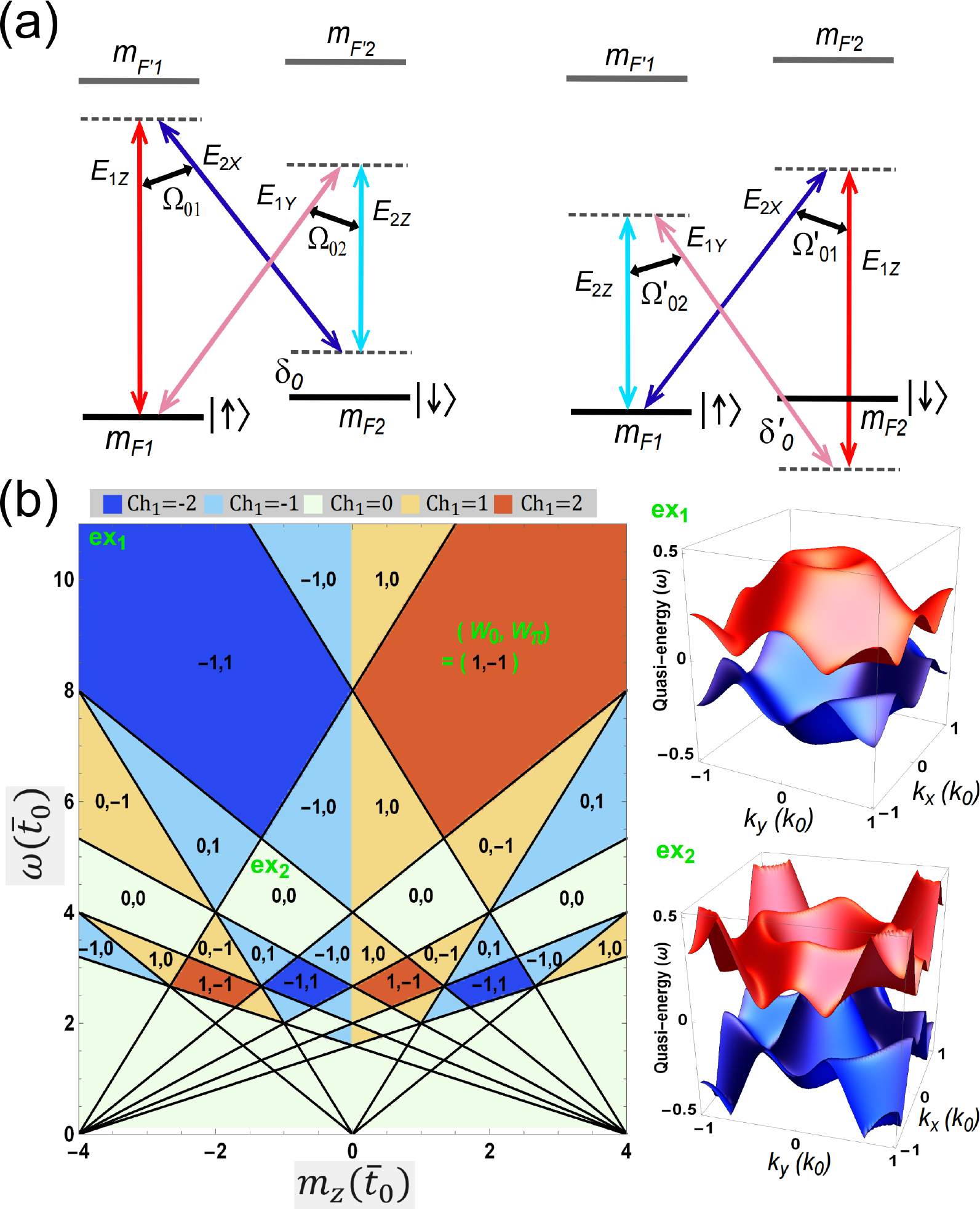}
	\caption{\label{fig:diagram}  Diagram of SO-coupled fermions beyond RWA. (a) Raman processes beyond RWA. Left: processes coupling $|\!\!\uparrow\rangle$ and $|\!\!\downarrow\rangle$ via absorbing a photon from Raman coupling beam 1 {\color{black}(red/light red)} and emitting one into beam 2 {\color{black}(blue/cyan)}. Right: processes via absorption from beam 2 and emission into beam 1. Incorporating all four Raman couplings yields two sets of SO couplings in Eq.~\eqref{eq:Ramancouplingmatrix}. (b) Topological phase diagram regarding the effective Zeeman splitting $m_{z}$ and modulation angular frequency $\omega$.
The anomalous topological phases are characterized by two winding numbers $W_{0,\pi}$.
Right panels show examples of quasi-energy bands in two phase regions.
	}
\end{figure}

\textit{The scheme.---}Our proposed scheme for realizing 2D SO couplings beyond RWA builds
on a spin-$1/2$ Fermi gas in a 2D square optical Raman lattice~\cite{liu14prl,liu18pra}, with a temporal modulation naturally emerging from two sets of Raman couplings, which is depicted by a time-dependent Hamiltonian {\color{black} for two spin states $|\!\!\uparrow\rangle$ and $|\!\!\downarrow\rangle$}:
 \begin{eqnarray}\label{eq:experimentalHamiltonian}
 \hat{H}(t) & = & \frac{\hbar^2\mathbf{k}^2}{2m} + V_{\mathrm{latt}}(x,y) + \Omega_R(x,y,t) + \frac{\delta_0}{2}\sigma_z.
 \end{eqnarray}
Here $t$ is the time, {\color{black}$\hbar$ is the reduced Planck's constant (set to 1 below), ${\bf k}$ is the atomic momentum}, $m$ is the atomic mass, $\delta_0$ denotes two-photon detuning, and  $\sigma_{x,y,z}$ are Pauli matrices. As illustrated in Fig.~\ref{fig:diagram}(a), the Raman coupling $\Omega_R$ beyond RWA
 \begin{eqnarray}\label{eq:Ramancouplingmatrix}
&& \Omega_R(x,y,t)  =  \left( \begin{array}{cc}
 0 & \Omega+\Omega'e^{\mathrm{i}\omega t} \\
 \Omega^{*}+\Omega'^{*}e^{-\mathrm{i}\omega t} & 0
 \end{array}
 \right) 
 \end{eqnarray}
contains two sets of Raman potentials $\Omega$ and $\Omega'$: $\Omega(x,y)=\Omega_{01}\sin k_0x\cos k_0y+\Omega_{02}e^{i\delta\varphi}\cos k_0x\sin k_0y$ and  $\Omega'(x,y) =\Omega'_{01}\sin k_0x\cos k_0y-\Omega'_{02}e^{-i\delta\varphi}\cos k_0x\sin k_0y$,
with each set including two Raman couplings to drive SO couplings in $X$ and $Y$ directions.
Here $k_0 = \pi/a$ relates to lattice spacing $a$, $\omega/2\pi$ equals to twice the frequency difference
between two Raman coupling beams (say $E_{1Z}$ and $E_{2X}$), $\delta\varphi$ is a tunable relative phase,
and $\Omega_{01(02)}$($\Omega_{01(02)}'$) denote the amplitudes of Raman couplings.
{\color{black} The two-photon detuning for Raman couplings in $\Omega'$ is given by $\delta'_0=\delta_0-\omega$.}
The lattice potential is $V_{\mathrm{latt}}(x,y) = \frac{V_{\mathrm{latt}\uparrow}+V_{\mathrm{latt}\downarrow}}{2}\mathbb{I} + \frac{V_{\mathrm{latt}\uparrow}-V_{\mathrm{latt}\downarrow}}{2}\sigma_z$, with $V_{\mathrm{latt}\uparrow,\downarrow}(x,y) = V_{0X\uparrow,\downarrow}\cos^2 k_0 x + V_{0Y\uparrow,\downarrow}\cos^2 k_0 y$, where $\mathbb{I}$ is the identity matrix, 
and $V_{0X\uparrow,\downarrow}$($V_{0Y\uparrow,\downarrow}$) denote the optical lattice depths along the $X$($Y$) direction for the two spin states.
An effective Zeeman splitting is further defined as $m_z \equiv \delta_0/2+(\epsilon_{_\uparrow} - \epsilon_{_\downarrow})/2$, where $\epsilon_{_{\uparrow(\downarrow)}}$ is the onsite energy of the $|\!\!\uparrow\rangle$ ($|\!\!\downarrow\rangle$) Wannier function at $\delta_0 = 0$.
Note that in the large-$\omega$ limit, the $\Omega'$ term in Eq.~\eqref{eq:Ramancouplingmatrix} can be neglected, known as the RWA, and Eq.~\eqref{eq:experimentalHamiltonian} realizes the Qi-Wu-Zhang model~\cite{QiWuZhang06prb,liu14prl,Liang21arxiv}.

Equations~\eqref{eq:experimentalHamiltonian} and \eqref{eq:Ramancouplingmatrix} describe the new scheme with modulated SO couplings. The main results can be captured by considering tight-binding regime with only the nearest-neighbor hopping terms.
We obtain for $\delta\varphi=\pi/2$ the temporally modulated Bloch Hamiltonian as
\begin{eqnarray}\label{eq:BlochHamiltonian}
\hat{H}(\mathbf{q},t) & = & \sum_{i=x,y,z}h^i(\mathbf{q},t)\sigma_i +U_0(\mathbf{q})\mathbb{I},
\end{eqnarray}
where
$\mathbf{q}$ is the Bloch wavevector and  $U_0(\mathbf{q})$ is an overall energy shift. Here, the transverse couplings $h^{x}$ and $h^{y}$ are temporally modulated and defined by
$h^x + \mathrm{i}h^y \equiv (h^x_0 + \mathrm{i}h^y_0)(1+e^{-\mathrm{i}\omega t})$,
leading to $h^{x}(\mathbf{q},t) = h^{x}_0[1+\cos(\omega t)] + h^y_0\sin(\omega t)$, $h^{y}(\mathbf{q},t) = h^y_0[1+\cos(\omega t)]-h^x_0 \sin(\omega t)$, with $h^{x/y}_0 = 2t_{_\mathrm{SO}}\sin (q_{y/x} a)$ and $\omega$ acting as the modulation angular frequency. The time-independent $h^z$ is given by $h^z(\mathbf{q}) = m_z - 2\bar{t}_0[\cos(q_x a)+\cos(q_y a)]$, where
 $t_{_\mathrm{SO}}$ and $\bar{t}_0 = (t_{0\uparrow} + t_{0\downarrow})/2$ represent the spin-flip and mean value of spin-conserved ($t_{0\uparrow,\downarrow}$) hopping coefficients, respectively.

\begin{figure}[t]
	\includegraphics[width = 8.6cm]{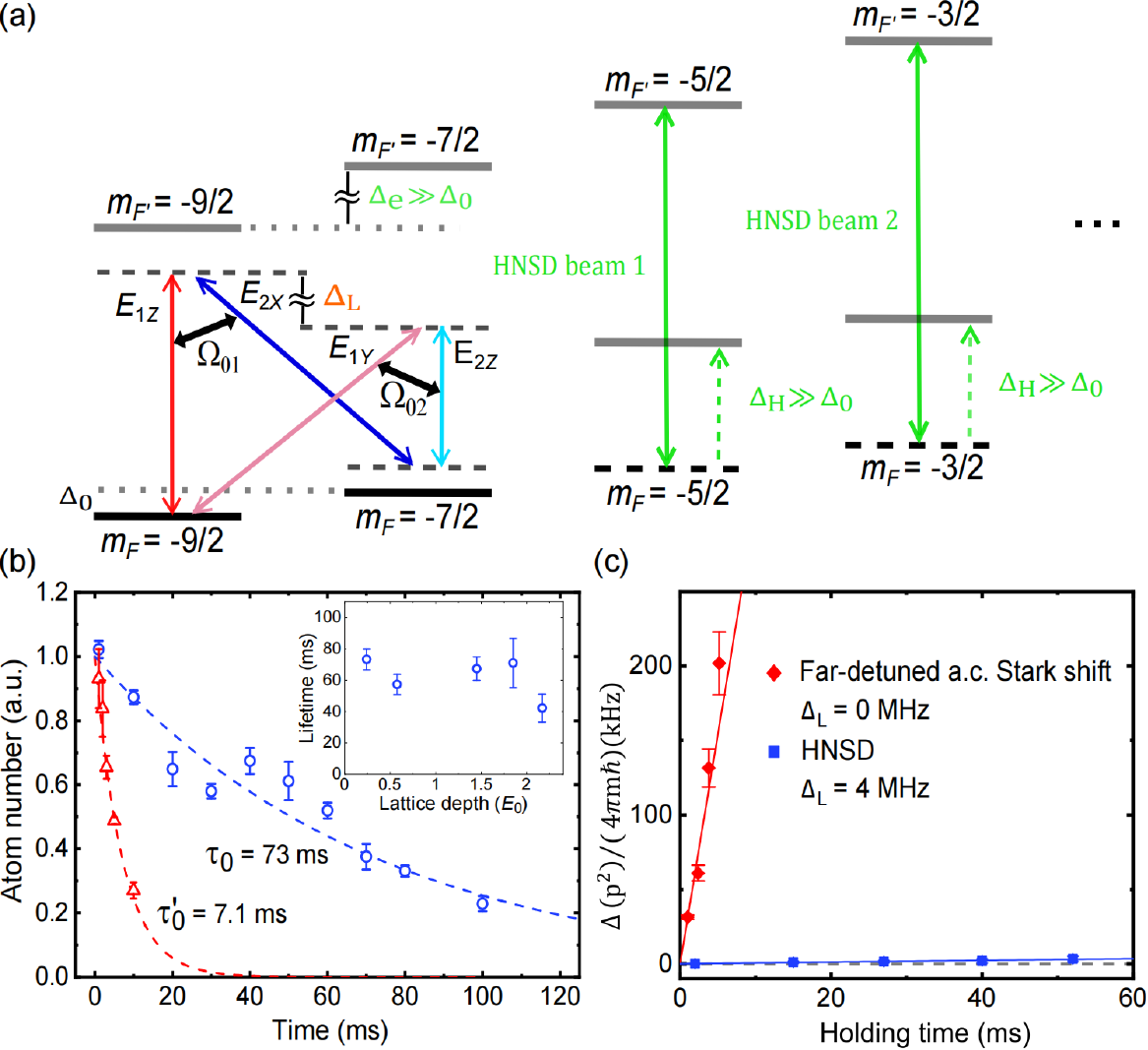}
	\caption{\label{fig:longlifetime}   Experimental realization. (a) Level diagram and the highly nonlinear spin-discriminating (HNSD) method. Two $\pi$-polarized HNSD beams (green) are applied for isolating the $|\!\!\uparrow\rangle$ and $|\!\!\downarrow\rangle$ states  ($m_{_F} = -\frac{9}{2}$,$-\frac{7}{2}$). For simplicity of illustration, only one set of Raman couplings are shown [see also Fig.~\ref{fig:diagram}(a)]. (b) Lifetime of 2D-SO-coupled fermions, measured under the  HNSD method (circles) or under the conventional {\color{black}far-detuned} a.c. Stark shift method (triangles). Dashed lines are exponential fits.
 Here we set 
$\Omega_{01}=\Omega'_{02}=0.61 E_0$, $\Omega_{02} = \Omega'_{01} = 0.23 E_0$, $V_{0X\uparrow}=V_{0Y\uparrow} = 0.44E_0$, $V_{0X\downarrow}=V_{0Y\downarrow}=0.06E_0$.
Inset: lifetime versus average lattice depth under similar Raman coupling strength.
	(c) Normalized variance of atomic momentum distribution versus holding time, reflecting the heating due to optical Raman lattices. 
	Solid lines are linear fits with slopes of 31~kHz/ms (red) or 0.06~kHz/ms (blue).	Error bars represent $1\sigma$ statistical uncertainties.
	}
\end{figure}

The above Bloch Hamiltonian~\eqref{eq:BlochHamiltonian} characterizes a novel {\color{black}and robust} Floquet topological system resulted from SO couplings beyond RWA, which is different from the counterparts
realized using photonic~\cite{White12nc,Chong15prx,Szameit17nc,Thomson17nc,Marrucci20optica} and phononic~\cite{Zhu16nc} systems and ultracold bosons~\cite{Bloch20natphysAF,Shuai22FloquetArxiv}.
The modulation of {\color{black}near-resonant} Raman couplings can drive new types of effective SO couplings from the Floquet high-order effects,
which may {\color{black}further contribute to novel topological physics}.
Fig.~\ref{fig:diagram}(b) shows a  diagram with rich anomalous topological phases characterized by two winding numbers $W_{0,\pi}$, which count the numbers of chiral edge modes inside the associated quasi-energy gaps and determine the Chern number by ${\rm Ch}_1=W_0-W_\pi$~\cite{Demler10prb,Levin13prx,Rudner15njp,Harper17prb,Rudner19prb,Rudner20nrp,Eckardt17rmp, LongZhang20prl,LongZhang22prxQ}.
Using quasi-energy band analysis~\cite{Floquet1883, Shirley65pr, Eckardt17rmp, LongZhang20prl,LongZhang22prxQ}, we determine the topological phase boundaries by identifying two types of band-inversion surfaces (BISs), the $0$-BIS and $\pi$-BIS, which are rings in 2D systems and depict where band crossing occurs, and originate from both the static ($\Omega$) and modulated ($\Omega'$) Raman couplings~\cite{LongZhang20prl,LongZhang22prxQ}. The winding number $W_{0}$ ($W_{\pi}$)
can be reduced to contributions of all the $0$-BISs ($\pi$-BISs)~\cite{LongZhang20prl,LongZhang22prxQ}.
Thus, via regulating $m_{z}$ and $\omega$, the system will enter a new topological phase whenever one BIS emerges or disappears.
Note that under time-reversal transformation, both $m_z$ and $\omega$ reverse sign, and topological invariants change sign accordingly. Thus the $\omega < 0$ part is not shown here but can be easily obtained.

\textit{Experimental setup.---}We realize the beyond-RWA scheme in a Fermi gas of $^{87}$Sr atoms with
$|\!\!\uparrow\rangle \equiv {}^1\mathrm{S}_0 |F=\frac{9}{2},m_{_F} = -\frac{9}{2}\rangle$ and
$|\!\!\downarrow\rangle \equiv |\frac{9}{2}, -\frac{7}{2}\rangle$; {\color{black}see Fig.~\ref{fig:longlifetime}(a) and the Supplemental Material~\cite{SM22HanZhang} for the setup}.
Compared to previous setups~\cite{liu14prl, shuai16science, shuai18prlrobust, jo19natphys,xjl21science,Liang21arxiv},
the two-level energy splitting of the present system is reduced to near zero, such that even moderate SO coupling strength is comparable to the energy splitting, bringing the system into the beyond-RWA regime in a unique manner that is advantageous for reaching a long lifetime.

{\color{black}We make two improvements for the experimental realization.}
First, we develop a highly nonlinear spin-discriminating (HNSD) method, by which $|\!\!\uparrow\rangle$ and  $|\!\!\downarrow\rangle$ are very well isolated from nearby states. As shown in Fig.~\ref{fig:longlifetime}(a), adjacent ${}^3$P${}_1$ spin excited states (with $F'=\frac{11}{2}$) are energetically separated by $\Delta_e \approx 13$~MHz under a 35-G magnetic field, whereas spin ground states are barely separated ($\Delta_0 \ll \Delta_e$). 
Two $\pi$-polarized HNSD lasers are applied with frequencies {\color{black}that are near resonance of the $m_{_F}$-conserving $\pi$-transitions for  the unwanted states (with $m_{_F}=-\frac{5}{2}$ and $-\frac{3}{2}$) but detuned from $\pi$-transitions for $|\!\!\uparrow\rangle$ and $|\!\!\downarrow\rangle$. Unlike the far-detuned a.c. Stark shift method where $|\!\!\uparrow\rangle$ and $|\!\!\downarrow\rangle$ experience large shifts~\cite{Jo16pra, jo19natphys,Liang21arxiv}, the HNSD method poses minimal perturbation to $|\!\!\uparrow\rangle$ and $|\!\!\downarrow\rangle$ (with only kHz-scale shifts), but strongly up-shifts the unwanted states by $\Delta_{\mathrm{H}}$ of more than 100~kHz}~\cite{SM22HanZhang} [see Fig.~\ref{fig:longlifetime}(a)]. {\color{black}Thus, the HNSD method realizes a well isolated effective spin-$\frac{1}{2}$ manifold of energetically nearly degenerate $|\!\!\uparrow\rangle$ and $|\!\!\downarrow\rangle$ states in a beyond-RWA regime.}
Second, to minimize detrimental interference effects~\cite{Liang21arxiv}, we implement a large frequency difference $\Delta_{\mathrm{L}}$ between similarly polarized beam components ($E_{1Z,2Z}$ or $E_{1Y,2X}$)~\cite{SM22HanZhang}, such that optical lattices remain essentially static and atomic heating due to moving lattice potentials~\cite{Liang21arxiv} is vastly suppressed.

\textit{Long-lived 2D-SO-coupled fermions.---}Under the new setup, we enhance the lifetime of 2D-SO-coupled fermions by an order of magnitude.
Figure~\ref{fig:longlifetime}(b) shows a long lifetime $\tau_0$ that reaches 73~ms, as compared with
$\tau_0' = 7.1$~ms based on the conventional {\color{black}far-detuned a.c. Stark shift method}~\cite{Jo16pra,Liang21arxiv}
under similar lattice conditions.
With a reference measurement under far-off-resonance SO couplings, we further extract a SO-coupling-restrained characteristic $1/e$ decay time $\sim$101~ms, {\color{black}showing the potential of our system for future improvements~\cite{SM22HanZhang}}.
Figure~\ref{fig:longlifetime}(b) inset shows that the lifetime remains fairly long under various average
lattice depths $(V_{0\uparrow}+V_{0\downarrow})/2$ up to about $2E_0$ (limited by available optical power), where $E_0\equiv\hbar^2 k_0^2/(2m)$ is the recoil energy.
Furthermore, we measure the variance of atomic momentum distribution by time-of-flight (TOF) imaging
after the fermions are held in the optical Raman lattice for various amount of time [Fig.~\ref{fig:longlifetime}(c)],
revealing a significant reduction of  heating by a factor of 500.
These results demonstrate a long-lived platform of 2D-SO-coupled fermions.

\textit{Verification of SO couplings beyond RWA.---}A key feature of our beyond-RWA scheme is 
the simultaneous generation of
two sets of SO couplings naturally related to each other.
From Eq.~\eqref{eq:Ramancouplingmatrix}, the sum of two relative phases for SO couplings driven by Raman potentials $\Omega$ and $\Omega'$ equals {\color{black}an intrinsically} fixed value of $\delta\varphi +(\pi - \delta\varphi) = \pi$~\cite{SM22HanZhang}.
 Thus, when one SO coupling is tuned to be quasi-1D in the $\hat{X}+\hat{Y}$ direction,
the other will be quasi-1D in the orthogonal $\hat{X}-\hat{Y}$ direction, {\color{black}
 providing a characteristic signature of SO couplings beyond RWA.}

\begin{figure}[t]
	\includegraphics[width = 8.7cm]{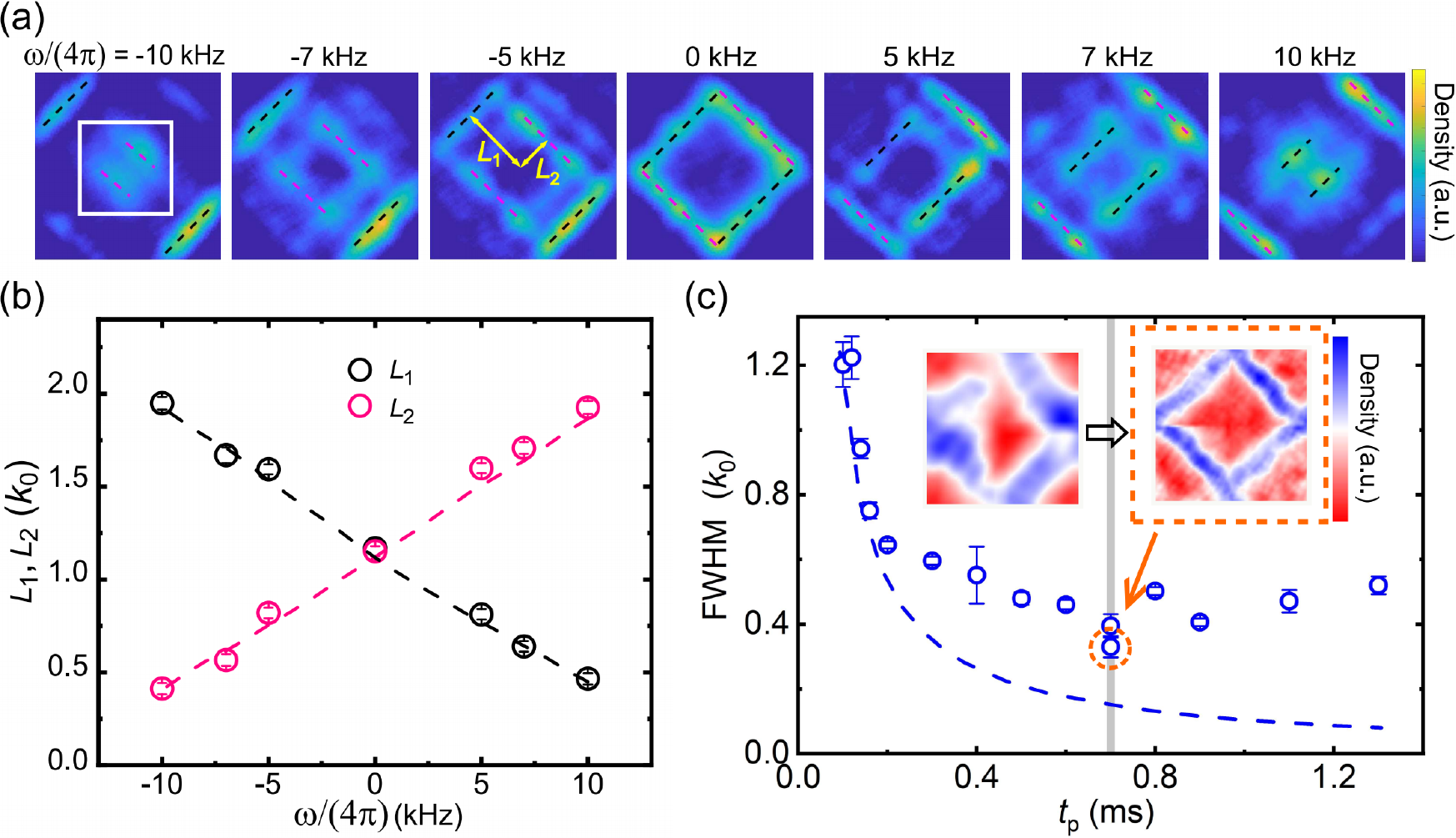}
	\caption{\label{fig:Floquet1DSOC} Verification of {\color{black}the beyond-RWA scheme using fermions with quasi-1D SO couplings}. (a) Pump-probe quench measurement (PPQM) of momentum distribution of $|\!\!\downarrow\rangle$ atoms at various $\omega$. As $\omega$ increases, two groups of atoms (marked by black and magenta dashed lines) move toward or away from the center of the FBZ (white square) in opposite manners, revealing two quasi-1D SO couplings in orthogonal directions of $\hat{X}\mp\hat{Y}$. Accordingly, the two relative phases for SO couplings are $\pi$ and 0, showing the key feature of a phase sum of $\pi$ for beyond-RWA SO couplings.  {\color{black} Here we set 
		$\Omega_{01}=\Omega'_{02}=0.36 E_0$, $\Omega_{02} = \Omega'_{01} = 0.14 E_0$, $V_{0X\uparrow}=V_{0Y\uparrow} = 0.26E_0$, $V_{0X\downarrow}=V_{0Y\downarrow}=0.03E_0$.  (b) Measured movement of peak positions $L_{1,2}$ of $|\!\!\downarrow\rangle$ atoms (circles) with $\omega$}, in comparison to numerical results (dashed lines). (c) Width of pumped atomic ring, measured as a function of the PPQM pulse length $t_{\mathrm{p}}$ under a fixed pulse area.  
Blue dashed line denotes the Fourier limit. Insets: sample atomic rings measured under $t_{\mathrm{p}} \approx 200$~$\mu$s (left) and 700~$\mu$s (right).		  
Error bars represent $1\sigma$ statistical uncertainties.}	
\end{figure}

For verification, we probe fermions with quasi-1D SO couplings using the recently developed pump-probe quench measurement (PPQM) method~\cite{Liang21arxiv}.
Atoms initially prepared in the $|\!\!\uparrow\rangle$ state experience SO-coupling-induced, quasi-momentum-dependent spin flip into  $|\!\!\downarrow\rangle$ under a short pulse of optical Raman lattice, and are subsequently probed by spin-resolved TOF measurements.
Fig.~\ref{fig:Floquet1DSOC}(a) shows TOF images
measured for various $\omega$.
As $\omega$ increases, we observe two groups of $|\!\!\downarrow\rangle$ atoms:
one group (black dashed lines) moves toward the center of the first Brillouin zone (FBZ) along the $\hat{X}-\hat{Y}$ direction; the other (magenta dashed lines) moves away from the center along  $\hat{X}+\hat{Y}$.
The two-photon detunings $\delta_0\sim \omega/2$ and $\delta'_0 = \delta_0 - \omega \sim -\omega/2$ [see Fig.~\ref{fig:diagram}(a)] indicate that the effective Zeeman splittings for the two sets of SO couplings depend on $\omega$
in reverse ways.
We identify the two groups as  pumped respectively
by the two quasi-1D SO couplings in the orthogonal directions of $\hat{X}\mp\hat{Y}$, corresponding to relative phases of $\pi$ and 0 for the two SO couplings (with a sum of $\pi$).
{\color{black} The measured highly symmetric movements (with respect to $\omega$) of the two atomic groups with opposite effective Zeeman splittings show unambiguously the realization of SO couplings beyond RWA, as further supported by agreement between measurements and numerical results {\color{black}by exact diagonalization in the non-tight-binding regime}~\cite{Liang21arxiv,SM22HanZhang} (Fig.~\ref{fig:Floquet1DSOC}(b)).}

The PPQM resolution can be sharpened by optimizing the pulse length $t_{\mathrm{p}}$ under a fixed pulse area~\cite{SM22HanZhang}.
Figure~\ref{fig:Floquet1DSOC}(c) shows a minimum atomic ring width at $t_{\mathrm{p}} \approx 700~\mu$s that is used in subsequent measurements.

\begin{figure}[t]
	\includegraphics[width = 8.6cm]{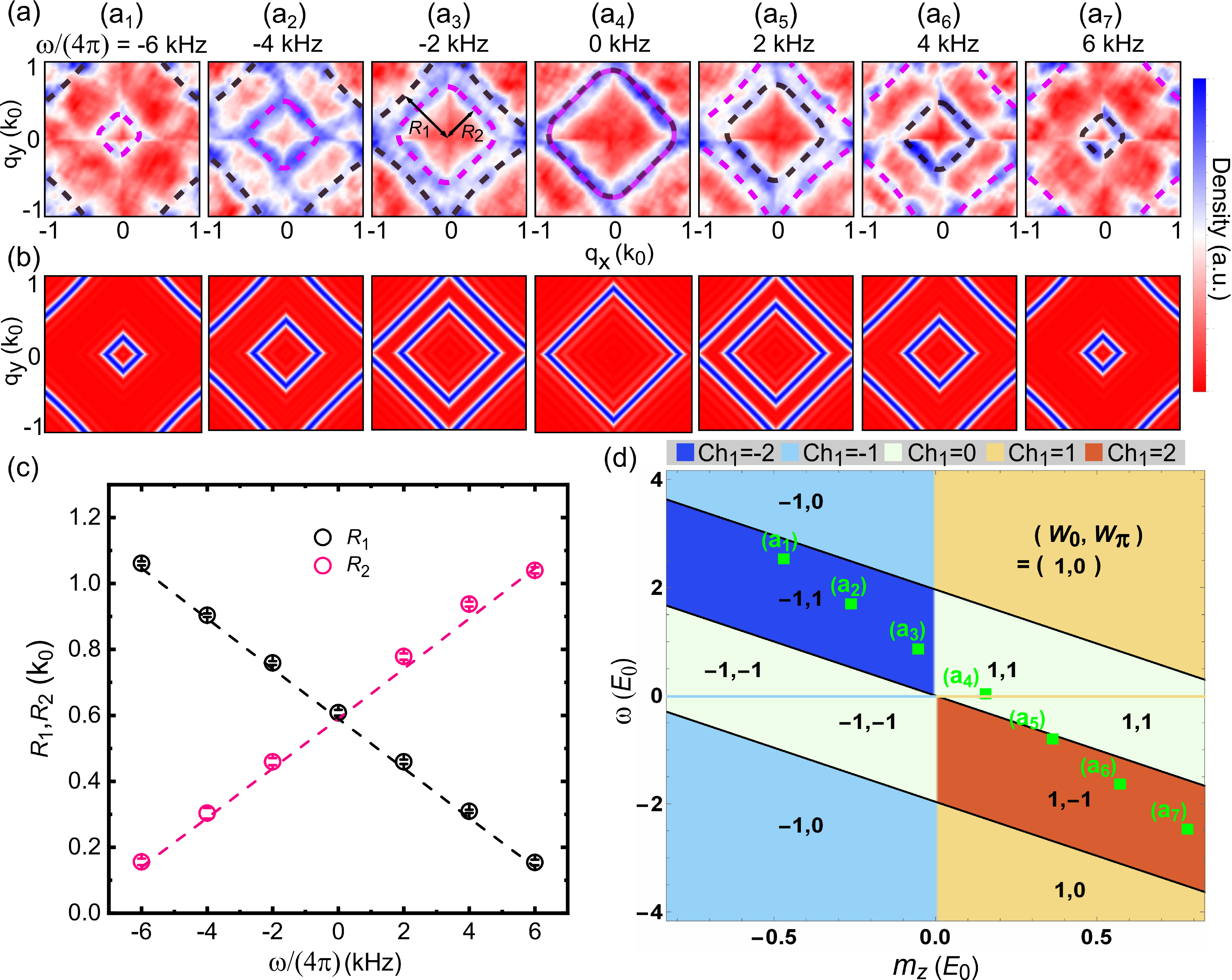}
	\caption{\label{fig:Floquet2DSOC} Experimental exploration of a Floquet topological phase diagram under optimum 2D SO couplings.
		(a) Observation of two BISs (rings) in the FBZ based on the PPQM method at various $\omega$ values. Black and magenta dashed guidelines are plotted for the two BISs.		(b) Numerical computations.
		(c) Measured sizes of the two BISs $R_{1,2}$ (circles) versus $\omega$, in comparison to numerical results (dashed lines).
		(d) A simplified phase diagram based on leading-order SO couplings only, with
		 the measurements in (a) marked by squares.
		Here we set  $\Omega_{01} = \Omega'_{02} = 0.17 E_0$, $\Omega_{02} = \Omega'_{01} = 0.07 E_0$, $V_{0X\uparrow} = V_{0Y\uparrow} = 0.13 E_0$, $V_{0X\downarrow} = V_{0Y\downarrow} = 0.02 E_0$. Error bars represent $1\sigma$ statistical uncertainties.
	}
\end{figure}

\textit{Experimental exploration of Floquet topological phase diagram.---}We explore the Floquet topological phase diagram by a systematic measurement of BIS configurations.
The noninteracting topological phases can be generically characterized by the dimension-reduced topological invariants on BISs~\cite{zhanglin18scibull,lee20prr,LongZhang20prl,ye20pra,gong21scibull,LongZhang22prxQ,Zhanglin2022scibull}, which provides feasible schemes to detect bulk topology~\cite{shuai18prltopo,hu20prl,liu21prxqu,jo19natphys,shuai19prl,du19pra, du20prl,Lu20prl,Liang21arxiv,Shuai22FloquetArxiv}.
For Chern bands in 2D optical Raman lattices, measuring the BIS configuration alone can yield precise information of the bulk topology~\cite{shuai18prltopo,Liang21arxiv,Shuai22FloquetArxiv}.
In this setup, Raman couplings and optical lattices are relatively weak,
such that the relevant topology is dominated by leading-order SO couplings, while the effect of higher-order couplings can be ignored~\cite{SM22HanZhang}.
Below we shall show that leading-order SO couplings alone lead to nontrivial topology.

We tune the SO coupling beyond RWA to the optimum 2D configuration by setting $\delta\varphi=\pi/2$, 
and then perform a systematic PPQM study 
by varying $\omega$.
In Fig.~\ref{fig:Floquet2DSOC}(a), we observe two BISs that emerge from M ($\pm k_0,\pm k_0$) and $\Gamma$ $(0,0)$ points in the FBZ, evolving toward each other, switching positions at $\omega=0$, and finally shrink at $\Gamma$ and M, respectively~\cite{note2}.
This observation agrees with numerical simulation in Fig.~\ref{fig:Floquet2DSOC}(b)~\cite{SM22HanZhang}.
In Fig.~\ref{fig:Floquet2DSOC}(c), we quantify the observation by measuring sizes of the two BISs versus $\omega$, which also shows  agreement with numerical results. The systematic measurements {\color{black}of BIS configurations,} which match well with the numerical study, unveil the underlying nontrivial topological regimes realized in the experiment, as we elaborate below.

{\color{black}In Fig.~\ref{fig:Floquet2DSOC}(d), we compute a simplified phase diagram with experimental parameters by considering only leading-order SO couplings~\footnote{{\color{black}The sub-leading-order SO coupling, which is a third-order coupling, has a strength estimated to be on the scale of $h \times 40$~Hz, which is not relevant in the current measurements; see more details in the Supplemental Material~\cite{SM22HanZhang}.}}, 
and compare it with the systematic measurements in Fig.~\ref{fig:Floquet2DSOC}(a), 
which scan through the high-Chern-number regime with $\mathrm{Ch}_1 = \pm 2$~\cite{SM22HanZhang}}. This regime is characterized by that the two BISs observed in Fig.~\ref{fig:Floquet2DSOC}(a) correspond to a 0-BIS (black) and a $\pi$-BIS (magenta), and are induced by two sets of SO couplings, respectively~\cite{SM22HanZhang}. Such two BISs
correspond to opposite effective Zeeman splittings and carry opposite winding numbers according to previous studies~\cite{LongZhang20prl,LongZhang22prxQ}, yielding $W_{0}=-W_{\pi}=\pm1$ and a high Chern number $\mathrm{Ch}_1=\pm 2$. At $\omega = 0$,
the measurement reduces to a single-ring BIS with $|\mathrm{Ch}_1| \le 1$ for a static Hamiltonian. These results indirectly show that the present experiment has achieved a fermionic anomalous Floquet topological system. A direct observation of topological invariants can be achieved by measuring winding numbers on BISs~\cite{LongZhang20prl,LongZhang22prxQ,shuai19prl} and will be presented in next studies.

\textit{Conclusion and discussion.---}In summary, we have realized 2D SO coupling beyond RWA for ultracold fermions with a long lifetime and observed {\color{black}various BIS configurations that provide an indirect measurement of the nontrivial topology engineered in the current system.} The SO coupling beyond RWA renders an intrinsic temporal engineering, giving rise to a rich Floquet topological phase diagram with high-Chern-number states.
Together with the HNSD method developed in our experiment, the 2D-SO-coupled Floquet fermion system has a long lifetime, which {\color{black}can be further improved~\cite{SM22HanZhang} and fulfills} a key prerequisite for future studies of novel correlated topological phases including the topological superfluid, dynamical gauge fields, and quantum magnetism {\color{black}in the interacting regimes}~\cite{liu14prl,bermudez20arxiv,bermudez22aop, Galitski12prlmagnetic,Rachel12prb}.
The present scheme {\color{black}can be naturally generalized to engineer} an isolated manifold of an arbitrary number of spin states for SO-coupled Fermi gases~\cite{Galitski12prlsu3,barreiro22prl}. Applying the beyond-RWA scheme to such large-spin systems can lead to fundamentally new type of SO couplings and can potentially bring about profound spin-orbit and topological physics, which deserves efforts in future studies.

\emph{Acknowledgments}.---We are grateful to Murray Barrett for very insightful discussions. We thank Shize Du, Tao Deng, Wei Qi, Yu-Dong Wei for discussions and technical support. This work is supported by the Chinese Academy of Sciences Strategic Priority Research Program under Grant No.~XDB35020100, the National Key Research and Development Program of China under Grant Nos~2018YFA0305601 and 2021YFA1400900, the National Natural Science Foundation of China (No.~11874073, 11825401, 12204187, 12261160368), the Open Project of Shenzhen Institute of Quantum Science and Engineering (Grant No.~SIQSE202003), the Hefei National Laboratory, and the Scientific and Technological Innovation
2030 Key Program of Quantum Communication and Quantum Computing under Grant Nos. 2021ZD0301903 and 2021ZD0302000.  X.Z. and X.-J.L. conceived the project.  H.Z., W.-W.W., C.Q. performed the experiments.  C.Q., L.Z., H.Z., W.-W.W. performed the numerical computations. All authors contribute to the data analysis and the writing of this manuscript.

\bibliography{SOCFermionsBeyondRWARef}

\providecommand{\noopsort}[1]{}\providecommand{\singleletter}[1]{#1}%
\begin{thebibliography}{116}%
\makeatletter
\providecommand \@ifxundefined [1]{%
 \@ifx{#1\undefined}
}%
\providecommand \@ifnum [1]{%
 \ifnum #1\expandafter \@firstoftwo
 \else \expandafter \@secondoftwo
 \fi
}%
\providecommand \@ifx [1]{%
 \ifx #1\expandafter \@firstoftwo
 \else \expandafter \@secondoftwo
 \fi
}%
\providecommand \natexlab [1]{#1}%
\providecommand \enquote  [1]{``#1''}%
\providecommand \bibnamefont  [1]{#1}%
\providecommand \bibfnamefont [1]{#1}%
\providecommand \citenamefont [1]{#1}%
\providecommand \href@noop [0]{\@secondoftwo}%
\providecommand \href [0]{\begingroup \@sanitize@url \@href}%
\providecommand \@href[1]{\@@startlink{#1}\@@href}%
\providecommand \@@href[1]{\endgroup#1\@@endlink}%
\providecommand \@sanitize@url [0]{\catcode `\\12\catcode `\$12\catcode
  `\&12\catcode `\#12\catcode `\^12\catcode `\_12\catcode `\%12\relax}%
\providecommand \@@startlink[1]{}%
\providecommand \@@endlink[0]{}%
\providecommand \url  [0]{\begingroup\@sanitize@url \@url }%
\providecommand \@url [1]{\endgroup\@href {#1}{\urlprefix }}%
\providecommand \urlprefix  [0]{URL }%
\providecommand \Eprint [0]{\href }%
\providecommand \doibase [0]{http://dx.doi.org/}%
\providecommand \selectlanguage [0]{\@gobble}%
\providecommand \bibinfo  [0]{\@secondoftwo}%
\providecommand \bibfield  [0]{\@secondoftwo}%
\providecommand \translation [1]{[#1]}%
\providecommand \BibitemOpen [0]{}%
\providecommand \bibitemStop [0]{}%
\providecommand \bibitemNoStop [0]{.\EOS\space}%
\providecommand \EOS [0]{\spacefactor3000\relax}%
\providecommand \BibitemShut  [1]{\csname bibitem#1\endcsname}%
\let\auto@bib@innerbib\@empty
\bibitem [{\citenamefont {Koralek}\ \emph {et~al.}(2009)\citenamefont
  {Koralek}, \citenamefont {Weber}, \citenamefont {Orenstein}, \citenamefont
  {Bernevig}, \citenamefont {Zhang}, \citenamefont {Mack},\ and\ \citenamefont
  {Awschalom}}]{Awschalom09nature}%
  \BibitemOpen
  \bibfield  {author} {\bibinfo {author} {\bibfnamefont {J.~D.}\ \bibnamefont
  {Koralek}}, \bibinfo {author} {\bibfnamefont {C.~P.}\ \bibnamefont {Weber}},
  \bibinfo {author} {\bibfnamefont {J.}~\bibnamefont {Orenstein}}, \bibinfo
  {author} {\bibfnamefont {B.~A.}\ \bibnamefont {Bernevig}}, \bibinfo {author}
  {\bibfnamefont {S.-C.}\ \bibnamefont {Zhang}}, \bibinfo {author}
  {\bibfnamefont {S.}~\bibnamefont {Mack}}, \ and\ \bibinfo {author}
  {\bibfnamefont {D.~D.}\ \bibnamefont {Awschalom}},\ }\href@noop {} {\bibfield
   {journal} {\bibinfo  {journal} {Nature}\ }\textbf {\bibinfo {volume}
  {458}},\ \bibinfo {pages} {610} (\bibinfo {year} {2009})}\BibitemShut
  {NoStop}%
\bibitem [{\citenamefont {Bernevig}\ \emph {et~al.}(2006)\citenamefont
  {Bernevig}, \citenamefont {Hughes},\ and\ \citenamefont
  {Zhang}}]{bhz06science}%
  \BibitemOpen
  \bibfield  {author} {\bibinfo {author} {\bibfnamefont {B.~A.}\ \bibnamefont
  {Bernevig}}, \bibinfo {author} {\bibfnamefont {T.~L.}\ \bibnamefont
  {Hughes}}, \ and\ \bibinfo {author} {\bibfnamefont {S.-C.}\ \bibnamefont
  {Zhang}},\ }\href@noop {} {\bibfield  {journal} {\bibinfo  {journal}
  {Science}\ }\textbf {\bibinfo {volume} {314}},\ \bibinfo {pages} {1757}
  (\bibinfo {year} {2006})}\BibitemShut {NoStop}%
\bibitem [{\citenamefont {Konig}\ \emph {et~al.}(2007)\citenamefont {Konig},
  \citenamefont {Wiedmann}, \citenamefont {Brune}, \citenamefont {Roth},
  \citenamefont {Buhmann}, \citenamefont {Molenkamp}, \citenamefont {Qi},\ and\
  \citenamefont {Zhang}}]{qsh07science}%
  \BibitemOpen
  \bibfield  {author} {\bibinfo {author} {\bibfnamefont {M.}~\bibnamefont
  {Konig}}, \bibinfo {author} {\bibfnamefont {S.}~\bibnamefont {Wiedmann}},
  \bibinfo {author} {\bibfnamefont {C.}~\bibnamefont {Brune}}, \bibinfo
  {author} {\bibfnamefont {A.}~\bibnamefont {Roth}}, \bibinfo {author}
  {\bibfnamefont {H.}~\bibnamefont {Buhmann}}, \bibinfo {author} {\bibfnamefont
  {L.~W.}\ \bibnamefont {Molenkamp}}, \bibinfo {author} {\bibfnamefont {X.-L.}\
  \bibnamefont {Qi}}, \ and\ \bibinfo {author} {\bibfnamefont {S.-C.}\
  \bibnamefont {Zhang}},\ }\href@noop {} {\bibfield  {journal} {\bibinfo
  {journal} {Science}\ }\textbf {\bibinfo {volume} {318}},\ \bibinfo {pages}
  {766} (\bibinfo {year} {2007})}\BibitemShut {NoStop}%
\bibitem [{\citenamefont {Qi}\ \emph {et~al.}(2006)\citenamefont {Qi},
  \citenamefont {Wu},\ and\ \citenamefont {Zhang}}]{QiWuZhang06prb}%
  \BibitemOpen
  \bibfield  {author} {\bibinfo {author} {\bibfnamefont {X.-L.}\ \bibnamefont
  {Qi}}, \bibinfo {author} {\bibfnamefont {Y.-S.}\ \bibnamefont {Wu}}, \ and\
  \bibinfo {author} {\bibfnamefont {S.-C.}\ \bibnamefont {Zhang}},\ }\href@noop
  {} {\bibfield  {journal} {\bibinfo  {journal} {Physical Review B}\ }\textbf
  {\bibinfo {volume} {74}},\ \bibinfo {pages} {085308} (\bibinfo {year}
  {2006})}\BibitemShut {NoStop}%
\bibitem [{\citenamefont {Yu}\ \emph {et~al.}(2010)\citenamefont {Yu},
  \citenamefont {Zhang}, \citenamefont {Zhang}, \citenamefont {Zhang},
  \citenamefont {Dai},\ and\ \citenamefont {Fang}}]{yufang10science}%
  \BibitemOpen
  \bibfield  {author} {\bibinfo {author} {\bibfnamefont {R.}~\bibnamefont
  {Yu}}, \bibinfo {author} {\bibfnamefont {W.}~\bibnamefont {Zhang}}, \bibinfo
  {author} {\bibfnamefont {H.-J.}\ \bibnamefont {Zhang}}, \bibinfo {author}
  {\bibfnamefont {S.-C.}\ \bibnamefont {Zhang}}, \bibinfo {author}
  {\bibfnamefont {X.}~\bibnamefont {Dai}}, \ and\ \bibinfo {author}
  {\bibfnamefont {Z.}~\bibnamefont {Fang}},\ }\href@noop {} {\bibfield
  {journal} {\bibinfo  {journal} {Science}\ }\textbf {\bibinfo {volume}
  {329}},\ \bibinfo {pages} {61} (\bibinfo {year} {2010})}\BibitemShut
  {NoStop}%
\bibitem [{\citenamefont {Chang}\ \emph {et~al.}(2013)\citenamefont {Chang},
  \citenamefont {Zhang}, \citenamefont {Feng}, \citenamefont {Shen},
  \citenamefont {Zhang}, \citenamefont {Guo}, \citenamefont {Li}, \citenamefont
  {Ou}, \citenamefont {Wei}, \citenamefont {Wang}, \citenamefont {Ji},
  \citenamefont {Feng}, \citenamefont {Ji}, \citenamefont {Chen}, \citenamefont
  {Jia}, \citenamefont {Dai}, \citenamefont {Fang}, \citenamefont {Zhang},
  \citenamefont {He}, \citenamefont {Wang}, \citenamefont {Lu}, \citenamefont
  {Ma},\ and\ \citenamefont {Xue}}]{Xue13science}%
  \BibitemOpen
  \bibfield  {author} {\bibinfo {author} {\bibfnamefont {C.-Z.}\ \bibnamefont
  {Chang}}, \bibinfo {author} {\bibfnamefont {J.}~\bibnamefont {Zhang}},
  \bibinfo {author} {\bibfnamefont {X.}~\bibnamefont {Feng}}, \bibinfo {author}
  {\bibfnamefont {J.}~\bibnamefont {Shen}}, \bibinfo {author} {\bibfnamefont
  {Z.}~\bibnamefont {Zhang}}, \bibinfo {author} {\bibfnamefont
  {M.}~\bibnamefont {Guo}}, \bibinfo {author} {\bibfnamefont {K.}~\bibnamefont
  {Li}}, \bibinfo {author} {\bibfnamefont {Y.}~\bibnamefont {Ou}}, \bibinfo
  {author} {\bibfnamefont {P.}~\bibnamefont {Wei}}, \bibinfo {author}
  {\bibfnamefont {L.-L.}\ \bibnamefont {Wang}}, \bibinfo {author}
  {\bibfnamefont {Z.-Q.}\ \bibnamefont {Ji}}, \bibinfo {author} {\bibfnamefont
  {Y.}~\bibnamefont {Feng}}, \bibinfo {author} {\bibfnamefont {S.}~\bibnamefont
  {Ji}}, \bibinfo {author} {\bibfnamefont {X.}~\bibnamefont {Chen}}, \bibinfo
  {author} {\bibfnamefont {J.}~\bibnamefont {Jia}}, \bibinfo {author}
  {\bibfnamefont {X.}~\bibnamefont {Dai}}, \bibinfo {author} {\bibfnamefont
  {Z.}~\bibnamefont {Fang}}, \bibinfo {author} {\bibfnamefont {S.-C.}\
  \bibnamefont {Zhang}}, \bibinfo {author} {\bibfnamefont {K.}~\bibnamefont
  {He}}, \bibinfo {author} {\bibfnamefont {Y.}~\bibnamefont {Wang}}, \bibinfo
  {author} {\bibfnamefont {L.}~\bibnamefont {Lu}}, \bibinfo {author}
  {\bibfnamefont {X.-C.}\ \bibnamefont {Ma}}, \ and\ \bibinfo {author}
  {\bibfnamefont {Q.-K.}\ \bibnamefont {Xue}},\ }\href@noop {} {\bibfield
  {journal} {\bibinfo  {journal} {Science}\ }\textbf {\bibinfo {volume}
  {340}},\ \bibinfo {pages} {167} (\bibinfo {year} {2013})}\BibitemShut
  {NoStop}%
\bibitem [{\citenamefont {Checkelsky}\ \emph {et~al.}(2014)\citenamefont
  {Checkelsky}, \citenamefont {Yoshimi}, \citenamefont {Tsukazaki},
  \citenamefont {Takahashi}, \citenamefont {Kozuka}, \citenamefont {Falson},
  \citenamefont {Kawasaki},\ and\ \citenamefont {Tokura}}]{qah14natphys}%
  \BibitemOpen
  \bibfield  {author} {\bibinfo {author} {\bibfnamefont {J.~G.}\ \bibnamefont
  {Checkelsky}}, \bibinfo {author} {\bibfnamefont {R.}~\bibnamefont {Yoshimi}},
  \bibinfo {author} {\bibfnamefont {A.}~\bibnamefont {Tsukazaki}}, \bibinfo
  {author} {\bibfnamefont {K.~S.}\ \bibnamefont {Takahashi}}, \bibinfo {author}
  {\bibfnamefont {Y.}~\bibnamefont {Kozuka}}, \bibinfo {author} {\bibfnamefont
  {J.}~\bibnamefont {Falson}}, \bibinfo {author} {\bibfnamefont
  {M.}~\bibnamefont {Kawasaki}}, \ and\ \bibinfo {author} {\bibfnamefont
  {Y.}~\bibnamefont {Tokura}},\ }\href@noop {} {\bibfield  {journal} {\bibinfo
  {journal} {Nature Physics}\ }\textbf {\bibinfo {volume} {10}},\ \bibinfo
  {pages} {731} (\bibinfo {year} {2014})}\BibitemShut {NoStop}%
\bibitem [{\citenamefont {Kou}\ \emph {et~al.}(2014)\citenamefont {Kou},
  \citenamefont {Guo}, \citenamefont {Fan}, \citenamefont {Pan}, \citenamefont
  {Lang}, \citenamefont {Jiang}, \citenamefont {Shao}, \citenamefont {Nie},
  \citenamefont {Murata}, \citenamefont {Tang}, \citenamefont {Wang},
  \citenamefont {He}, \citenamefont {Lee}, \citenamefont {Lee},\ and\
  \citenamefont {Wang}}]{qah14prl}%
  \BibitemOpen
  \bibfield  {author} {\bibinfo {author} {\bibfnamefont {X.}~\bibnamefont
  {Kou}}, \bibinfo {author} {\bibfnamefont {S.-T.}\ \bibnamefont {Guo}},
  \bibinfo {author} {\bibfnamefont {Y.}~\bibnamefont {Fan}}, \bibinfo {author}
  {\bibfnamefont {L.}~\bibnamefont {Pan}}, \bibinfo {author} {\bibfnamefont
  {M.}~\bibnamefont {Lang}}, \bibinfo {author} {\bibfnamefont {Y.}~\bibnamefont
  {Jiang}}, \bibinfo {author} {\bibfnamefont {Q.}~\bibnamefont {Shao}},
  \bibinfo {author} {\bibfnamefont {T.}~\bibnamefont {Nie}}, \bibinfo {author}
  {\bibfnamefont {K.}~\bibnamefont {Murata}}, \bibinfo {author} {\bibfnamefont
  {J.}~\bibnamefont {Tang}}, \bibinfo {author} {\bibfnamefont {Y.}~\bibnamefont
  {Wang}}, \bibinfo {author} {\bibfnamefont {L.}~\bibnamefont {He}}, \bibinfo
  {author} {\bibfnamefont {T.-K.}\ \bibnamefont {Lee}}, \bibinfo {author}
  {\bibfnamefont {W.-L.}\ \bibnamefont {Lee}}, \ and\ \bibinfo {author}
  {\bibfnamefont {K.~L.}\ \bibnamefont {Wang}},\ }\href@noop {} {\bibfield
  {journal} {\bibinfo  {journal} {Physical Review Letters}\ }\textbf {\bibinfo
  {volume} {113}},\ \bibinfo {pages} {137201} (\bibinfo {year}
  {2014})}\BibitemShut {NoStop}%
\bibitem [{\citenamefont {He}\ \emph {et~al.}(2014)\citenamefont {He},
  \citenamefont {Wang},\ and\ \citenamefont {Xue}}]{Xue14nsr}%
  \BibitemOpen
  \bibfield  {author} {\bibinfo {author} {\bibfnamefont {K.}~\bibnamefont
  {He}}, \bibinfo {author} {\bibfnamefont {Y.}~\bibnamefont {Wang}}, \ and\
  \bibinfo {author} {\bibfnamefont {Q.-K.}\ \bibnamefont {Xue}},\ }\href@noop
  {} {\bibfield  {journal} {\bibinfo  {journal} {National Science Review}\
  }\textbf {\bibinfo {volume} {1}},\ \bibinfo {pages} {38} (\bibinfo {year}
  {2014})}\BibitemShut {NoStop}%
\bibitem [{\citenamefont {Hasan}\ and\ \citenamefont {Kane}(2010)}]{Kane10rmp}%
  \BibitemOpen
  \bibfield  {author} {\bibinfo {author} {\bibfnamefont {M.~Z.}\ \bibnamefont
  {Hasan}}\ and\ \bibinfo {author} {\bibfnamefont {C.~L.}\ \bibnamefont
  {Kane}},\ }\href@noop {} {\bibfield  {journal} {\bibinfo  {journal} {Reviews
  of Modern Physics}\ }\textbf {\bibinfo {volume} {82}},\ \bibinfo {pages}
  {3045} (\bibinfo {year} {2010})}\BibitemShut {NoStop}%
\bibitem [{\citenamefont {Qi}\ and\ \citenamefont
  {Zhang}(2011)}]{sczhang11rmp}%
  \BibitemOpen
  \bibfield  {author} {\bibinfo {author} {\bibfnamefont {X.-L.}\ \bibnamefont
  {Qi}}\ and\ \bibinfo {author} {\bibfnamefont {S.-C.}\ \bibnamefont {Zhang}},\
  }\href@noop {} {\bibfield  {journal} {\bibinfo  {journal} {Reviews of Modern
  Physics}\ }\textbf {\bibinfo {volume} {83}},\ \bibinfo {pages} {1057}
  (\bibinfo {year} {2011})}\BibitemShut {NoStop}%
\bibitem [{\citenamefont {Asb{\'o}th}\ \emph {et~al.}(2016)\citenamefont
  {Asb{\'o}th}, \citenamefont {Oroszl{\'a}ny},\ and\ \citenamefont
  {P{\'a}lyi}}]{asboth2016short}%
  \BibitemOpen
  \bibfield  {author} {\bibinfo {author} {\bibfnamefont {J.~K.}\ \bibnamefont
  {Asb{\'o}th}}, \bibinfo {author} {\bibfnamefont {L.}~\bibnamefont
  {Oroszl{\'a}ny}}, \ and\ \bibinfo {author} {\bibfnamefont {A.}~\bibnamefont
  {P{\'a}lyi}},\ }\href@noop {} {\emph {\bibinfo {title} {A Short Course on
  Topological Insulators}}}\ (\bibinfo  {publisher} {Springer},\ \bibinfo
  {year} {2016})\BibitemShut {NoStop}%
\bibitem [{\citenamefont {Yan}\ and\ \citenamefont
  {Felser}(2017)}]{Yan17ARCMP}%
  \BibitemOpen
  \bibfield  {author} {\bibinfo {author} {\bibfnamefont {B.}~\bibnamefont
  {Yan}}\ and\ \bibinfo {author} {\bibfnamefont {C.}~\bibnamefont {Felser}},\
  }\href@noop {} {\bibfield  {journal} {\bibinfo  {journal} {Annu. Rev.
  Condens. Matter Phys.}\ }\textbf {\bibinfo {volume} {8}},\ \bibinfo {pages}
  {337} (\bibinfo {year} {2017})}\BibitemShut {NoStop}%
\bibitem [{\citenamefont {Armitage}\ \emph {et~al.}(2018)\citenamefont
  {Armitage}, \citenamefont {Mele},\ and\ \citenamefont
  {Vishwanath}}]{Armitage18rmp}%
  \BibitemOpen
  \bibfield  {author} {\bibinfo {author} {\bibfnamefont {N.~P.}\ \bibnamefont
  {Armitage}}, \bibinfo {author} {\bibfnamefont {E.~J.}\ \bibnamefont {Mele}},
  \ and\ \bibinfo {author} {\bibfnamefont {A.}~\bibnamefont {Vishwanath}},\
  }\href@noop {} {\bibfield  {journal} {\bibinfo  {journal} {Reviews of Modern
  Physics}\ }\textbf {\bibinfo {volume} {90}},\ \bibinfo {pages} {015001}
  (\bibinfo {year} {2018})}\BibitemShut {NoStop}%
\bibitem [{\citenamefont {Lv}\ \emph {et~al.}(2021)\citenamefont {Lv},
  \citenamefont {Qian},\ and\ \citenamefont {Ding}}]{Lv21rmp}%
  \BibitemOpen
  \bibfield  {author} {\bibinfo {author} {\bibfnamefont {B.~Q.}\ \bibnamefont
  {Lv}}, \bibinfo {author} {\bibfnamefont {T.}~\bibnamefont {Qian}}, \ and\
  \bibinfo {author} {\bibfnamefont {H.}~\bibnamefont {Ding}},\ }\href@noop {}
  {\bibfield  {journal} {\bibinfo  {journal} {Reviews of Modern Physics}\
  }\textbf {\bibinfo {volume} {93}},\ \bibinfo {pages} {025002} (\bibinfo
  {year} {2021})}\BibitemShut {NoStop}%
\bibitem [{\citenamefont {Sato}\ and\ \citenamefont {Ando}(2017)}]{Sato17rpp}%
  \BibitemOpen
  \bibfield  {author} {\bibinfo {author} {\bibfnamefont {M.}~\bibnamefont
  {Sato}}\ and\ \bibinfo {author} {\bibfnamefont {Y.}~\bibnamefont {Ando}},\
  }\href@noop {} {\bibfield  {journal} {\bibinfo  {journal} {Reports on
  Progress in Physics}\ }\textbf {\bibinfo {volume} {80}},\ \bibinfo {pages}
  {076501} (\bibinfo {year} {2017})}\BibitemShut {NoStop}%
\bibitem [{\citenamefont {Sharma}\ \emph {et~al.}(2022)\citenamefont {Sharma},
  \citenamefont {Sharma}, \citenamefont {Karn},\ and\ \citenamefont
  {Awana}}]{Sharma22sst}%
  \BibitemOpen
  \bibfield  {author} {\bibinfo {author} {\bibfnamefont {M.~M.}\ \bibnamefont
  {Sharma}}, \bibinfo {author} {\bibfnamefont {P.}~\bibnamefont {Sharma}},
  \bibinfo {author} {\bibfnamefont {N.~K.}\ \bibnamefont {Karn}}, \ and\
  \bibinfo {author} {\bibfnamefont {V.~P.~S.}\ \bibnamefont {Awana}},\
  }\href@noop {} {\bibfield  {journal} {\bibinfo  {journal} {Superconductor
  Science and Technology}\ }\textbf {\bibinfo {volume} {35}},\ \bibinfo {pages}
  {083003} (\bibinfo {year} {2022})}\BibitemShut {NoStop}%
\bibitem [{\citenamefont {Dalibard}\ \emph {et~al.}(2011)\citenamefont
  {Dalibard}, \citenamefont {Gerbier}, \citenamefont {Juzeliunas},\ and\
  \citenamefont {Ohberg}}]{Dalibard11rmp}%
  \BibitemOpen
  \bibfield  {author} {\bibinfo {author} {\bibfnamefont {J.}~\bibnamefont
  {Dalibard}}, \bibinfo {author} {\bibfnamefont {F.}~\bibnamefont {Gerbier}},
  \bibinfo {author} {\bibfnamefont {G.}~\bibnamefont {Juzeliunas}}, \ and\
  \bibinfo {author} {\bibfnamefont {P.}~\bibnamefont {Ohberg}},\ }\href@noop {}
  {\bibfield  {journal} {\bibinfo  {journal} {Reviews of Modern Physics}\
  }\textbf {\bibinfo {volume} {83}},\ \bibinfo {pages} {1523} (\bibinfo {year}
  {2011})}\BibitemShut {NoStop}%
\bibitem [{\citenamefont {Goldman}\ \emph {et~al.}(2014)\citenamefont
  {Goldman}, \citenamefont {Juzeliunas}, \citenamefont {Ohberg},\ and\
  \citenamefont {Spielman}}]{Goldman14rpp}%
  \BibitemOpen
  \bibfield  {author} {\bibinfo {author} {\bibfnamefont {N.}~\bibnamefont
  {Goldman}}, \bibinfo {author} {\bibfnamefont {G.}~\bibnamefont {Juzeliunas}},
  \bibinfo {author} {\bibfnamefont {P.}~\bibnamefont {Ohberg}}, \ and\ \bibinfo
  {author} {\bibfnamefont {I.~B.}\ \bibnamefont {Spielman}},\ }\href@noop {}
  {\bibfield  {journal} {\bibinfo  {journal} {Reports on Progress in Physics}\
  }\textbf {\bibinfo {volume} {77}},\ \bibinfo {pages} {126401} (\bibinfo
  {year} {2014})}\BibitemShut {NoStop}%
\bibitem [{\citenamefont {Zhai}(2015)}]{Zhai15rpp}%
  \BibitemOpen
  \bibfield  {author} {\bibinfo {author} {\bibfnamefont {H.}~\bibnamefont
  {Zhai}},\ }\href@noop {} {\bibfield  {journal} {\bibinfo  {journal} {Reports
  on Progress in Physics}\ }\textbf {\bibinfo {volume} {78}},\ \bibinfo {pages}
  {026001} (\bibinfo {year} {2015})}\BibitemShut {NoStop}%
\bibitem [{\citenamefont {Zhang}\ and\ \citenamefont
  {Liu}(2018)}]{xjl18bookchapter}%
  \BibitemOpen
  \bibfield  {author} {\bibinfo {author} {\bibfnamefont {L.}~\bibnamefont
  {Zhang}}\ and\ \bibinfo {author} {\bibfnamefont {X.-J.}\ \bibnamefont
  {Liu}},\ }in\ \href@noop {} {\emph {\bibinfo {booktitle} {Synthetic
  Spin-Orbit Coupling in Cold Atoms}}},\ \bibinfo {editor} {edited by\ \bibinfo
  {editor} {\bibfnamefont {W.}~\bibnamefont {Zhang}}, \bibinfo {editor}
  {\bibfnamefont {W.}~\bibnamefont {Yi}}, \ and\ \bibinfo {editor}
  {\bibfnamefont {C.~A. R.~S.}\ \bibnamefont {de~Melo}}}\ (\bibinfo
  {publisher} {World Scientific, Singapore},\ \bibinfo {year} {2018})\ \bibinfo
  {type} {Chapter}~\bibinfo {chapter} {1}, pp.\ \bibinfo {pages}
  {1--87}\BibitemShut {NoStop}%
\bibitem [{\citenamefont {Lin}\ \emph {et~al.}(2011)\citenamefont {Lin},
  \citenamefont {Jimenez-Garcia},\ and\ \citenamefont
  {Spielman}}]{Spielman11nature}%
  \BibitemOpen
  \bibfield  {author} {\bibinfo {author} {\bibfnamefont {Y.-J.}\ \bibnamefont
  {Lin}}, \bibinfo {author} {\bibfnamefont {K.}~\bibnamefont {Jimenez-Garcia}},
  \ and\ \bibinfo {author} {\bibfnamefont {I.~B.}\ \bibnamefont {Spielman}},\
  }\href@noop {} {\bibfield  {journal} {\bibinfo  {journal} {Nature}\ }\textbf
  {\bibinfo {volume} {471}},\ \bibinfo {pages} {83} (\bibinfo {year}
  {2011})}\BibitemShut {NoStop}%
\bibitem [{\citenamefont {Zhang}\ \emph {et~al.}(2012)\citenamefont {Zhang},
  \citenamefont {Ji}, \citenamefont {Chen}, \citenamefont {Zhang},
  \citenamefont {Du}, \citenamefont {Yan}, \citenamefont {Pan}, \citenamefont
  {Zhao}, \citenamefont {Deng}, \citenamefont {Zhai}, \citenamefont {Chen},\
  and\ \citenamefont {Pan}}]{Pan12prl}%
  \BibitemOpen
  \bibfield  {author} {\bibinfo {author} {\bibfnamefont {J.-Y.}\ \bibnamefont
  {Zhang}}, \bibinfo {author} {\bibfnamefont {S.-C.}\ \bibnamefont {Ji}},
  \bibinfo {author} {\bibfnamefont {Z.}~\bibnamefont {Chen}}, \bibinfo {author}
  {\bibfnamefont {L.}~\bibnamefont {Zhang}}, \bibinfo {author} {\bibfnamefont
  {Z.-D.}\ \bibnamefont {Du}}, \bibinfo {author} {\bibfnamefont
  {B.}~\bibnamefont {Yan}}, \bibinfo {author} {\bibfnamefont {G.-S.}\
  \bibnamefont {Pan}}, \bibinfo {author} {\bibfnamefont {B.}~\bibnamefont
  {Zhao}}, \bibinfo {author} {\bibfnamefont {Y.-J.}\ \bibnamefont {Deng}},
  \bibinfo {author} {\bibfnamefont {H.}~\bibnamefont {Zhai}}, \bibinfo {author}
  {\bibfnamefont {S.}~\bibnamefont {Chen}}, \ and\ \bibinfo {author}
  {\bibfnamefont {J.-W.}\ \bibnamefont {Pan}},\ }\href@noop {} {\bibfield
  {journal} {\bibinfo  {journal} {Physical Review Letters}\ }\textbf {\bibinfo
  {volume} {109}},\ \bibinfo {pages} {115301} (\bibinfo {year}
  {2012})}\BibitemShut {NoStop}%
\bibitem [{\citenamefont {Beeler}\ \emph {et~al.}(2013)\citenamefont {Beeler},
  \citenamefont {Williams}, \citenamefont {Jimenez-Garcia}, \citenamefont
  {LeBlanc}, \citenamefont {Perry},\ and\ \citenamefont
  {Spielman}}]{Spielman13natureSHE}%
  \BibitemOpen
  \bibfield  {author} {\bibinfo {author} {\bibfnamefont {M.~C.}\ \bibnamefont
  {Beeler}}, \bibinfo {author} {\bibfnamefont {R.~A.}\ \bibnamefont
  {Williams}}, \bibinfo {author} {\bibfnamefont {K.}~\bibnamefont
  {Jimenez-Garcia}}, \bibinfo {author} {\bibfnamefont {L.~J.}\ \bibnamefont
  {LeBlanc}}, \bibinfo {author} {\bibfnamefont {A.~R.}\ \bibnamefont {Perry}},
  \ and\ \bibinfo {author} {\bibfnamefont {I.~B.}\ \bibnamefont {Spielman}},\
  }\href@noop {} {\bibfield  {journal} {\bibinfo  {journal} {Nature}\ }\textbf
  {\bibinfo {volume} {498}},\ \bibinfo {pages} {201} (\bibinfo {year}
  {2013})}\BibitemShut {NoStop}%
\bibitem [{\citenamefont {Ji}\ \emph {et~al.}(2014)\citenamefont {Ji},
  \citenamefont {Zhang}, \citenamefont {Zhang}, \citenamefont {Du},
  \citenamefont {Zhang}, \citenamefont {Deng}, \citenamefont {Zhai},
  \citenamefont {Chen},\ and\ \citenamefont {Pan}}]{Pan14np}%
  \BibitemOpen
  \bibfield  {author} {\bibinfo {author} {\bibfnamefont {S.-C.}\ \bibnamefont
  {Ji}}, \bibinfo {author} {\bibfnamefont {J.-Y.}\ \bibnamefont {Zhang}},
  \bibinfo {author} {\bibfnamefont {L.}~\bibnamefont {Zhang}}, \bibinfo
  {author} {\bibfnamefont {Z.-D.}\ \bibnamefont {Du}}, \bibinfo {author}
  {\bibfnamefont {W.}~\bibnamefont {Zhang}}, \bibinfo {author} {\bibfnamefont
  {Y.-J.}\ \bibnamefont {Deng}}, \bibinfo {author} {\bibfnamefont
  {H.}~\bibnamefont {Zhai}}, \bibinfo {author} {\bibfnamefont {S.}~\bibnamefont
  {Chen}}, \ and\ \bibinfo {author} {\bibfnamefont {J.-W.}\ \bibnamefont
  {Pan}},\ }\href@noop {} {\bibfield  {journal} {\bibinfo  {journal} {Nature
  Physics}\ }\textbf {\bibinfo {volume} {10}},\ \bibinfo {pages} {314}
  (\bibinfo {year} {2014})}\BibitemShut {NoStop}%
\bibitem [{\citenamefont {Hamner}\ \emph {et~al.}(2014)\citenamefont {Hamner},
  \citenamefont {Qu}, \citenamefont {Zhang}, \citenamefont {Chang},
  \citenamefont {Gong}, \citenamefont {Zhang},\ and\ \citenamefont
  {Engels}}]{Engels14nc}%
  \BibitemOpen
  \bibfield  {author} {\bibinfo {author} {\bibfnamefont {C.}~\bibnamefont
  {Hamner}}, \bibinfo {author} {\bibfnamefont {C.}~\bibnamefont {Qu}}, \bibinfo
  {author} {\bibfnamefont {Y.}~\bibnamefont {Zhang}}, \bibinfo {author}
  {\bibfnamefont {J.}~\bibnamefont {Chang}}, \bibinfo {author} {\bibfnamefont
  {M.}~\bibnamefont {Gong}}, \bibinfo {author} {\bibfnamefont {C.}~\bibnamefont
  {Zhang}}, \ and\ \bibinfo {author} {\bibfnamefont {P.}~\bibnamefont
  {Engels}},\ }\href@noop {} {\bibfield  {journal} {\bibinfo  {journal} {Nature
  Communications}\ }\textbf {\bibinfo {volume} {5}},\ \bibinfo {pages} {4023}
  (\bibinfo {year} {2014})}\BibitemShut {NoStop}%
\bibitem [{\citenamefont {Olson}\ \emph {et~al.}(2014)\citenamefont {Olson},
  \citenamefont {Wang}, \citenamefont {Niffenegger}, \citenamefont {Li},
  \citenamefont {Greene},\ and\ \citenamefont {Chen}}]{ChenYong14pra}%
  \BibitemOpen
  \bibfield  {author} {\bibinfo {author} {\bibfnamefont {A.~J.}\ \bibnamefont
  {Olson}}, \bibinfo {author} {\bibfnamefont {S.-J.}\ \bibnamefont {Wang}},
  \bibinfo {author} {\bibfnamefont {R.~J.}\ \bibnamefont {Niffenegger}},
  \bibinfo {author} {\bibfnamefont {C.-H.}\ \bibnamefont {Li}}, \bibinfo
  {author} {\bibfnamefont {C.~H.}\ \bibnamefont {Greene}}, \ and\ \bibinfo
  {author} {\bibfnamefont {Y.~P.}\ \bibnamefont {Chen}},\ }\href@noop {}
  {\bibfield  {journal} {\bibinfo  {journal} {Physical Review A}\ }\textbf
  {\bibinfo {volume} {90}},\ \bibinfo {pages} {013616} (\bibinfo {year}
  {2014})}\BibitemShut {NoStop}%
\bibitem [{\citenamefont {Atala}\ \emph {et~al.}(2014)\citenamefont {Atala},
  \citenamefont {Aidelsburger}, \citenamefont {Lohse}, \citenamefont
  {Barreiro}, \citenamefont {Paredes},\ and\ \citenamefont
  {Bloch}}]{Bloch14natphys}%
  \BibitemOpen
  \bibfield  {author} {\bibinfo {author} {\bibfnamefont {M.}~\bibnamefont
  {Atala}}, \bibinfo {author} {\bibfnamefont {M.}~\bibnamefont {Aidelsburger}},
  \bibinfo {author} {\bibfnamefont {M.}~\bibnamefont {Lohse}}, \bibinfo
  {author} {\bibfnamefont {J.~T.}\ \bibnamefont {Barreiro}}, \bibinfo {author}
  {\bibfnamefont {B.}~\bibnamefont {Paredes}}, \ and\ \bibinfo {author}
  {\bibfnamefont {I.}~\bibnamefont {Bloch}},\ }\href@noop {} {\bibfield
  {journal} {\bibinfo  {journal} {Nature Physics}\ }\textbf {\bibinfo {volume}
  {10}},\ \bibinfo {pages} {588} (\bibinfo {year} {2014})}\BibitemShut
  {NoStop}%
\bibitem [{\citenamefont {Li}\ \emph {et~al.}(2016)\citenamefont {Li},
  \citenamefont {Huang}, \citenamefont {Shteynas}, \citenamefont {Burchesky},
  \citenamefont {Top}, \citenamefont {Su}, \citenamefont {Lee}, \citenamefont
  {Jamison},\ and\ \citenamefont {Ketterle}}]{Ketterle16prl}%
  \BibitemOpen
  \bibfield  {author} {\bibinfo {author} {\bibfnamefont {J.}~\bibnamefont
  {Li}}, \bibinfo {author} {\bibfnamefont {W.}~\bibnamefont {Huang}}, \bibinfo
  {author} {\bibfnamefont {B.}~\bibnamefont {Shteynas}}, \bibinfo {author}
  {\bibfnamefont {S.}~\bibnamefont {Burchesky}}, \bibinfo {author}
  {\bibfnamefont {F.~C.}\ \bibnamefont {Top}}, \bibinfo {author} {\bibfnamefont
  {E.}~\bibnamefont {Su}}, \bibinfo {author} {\bibfnamefont {J.}~\bibnamefont
  {Lee}}, \bibinfo {author} {\bibfnamefont {A.~O.}\ \bibnamefont {Jamison}}, \
  and\ \bibinfo {author} {\bibfnamefont {W.}~\bibnamefont {Ketterle}},\
  }\href@noop {} {\bibfield  {journal} {\bibinfo  {journal} {Physical Review
  Letters}\ }\textbf {\bibinfo {volume} {117}},\ \bibinfo {pages} {185301}
  (\bibinfo {year} {2016})}\BibitemShut {NoStop}%
\bibitem [{\citenamefont {Li}\ \emph {et~al.}(2017)\citenamefont {Li},
  \citenamefont {Lee}, \citenamefont {Huang}, \citenamefont {Burchesky},
  \citenamefont {Shteynas}, \citenamefont {Top}, \citenamefont {Jamison},\ and\
  \citenamefont {Ketterle}}]{Ketterle17nature}%
  \BibitemOpen
  \bibfield  {author} {\bibinfo {author} {\bibfnamefont {J.-R.}\ \bibnamefont
  {Li}}, \bibinfo {author} {\bibfnamefont {J.}~\bibnamefont {Lee}}, \bibinfo
  {author} {\bibfnamefont {W.}~\bibnamefont {Huang}}, \bibinfo {author}
  {\bibfnamefont {S.}~\bibnamefont {Burchesky}}, \bibinfo {author}
  {\bibfnamefont {B.}~\bibnamefont {Shteynas}}, \bibinfo {author}
  {\bibfnamefont {F.~C.}\ \bibnamefont {Top}}, \bibinfo {author} {\bibfnamefont
  {A.~O.}\ \bibnamefont {Jamison}}, \ and\ \bibinfo {author} {\bibfnamefont
  {W.}~\bibnamefont {Ketterle}},\ }\href@noop {} {\bibfield  {journal}
  {\bibinfo  {journal} {Nature}\ }\textbf {\bibinfo {volume} {543}},\ \bibinfo
  {pages} {91} (\bibinfo {year} {2017})}\BibitemShut {NoStop}%
\bibitem [{\citenamefont {Valdes-Curiel}\ \emph {et~al.}(2017)\citenamefont
  {Valdes-Curiel}, \citenamefont {Trypogeorgos}, \citenamefont {Marshall},\
  and\ \citenamefont {Spielman}}]{Spielman17njp}%
  \BibitemOpen
  \bibfield  {author} {\bibinfo {author} {\bibfnamefont {A.}~\bibnamefont
  {Valdes-Curiel}}, \bibinfo {author} {\bibfnamefont {D.}~\bibnamefont
  {Trypogeorgos}}, \bibinfo {author} {\bibfnamefont {E.~E.}\ \bibnamefont
  {Marshall}}, \ and\ \bibinfo {author} {\bibfnamefont {I.~B.}\ \bibnamefont
  {Spielman}},\ }\href@noop {} {\bibfield  {journal} {\bibinfo  {journal} {New
  Journal of Physics}\ }\textbf {\bibinfo {volume} {19}},\ \bibinfo {pages}
  {033025} (\bibinfo {year} {2017})}\BibitemShut {NoStop}%
\bibitem [{\citenamefont {Chen}\ \emph {et~al.}(2018)\citenamefont {Chen},
  \citenamefont {Lin}, \citenamefont {Chen}, \citenamefont {Chiu},
  \citenamefont {Wang}, \citenamefont {Chen}, \citenamefont {Huang},
  \citenamefont {Yip}, \citenamefont {Kawaguchi},\ and\ \citenamefont
  {Lin}}]{Lin18prlsoac}%
  \BibitemOpen
  \bibfield  {author} {\bibinfo {author} {\bibfnamefont {H.-R.}\ \bibnamefont
  {Chen}}, \bibinfo {author} {\bibfnamefont {K.-Y.}\ \bibnamefont {Lin}},
  \bibinfo {author} {\bibfnamefont {P.-K.}\ \bibnamefont {Chen}}, \bibinfo
  {author} {\bibfnamefont {N.-C.}\ \bibnamefont {Chiu}}, \bibinfo {author}
  {\bibfnamefont {J.-B.}\ \bibnamefont {Wang}}, \bibinfo {author}
  {\bibfnamefont {C.-A.}\ \bibnamefont {Chen}}, \bibinfo {author}
  {\bibfnamefont {P.}~\bibnamefont {Huang}}, \bibinfo {author} {\bibfnamefont
  {S.-K.}\ \bibnamefont {Yip}}, \bibinfo {author} {\bibfnamefont
  {Y.}~\bibnamefont {Kawaguchi}}, \ and\ \bibinfo {author} {\bibfnamefont
  {Y.-J.}\ \bibnamefont {Lin}},\ }\href@noop {} {\bibfield  {journal} {\bibinfo
   {journal} {Physical Review Letters}\ }\textbf {\bibinfo {volume} {121}},\
  \bibinfo {pages} {113204} (\bibinfo {year} {2018})}\BibitemShut {NoStop}%
\bibitem [{\citenamefont {Zhang}\ \emph {et~al.}(2019)\citenamefont {Zhang},
  \citenamefont {Gao}, \citenamefont {Zou}, \citenamefont {Kong}, \citenamefont
  {Li}, \citenamefont {Shen}, \citenamefont {Chen}, \citenamefont {Peng},
  \citenamefont {Zhan}, \citenamefont {Pu},\ and\ \citenamefont
  {Jiang}}]{Jiang19prlsoac}%
  \BibitemOpen
  \bibfield  {author} {\bibinfo {author} {\bibfnamefont {D.}~\bibnamefont
  {Zhang}}, \bibinfo {author} {\bibfnamefont {T.}~\bibnamefont {Gao}}, \bibinfo
  {author} {\bibfnamefont {P.}~\bibnamefont {Zou}}, \bibinfo {author}
  {\bibfnamefont {L.}~\bibnamefont {Kong}}, \bibinfo {author} {\bibfnamefont
  {R.}~\bibnamefont {Li}}, \bibinfo {author} {\bibfnamefont {X.}~\bibnamefont
  {Shen}}, \bibinfo {author} {\bibfnamefont {X.-L.}\ \bibnamefont {Chen}},
  \bibinfo {author} {\bibfnamefont {S.-G.}\ \bibnamefont {Peng}}, \bibinfo
  {author} {\bibfnamefont {M.}~\bibnamefont {Zhan}}, \bibinfo {author}
  {\bibfnamefont {H.}~\bibnamefont {Pu}}, \ and\ \bibinfo {author}
  {\bibfnamefont {K.}~\bibnamefont {Jiang}},\ }\href@noop {} {\bibfield
  {journal} {\bibinfo  {journal} {Physical Review Letters}\ }\textbf {\bibinfo
  {volume} {122}},\ \bibinfo {pages} {110402} (\bibinfo {year}
  {2019})}\BibitemShut {NoStop}%
\bibitem [{\citenamefont {Wang}\ \emph {et~al.}(2012)\citenamefont {Wang},
  \citenamefont {Yu}, \citenamefont {Fu}, \citenamefont {Miao}, \citenamefont
  {Huang}, \citenamefont {Chai}, \citenamefont {Zhai},\ and\ \citenamefont
  {Zhang}}]{jing12prl}%
  \BibitemOpen
  \bibfield  {author} {\bibinfo {author} {\bibfnamefont {P.}~\bibnamefont
  {Wang}}, \bibinfo {author} {\bibfnamefont {Z.-Q.}\ \bibnamefont {Yu}},
  \bibinfo {author} {\bibfnamefont {Z.}~\bibnamefont {Fu}}, \bibinfo {author}
  {\bibfnamefont {J.}~\bibnamefont {Miao}}, \bibinfo {author} {\bibfnamefont
  {L.}~\bibnamefont {Huang}}, \bibinfo {author} {\bibfnamefont
  {S.}~\bibnamefont {Chai}}, \bibinfo {author} {\bibfnamefont {H.}~\bibnamefont
  {Zhai}}, \ and\ \bibinfo {author} {\bibfnamefont {J.}~\bibnamefont {Zhang}},\
  }\href@noop {} {\bibfield  {journal} {\bibinfo  {journal} {Physical Review
  Letters}\ }\textbf {\bibinfo {volume} {109}},\ \bibinfo {pages} {095301}
  (\bibinfo {year} {2012})}\BibitemShut {NoStop}%
\bibitem [{\citenamefont {Cheuk}\ \emph {et~al.}(2012)\citenamefont {Cheuk},
  \citenamefont {Sommer}, \citenamefont {Hadzibabic}, \citenamefont {Yefsah},
  \citenamefont {Bakr},\ and\ \citenamefont {Zwierlein}}]{mz12prl}%
  \BibitemOpen
  \bibfield  {author} {\bibinfo {author} {\bibfnamefont {L.~W.}\ \bibnamefont
  {Cheuk}}, \bibinfo {author} {\bibfnamefont {A.~T.}\ \bibnamefont {Sommer}},
  \bibinfo {author} {\bibfnamefont {Z.}~\bibnamefont {Hadzibabic}}, \bibinfo
  {author} {\bibfnamefont {T.}~\bibnamefont {Yefsah}}, \bibinfo {author}
  {\bibfnamefont {W.~S.}\ \bibnamefont {Bakr}}, \ and\ \bibinfo {author}
  {\bibfnamefont {M.~W.}\ \bibnamefont {Zwierlein}},\ }\href@noop {} {\bibfield
   {journal} {\bibinfo  {journal} {Physical Review Letters}\ }\textbf {\bibinfo
  {volume} {109}},\ \bibinfo {pages} {095302} (\bibinfo {year}
  {2012})}\BibitemShut {NoStop}%
\bibitem [{\citenamefont {Williams}\ \emph {et~al.}(2013)\citenamefont
  {Williams}, \citenamefont {Beeler}, \citenamefont {LeBlanc}, \citenamefont
  {Jimenez-Garcia},\ and\ \citenamefont {Spielman}}]{Spielman13prl}%
  \BibitemOpen
  \bibfield  {author} {\bibinfo {author} {\bibfnamefont {R.~A.}\ \bibnamefont
  {Williams}}, \bibinfo {author} {\bibfnamefont {M.~C.}\ \bibnamefont
  {Beeler}}, \bibinfo {author} {\bibfnamefont {L.~J.}\ \bibnamefont {LeBlanc}},
  \bibinfo {author} {\bibfnamefont {K.}~\bibnamefont {Jimenez-Garcia}}, \ and\
  \bibinfo {author} {\bibfnamefont {I.~B.}\ \bibnamefont {Spielman}},\
  }\href@noop {} {\bibfield  {journal} {\bibinfo  {journal} {Physical Review
  Letters}\ }\textbf {\bibinfo {volume} {111}},\ \bibinfo {pages} {095301}
  (\bibinfo {year} {2013})}\BibitemShut {NoStop}%
\bibitem [{\citenamefont {Zhang}\ \emph {et~al.}(2014)\citenamefont {Zhang},
  \citenamefont {Hu}, \citenamefont {Liu},\ and\ \citenamefont
  {Pu}}]{jingzhang14review}%
  \BibitemOpen
  \bibfield  {author} {\bibinfo {author} {\bibfnamefont {J.}~\bibnamefont
  {Zhang}}, \bibinfo {author} {\bibfnamefont {H.}~\bibnamefont {Hu}}, \bibinfo
  {author} {\bibfnamefont {X.-J.}\ \bibnamefont {Liu}}, \ and\ \bibinfo
  {author} {\bibfnamefont {H.}~\bibnamefont {Pu}},\ }\href@noop {} {\bibfield
  {journal} {\bibinfo  {journal} {Annual Review of Cold Atoms and Molecules}\
  }\textbf {\bibinfo {volume} {2}},\ \bibinfo {pages} {81} (\bibinfo {year}
  {2014})}\BibitemShut {NoStop}%
\bibitem [{\citenamefont {Burdick}\ \emph {et~al.}(2016)\citenamefont
  {Burdick}, \citenamefont {Tang},\ and\ \citenamefont {Lev}}]{Lev16prx}%
  \BibitemOpen
  \bibfield  {author} {\bibinfo {author} {\bibfnamefont {N.~Q.}\ \bibnamefont
  {Burdick}}, \bibinfo {author} {\bibfnamefont {Y.}~\bibnamefont {Tang}}, \
  and\ \bibinfo {author} {\bibfnamefont {B.~L.}\ \bibnamefont {Lev}},\
  }\href@noop {} {\bibfield  {journal} {\bibinfo  {journal} {Physical Review
  X}\ }\textbf {\bibinfo {volume} {6}},\ \bibinfo {pages} {031022} (\bibinfo
  {year} {2016})}\BibitemShut {NoStop}%
\bibitem [{\citenamefont {Livi}\ \emph {et~al.}(2016)\citenamefont {Livi},
  \citenamefont {Cappellini}, \citenamefont {Diem}, \citenamefont {Franchi},
  \citenamefont {Clivati}, \citenamefont {Frittelli}, \citenamefont {Levi},
  \citenamefont {Calonico}, \citenamefont {Catani}, \citenamefont {Inguscio},\
  and\ \citenamefont {Fallani}}]{livi16prl}%
  \BibitemOpen
  \bibfield  {author} {\bibinfo {author} {\bibfnamefont {L.~F.}\ \bibnamefont
  {Livi}}, \bibinfo {author} {\bibfnamefont {G.}~\bibnamefont {Cappellini}},
  \bibinfo {author} {\bibfnamefont {M.}~\bibnamefont {Diem}}, \bibinfo {author}
  {\bibfnamefont {L.}~\bibnamefont {Franchi}}, \bibinfo {author} {\bibfnamefont
  {C.}~\bibnamefont {Clivati}}, \bibinfo {author} {\bibfnamefont
  {M.}~\bibnamefont {Frittelli}}, \bibinfo {author} {\bibfnamefont
  {F.}~\bibnamefont {Levi}}, \bibinfo {author} {\bibfnamefont {D.}~\bibnamefont
  {Calonico}}, \bibinfo {author} {\bibfnamefont {J.}~\bibnamefont {Catani}},
  \bibinfo {author} {\bibfnamefont {M.}~\bibnamefont {Inguscio}}, \ and\
  \bibinfo {author} {\bibfnamefont {L.}~\bibnamefont {Fallani}},\ }\href@noop
  {} {\bibfield  {journal} {\bibinfo  {journal} {Physical Review Letters}\
  }\textbf {\bibinfo {volume} {117}},\ \bibinfo {pages} {220401} (\bibinfo
  {year} {2016})}\BibitemShut {NoStop}%
\bibitem [{\citenamefont {Song}\ \emph {et~al.}(2016)\citenamefont {Song},
  \citenamefont {He}, \citenamefont {Zhang}, \citenamefont {Hajiyev},
  \citenamefont {Huang}, \citenamefont {Liu},\ and\ \citenamefont
  {Jo}}]{Jo16pra}%
  \BibitemOpen
  \bibfield  {author} {\bibinfo {author} {\bibfnamefont {B.}~\bibnamefont
  {Song}}, \bibinfo {author} {\bibfnamefont {C.}~\bibnamefont {He}}, \bibinfo
  {author} {\bibfnamefont {S.}~\bibnamefont {Zhang}}, \bibinfo {author}
  {\bibfnamefont {E.}~\bibnamefont {Hajiyev}}, \bibinfo {author} {\bibfnamefont
  {W.}~\bibnamefont {Huang}}, \bibinfo {author} {\bibfnamefont {X.-J.}\
  \bibnamefont {Liu}}, \ and\ \bibinfo {author} {\bibfnamefont {G.-B.}\
  \bibnamefont {Jo}},\ }\href@noop {} {\bibfield  {journal} {\bibinfo
  {journal} {Physical Review A}\ }\textbf {\bibinfo {volume} {94}},\ \bibinfo
  {pages} {061604(R)} (\bibinfo {year} {2016})}\BibitemShut {NoStop}%
\bibitem [{\citenamefont {Kolkowitz}\ \emph {et~al.}(2017)\citenamefont
  {Kolkowitz}, \citenamefont {Bromley}, \citenamefont {Bothwell}, \citenamefont
  {Wall}, \citenamefont {Marti}, \citenamefont {Koller}, \citenamefont {Zhang},
  \citenamefont {Rey},\ and\ \citenamefont {Ye}}]{shimon17nature}%
  \BibitemOpen
  \bibfield  {author} {\bibinfo {author} {\bibfnamefont {S.}~\bibnamefont
  {Kolkowitz}}, \bibinfo {author} {\bibfnamefont {S.~L.}\ \bibnamefont
  {Bromley}}, \bibinfo {author} {\bibfnamefont {T.}~\bibnamefont {Bothwell}},
  \bibinfo {author} {\bibfnamefont {M.~L.}\ \bibnamefont {Wall}}, \bibinfo
  {author} {\bibfnamefont {G.~E.}\ \bibnamefont {Marti}}, \bibinfo {author}
  {\bibfnamefont {A.~P.}\ \bibnamefont {Koller}}, \bibinfo {author}
  {\bibfnamefont {X.}~\bibnamefont {Zhang}}, \bibinfo {author} {\bibfnamefont
  {A.~M.}\ \bibnamefont {Rey}}, \ and\ \bibinfo {author} {\bibfnamefont
  {J.}~\bibnamefont {Ye}},\ }\href@noop {} {\bibfield  {journal} {\bibinfo
  {journal} {Nature}\ }\textbf {\bibinfo {volume} {542}},\ \bibinfo {pages}
  {66} (\bibinfo {year} {2017})}\BibitemShut {NoStop}%
\bibitem [{\citenamefont {Jaksch}\ and\ \citenamefont
  {Zoller}(2003)}]{Zoller03njp}%
  \BibitemOpen
  \bibfield  {author} {\bibinfo {author} {\bibfnamefont {D.}~\bibnamefont
  {Jaksch}}\ and\ \bibinfo {author} {\bibfnamefont {P.}~\bibnamefont
  {Zoller}},\ }\href@noop {} {\bibfield  {journal} {\bibinfo  {journal} {New
  Journal of Physics}\ }\textbf {\bibinfo {volume} {5}},\ \bibinfo {pages} {56}
  (\bibinfo {year} {2003})}\BibitemShut {NoStop}%
\bibitem [{\citenamefont {Juzeliunas}\ and\ \citenamefont
  {Ohberg}(2004)}]{Ohberg04prl}%
  \BibitemOpen
  \bibfield  {author} {\bibinfo {author} {\bibfnamefont {G.}~\bibnamefont
  {Juzeliunas}}\ and\ \bibinfo {author} {\bibfnamefont {P.}~\bibnamefont
  {Ohberg}},\ }\href@noop {} {\bibfield  {journal} {\bibinfo  {journal}
  {Physical Review Letters}\ }\textbf {\bibinfo {volume} {93}},\ \bibinfo
  {pages} {033602} (\bibinfo {year} {2004})}\BibitemShut {NoStop}%
\bibitem [{\citenamefont {Liu}\ \emph {et~al.}(2006)\citenamefont {Liu},
  \citenamefont {Jing}, \citenamefont {Liu},\ and\ \citenamefont
  {Ge}}]{xiongjunliu06epjd}%
  \BibitemOpen
  \bibfield  {author} {\bibinfo {author} {\bibfnamefont {X.-J.}\ \bibnamefont
  {Liu}}, \bibinfo {author} {\bibfnamefont {H.}~\bibnamefont {Jing}}, \bibinfo
  {author} {\bibfnamefont {X.}~\bibnamefont {Liu}}, \ and\ \bibinfo {author}
  {\bibfnamefont {M.-L.}\ \bibnamefont {Ge}},\ }\href@noop {} {\bibfield
  {journal} {\bibinfo  {journal} {The European Physical Journal D}\ }\textbf
  {\bibinfo {volume} {37}},\ \bibinfo {pages} {261} (\bibinfo {year}
  {2006})}\BibitemShut {NoStop}%
\bibitem [{\citenamefont {Zhu}\ \emph {et~al.}(2006)\citenamefont {Zhu},
  \citenamefont {Fu}, \citenamefont {Wu}, \citenamefont {Zhang},\ and\
  \citenamefont {Duan}}]{duan06prl}%
  \BibitemOpen
  \bibfield  {author} {\bibinfo {author} {\bibfnamefont {S.-L.}\ \bibnamefont
  {Zhu}}, \bibinfo {author} {\bibfnamefont {H.}~\bibnamefont {Fu}}, \bibinfo
  {author} {\bibfnamefont {C.-J.}\ \bibnamefont {Wu}}, \bibinfo {author}
  {\bibfnamefont {S.-C.}\ \bibnamefont {Zhang}}, \ and\ \bibinfo {author}
  {\bibfnamefont {L.-M.}\ \bibnamefont {Duan}},\ }\href@noop {} {\bibfield
  {journal} {\bibinfo  {journal} {Physical Review Letters}\ }\textbf {\bibinfo
  {volume} {97}},\ \bibinfo {pages} {240401} (\bibinfo {year}
  {2006})}\BibitemShut {NoStop}%
\bibitem [{\citenamefont {Liu}\ \emph {et~al.}(2007)\citenamefont {Liu},
  \citenamefont {Liu}, \citenamefont {Kwek},\ and\ \citenamefont
  {Oh}}]{xjl07prl}%
  \BibitemOpen
  \bibfield  {author} {\bibinfo {author} {\bibfnamefont {X.-J.}\ \bibnamefont
  {Liu}}, \bibinfo {author} {\bibfnamefont {X.}~\bibnamefont {Liu}}, \bibinfo
  {author} {\bibfnamefont {L.~C.}\ \bibnamefont {Kwek}}, \ and\ \bibinfo
  {author} {\bibfnamefont {C.~H.}\ \bibnamefont {Oh}},\ }\href@noop {}
  {\bibfield  {journal} {\bibinfo  {journal} {Physical Review Letters}\
  }\textbf {\bibinfo {volume} {98}},\ \bibinfo {pages} {026602} (\bibinfo
  {year} {2007})}\BibitemShut {NoStop}%
\bibitem [{\citenamefont {Liu}\ \emph {et~al.}(2009)\citenamefont {Liu},
  \citenamefont {Borunda}, \citenamefont {Liu},\ and\ \citenamefont
  {Sinova}}]{xjl09prl}%
  \BibitemOpen
  \bibfield  {author} {\bibinfo {author} {\bibfnamefont {X.-J.}\ \bibnamefont
  {Liu}}, \bibinfo {author} {\bibfnamefont {M.~F.}\ \bibnamefont {Borunda}},
  \bibinfo {author} {\bibfnamefont {X.}~\bibnamefont {Liu}}, \ and\ \bibinfo
  {author} {\bibfnamefont {J.}~\bibnamefont {Sinova}},\ }\href@noop {}
  {\bibfield  {journal} {\bibinfo  {journal} {Physical Review Letters}\
  }\textbf {\bibinfo {volume} {102}},\ \bibinfo {pages} {046402} (\bibinfo
  {year} {2009})}\BibitemShut {NoStop}%
\bibitem [{\citenamefont {Spielman}(2009)}]{Spielman09pra}%
  \BibitemOpen
  \bibfield  {author} {\bibinfo {author} {\bibfnamefont {I.~B.}\ \bibnamefont
  {Spielman}},\ }\href@noop {} {\bibfield  {journal} {\bibinfo  {journal}
  {Physical Review A}\ }\textbf {\bibinfo {volume} {79}},\ \bibinfo {pages}
  {063613} (\bibinfo {year} {2009})}\BibitemShut {NoStop}%
\bibitem [{\citenamefont {Lin}\ \emph {et~al.}(2009)\citenamefont {Lin},
  \citenamefont {Compton}, \citenamefont {Jimenez-Garcia}, \citenamefont
  {Porto},\ and\ \citenamefont {Spielman}}]{Spielman09nature}%
  \BibitemOpen
  \bibfield  {author} {\bibinfo {author} {\bibfnamefont {Y.-J.}\ \bibnamefont
  {Lin}}, \bibinfo {author} {\bibfnamefont {R.~L.}\ \bibnamefont {Compton}},
  \bibinfo {author} {\bibfnamefont {K.}~\bibnamefont {Jimenez-Garcia}},
  \bibinfo {author} {\bibfnamefont {J.~V.}\ \bibnamefont {Porto}}, \ and\
  \bibinfo {author} {\bibfnamefont {I.~B.}\ \bibnamefont {Spielman}},\
  }\href@noop {} {\bibfield  {journal} {\bibinfo  {journal} {Nature}\ }\textbf
  {\bibinfo {volume} {462}},\ \bibinfo {pages} {628} (\bibinfo {year}
  {2009})}\BibitemShut {NoStop}%
\bibitem [{\citenamefont {Campbell}\ \emph {et~al.}(2011)\citenamefont
  {Campbell}, \citenamefont {Juzeliunas},\ and\ \citenamefont
  {Spielman}}]{Spielman11pra}%
  \BibitemOpen
  \bibfield  {author} {\bibinfo {author} {\bibfnamefont {D.~L.}\ \bibnamefont
  {Campbell}}, \bibinfo {author} {\bibfnamefont {G.}~\bibnamefont
  {Juzeliunas}}, \ and\ \bibinfo {author} {\bibfnamefont {I.~B.}\ \bibnamefont
  {Spielman}},\ }\href@noop {} {\bibfield  {journal} {\bibinfo  {journal}
  {Physical Review A}\ }\textbf {\bibinfo {volume} {84}},\ \bibinfo {pages}
  {025602} (\bibinfo {year} {2011})}\BibitemShut {NoStop}%
\bibitem [{\citenamefont {Galitski}\ and\ \citenamefont
  {Spielman}(2013)}]{Spielman13nature}%
  \BibitemOpen
  \bibfield  {author} {\bibinfo {author} {\bibfnamefont {V.}~\bibnamefont
  {Galitski}}\ and\ \bibinfo {author} {\bibfnamefont {I.~B.}\ \bibnamefont
  {Spielman}},\ }\href@noop {} {\bibfield  {journal} {\bibinfo  {journal}
  {Nature}\ }\textbf {\bibinfo {volume} {494}},\ \bibinfo {pages} {49}
  (\bibinfo {year} {2013})}\BibitemShut {NoStop}%
\bibitem [{\citenamefont {Liu}\ \emph {et~al.}(2014)\citenamefont {Liu},
  \citenamefont {Law},\ and\ \citenamefont {Ng}}]{liu14prl}%
  \BibitemOpen
  \bibfield  {author} {\bibinfo {author} {\bibfnamefont {X.-J.}\ \bibnamefont
  {Liu}}, \bibinfo {author} {\bibfnamefont {K.~T.}\ \bibnamefont {Law}}, \ and\
  \bibinfo {author} {\bibfnamefont {T.~K.}\ \bibnamefont {Ng}},\ }\href@noop {}
  {\bibfield  {journal} {\bibinfo  {journal} {Physical Review Letters}\
  }\textbf {\bibinfo {volume} {112}},\ \bibinfo {pages} {086401} (\bibinfo
  {year} {2014})}\BibitemShut {NoStop}%
\bibitem [{\citenamefont {Mancini}\ \emph {et~al.}(2015)\citenamefont
  {Mancini}, \citenamefont {Pagano}, \citenamefont {Cappellini}, \citenamefont
  {Livi}, \citenamefont {Rider}, \citenamefont {Catani}, \citenamefont {Sias},
  \citenamefont {Zoller}, \citenamefont {Inguscio}, \citenamefont {Dalmonte},\
  and\ \citenamefont {Fallani}}]{Mancini15science}%
  \BibitemOpen
  \bibfield  {author} {\bibinfo {author} {\bibfnamefont {M.}~\bibnamefont
  {Mancini}}, \bibinfo {author} {\bibfnamefont {G.}~\bibnamefont {Pagano}},
  \bibinfo {author} {\bibfnamefont {G.}~\bibnamefont {Cappellini}}, \bibinfo
  {author} {\bibfnamefont {L.}~\bibnamefont {Livi}}, \bibinfo {author}
  {\bibfnamefont {M.}~\bibnamefont {Rider}}, \bibinfo {author} {\bibfnamefont
  {J.}~\bibnamefont {Catani}}, \bibinfo {author} {\bibfnamefont
  {C.}~\bibnamefont {Sias}}, \bibinfo {author} {\bibfnamefont {P.}~\bibnamefont
  {Zoller}}, \bibinfo {author} {\bibfnamefont {M.}~\bibnamefont {Inguscio}},
  \bibinfo {author} {\bibfnamefont {M.}~\bibnamefont {Dalmonte}}, \ and\
  \bibinfo {author} {\bibfnamefont {L.}~\bibnamefont {Fallani}},\ }\href@noop
  {} {\bibfield  {journal} {\bibinfo  {journal} {Science}\ }\textbf {\bibinfo
  {volume} {349}},\ \bibinfo {pages} {1510} (\bibinfo {year}
  {2015})}\BibitemShut {NoStop}%
\bibitem [{\citenamefont {Stuhl}\ \emph {et~al.}(2015)\citenamefont {Stuhl},
  \citenamefont {Lu}, \citenamefont {Aycock}, \citenamefont {Genkina},\ and\
  \citenamefont {Spielman}}]{Spielman15science}%
  \BibitemOpen
  \bibfield  {author} {\bibinfo {author} {\bibfnamefont {B.~K.}\ \bibnamefont
  {Stuhl}}, \bibinfo {author} {\bibfnamefont {H.-I.}\ \bibnamefont {Lu}},
  \bibinfo {author} {\bibfnamefont {L.~M.}\ \bibnamefont {Aycock}}, \bibinfo
  {author} {\bibfnamefont {D.}~\bibnamefont {Genkina}}, \ and\ \bibinfo
  {author} {\bibfnamefont {I.~B.}\ \bibnamefont {Spielman}},\ }\href@noop {}
  {\bibfield  {journal} {\bibinfo  {journal} {Science}\ }\textbf {\bibinfo
  {volume} {349}},\ \bibinfo {pages} {1514} (\bibinfo {year}
  {2015})}\BibitemShut {NoStop}%
\bibitem [{\citenamefont {An}\ \emph {et~al.}(2017)\citenamefont {An},
  \citenamefont {Meier},\ and\ \citenamefont {Gadway}}]{Gadway17sciadv}%
  \BibitemOpen
  \bibfield  {author} {\bibinfo {author} {\bibfnamefont {F.~A.}\ \bibnamefont
  {An}}, \bibinfo {author} {\bibfnamefont {E.~J.}\ \bibnamefont {Meier}}, \
  and\ \bibinfo {author} {\bibfnamefont {B.}~\bibnamefont {Gadway}},\
  }\href@noop {} {\bibfield  {journal} {\bibinfo  {journal} {Science Advances}\
  }\textbf {\bibinfo {volume} {3}},\ \bibinfo {pages} {e1602685} (\bibinfo
  {year} {2017})}\BibitemShut {NoStop}%
\bibitem [{\citenamefont {Wang}\ \emph {et~al.}(2018)\citenamefont {Wang},
  \citenamefont {Lu}, \citenamefont {Sun}, \citenamefont {Chen}, \citenamefont
  {Deng},\ and\ \citenamefont {Liu}}]{liu18pra}%
  \BibitemOpen
  \bibfield  {author} {\bibinfo {author} {\bibfnamefont {B.-Z.}\ \bibnamefont
  {Wang}}, \bibinfo {author} {\bibfnamefont {Y.-H.}\ \bibnamefont {Lu}},
  \bibinfo {author} {\bibfnamefont {W.}~\bibnamefont {Sun}}, \bibinfo {author}
  {\bibfnamefont {S.}~\bibnamefont {Chen}}, \bibinfo {author} {\bibfnamefont
  {Y.}~\bibnamefont {Deng}}, \ and\ \bibinfo {author} {\bibfnamefont {X.-J.}\
  \bibnamefont {Liu}},\ }\href@noop {} {\bibfield  {journal} {\bibinfo
  {journal} {Phys. Rev. A}\ }\textbf {\bibinfo {volume} {97}},\ \bibinfo
  {pages} {011605(R)} (\bibinfo {year} {2018})}\BibitemShut {NoStop}%
\bibitem [{\citenamefont {Peng}\ \emph {et~al.}(2022)\citenamefont {Peng},
  \citenamefont {Jiang}, \citenamefont {Chen}, \citenamefont {Chen},
  \citenamefont {Zou},\ and\ \citenamefont {He}}]{Peng22aappsb}%
  \BibitemOpen
  \bibfield  {author} {\bibinfo {author} {\bibfnamefont {S.-G.}\ \bibnamefont
  {Peng}}, \bibinfo {author} {\bibfnamefont {K.}~\bibnamefont {Jiang}},
  \bibinfo {author} {\bibfnamefont {X.-L.}\ \bibnamefont {Chen}}, \bibinfo
  {author} {\bibfnamefont {K.-J.}\ \bibnamefont {Chen}}, \bibinfo {author}
  {\bibfnamefont {P.}~\bibnamefont {Zou}}, \ and\ \bibinfo {author}
  {\bibfnamefont {L.}~\bibnamefont {He}},\ }\href@noop {} {\bibfield  {journal}
  {\bibinfo  {journal} {AAPPS Bulletin}\ }\textbf {\bibinfo {volume} {32}},\
  \bibinfo {pages} {36} (\bibinfo {year} {2022})}\BibitemShut {NoStop}%
\bibitem [{\citenamefont {Huang}\ \emph {et~al.}(2016)\citenamefont {Huang},
  \citenamefont {Meng}, \citenamefont {Wang}, \citenamefont {Peng},
  \citenamefont {Zhang}, \citenamefont {Chen}, \citenamefont {Li},
  \citenamefont {Zhou},\ and\ \citenamefont {Zhang}}]{JingZhang16natphys}%
  \BibitemOpen
  \bibfield  {author} {\bibinfo {author} {\bibfnamefont {L.}~\bibnamefont
  {Huang}}, \bibinfo {author} {\bibfnamefont {Z.}~\bibnamefont {Meng}},
  \bibinfo {author} {\bibfnamefont {P.}~\bibnamefont {Wang}}, \bibinfo {author}
  {\bibfnamefont {P.}~\bibnamefont {Peng}}, \bibinfo {author} {\bibfnamefont
  {S.-L.}\ \bibnamefont {Zhang}}, \bibinfo {author} {\bibfnamefont
  {L.}~\bibnamefont {Chen}}, \bibinfo {author} {\bibfnamefont {D.}~\bibnamefont
  {Li}}, \bibinfo {author} {\bibfnamefont {Q.}~\bibnamefont {Zhou}}, \ and\
  \bibinfo {author} {\bibfnamefont {J.}~\bibnamefont {Zhang}},\ }\href@noop {}
  {\bibfield  {journal} {\bibinfo  {journal} {Nature Physics}\ }\textbf
  {\bibinfo {volume} {12}},\ \bibinfo {pages} {540} (\bibinfo {year}
  {2016})}\BibitemShut {NoStop}%
\bibitem [{\citenamefont {Wu}\ \emph {et~al.}(2016)\citenamefont {Wu},
  \citenamefont {Zhang}, \citenamefont {Sun}, \citenamefont {Xu}, \citenamefont
  {Wang}, \citenamefont {Ji}, \citenamefont {Deng}, \citenamefont {Chen},
  \citenamefont {Liu},\ and\ \citenamefont {Pan}}]{shuai16science}%
  \BibitemOpen
  \bibfield  {author} {\bibinfo {author} {\bibfnamefont {Z.}~\bibnamefont
  {Wu}}, \bibinfo {author} {\bibfnamefont {L.}~\bibnamefont {Zhang}}, \bibinfo
  {author} {\bibfnamefont {W.}~\bibnamefont {Sun}}, \bibinfo {author}
  {\bibfnamefont {X.-T.}\ \bibnamefont {Xu}}, \bibinfo {author} {\bibfnamefont
  {B.-Z.}\ \bibnamefont {Wang}}, \bibinfo {author} {\bibfnamefont {S.-C.}\
  \bibnamefont {Ji}}, \bibinfo {author} {\bibfnamefont {Y.}~\bibnamefont
  {Deng}}, \bibinfo {author} {\bibfnamefont {S.}~\bibnamefont {Chen}}, \bibinfo
  {author} {\bibfnamefont {X.-J.}\ \bibnamefont {Liu}}, \ and\ \bibinfo
  {author} {\bibfnamefont {J.-W.}\ \bibnamefont {Pan}},\ }\href@noop {}
  {\bibfield  {journal} {\bibinfo  {journal} {Science}\ }\textbf {\bibinfo
  {volume} {354}},\ \bibinfo {pages} {83} (\bibinfo {year} {2016})}\BibitemShut
  {NoStop}%
\bibitem [{\citenamefont {Meng}\ \emph {et~al.}(2016)\citenamefont {Meng},
  \citenamefont {Huang}, \citenamefont {Peng}, \citenamefont {Li},
  \citenamefont {Chen}, \citenamefont {Xu}, \citenamefont {Zhang},
  \citenamefont {Wang},\ and\ \citenamefont {Zhang}}]{JingZhang16prl}%
  \BibitemOpen
  \bibfield  {author} {\bibinfo {author} {\bibfnamefont {Z.}~\bibnamefont
  {Meng}}, \bibinfo {author} {\bibfnamefont {L.}~\bibnamefont {Huang}},
  \bibinfo {author} {\bibfnamefont {P.}~\bibnamefont {Peng}}, \bibinfo {author}
  {\bibfnamefont {D.}~\bibnamefont {Li}}, \bibinfo {author} {\bibfnamefont
  {L.}~\bibnamefont {Chen}}, \bibinfo {author} {\bibfnamefont {Y.}~\bibnamefont
  {Xu}}, \bibinfo {author} {\bibfnamefont {C.}~\bibnamefont {Zhang}}, \bibinfo
  {author} {\bibfnamefont {P.}~\bibnamefont {Wang}}, \ and\ \bibinfo {author}
  {\bibfnamefont {J.}~\bibnamefont {Zhang}},\ }\href@noop {} {\bibfield
  {journal} {\bibinfo  {journal} {Physical Review Letters}\ }\textbf {\bibinfo
  {volume} {117}},\ \bibinfo {pages} {235304} (\bibinfo {year}
  {2016})}\BibitemShut {NoStop}%
\bibitem [{\citenamefont {Song}\ \emph {et~al.}(2019)\citenamefont {Song},
  \citenamefont {He}, \citenamefont {Niu}, \citenamefont {Zhang}, \citenamefont
  {Ren}, \citenamefont {Liu},\ and\ \citenamefont {Jo}}]{jo19natphys}%
  \BibitemOpen
  \bibfield  {author} {\bibinfo {author} {\bibfnamefont {B.}~\bibnamefont
  {Song}}, \bibinfo {author} {\bibfnamefont {C.}~\bibnamefont {He}}, \bibinfo
  {author} {\bibfnamefont {S.}~\bibnamefont {Niu}}, \bibinfo {author}
  {\bibfnamefont {L.}~\bibnamefont {Zhang}}, \bibinfo {author} {\bibfnamefont
  {Z.}~\bibnamefont {Ren}}, \bibinfo {author} {\bibfnamefont {X.-J.}\
  \bibnamefont {Liu}}, \ and\ \bibinfo {author} {\bibfnamefont {G.-B.}\
  \bibnamefont {Jo}},\ }\href@noop {} {\bibfield  {journal} {\bibinfo
  {journal} {Nature Physics}\ }\textbf {\bibinfo {volume} {15}},\ \bibinfo
  {pages} {911} (\bibinfo {year} {2019})}\BibitemShut {NoStop}%
\bibitem [{\citenamefont {Wang}\ \emph {et~al.}(2021)\citenamefont {Wang},
  \citenamefont {Cheng}, \citenamefont {Wang}, \citenamefont {Zhang},
  \citenamefont {Lu}, \citenamefont {Yi}, \citenamefont {Niu}, \citenamefont
  {Deng}, \citenamefont {Liu}, \citenamefont {Chen},\ and\ \citenamefont
  {Pan}}]{xjl21science}%
  \BibitemOpen
  \bibfield  {author} {\bibinfo {author} {\bibfnamefont {Z.-Y.}\ \bibnamefont
  {Wang}}, \bibinfo {author} {\bibfnamefont {X.-C.}\ \bibnamefont {Cheng}},
  \bibinfo {author} {\bibfnamefont {B.-Z.}\ \bibnamefont {Wang}}, \bibinfo
  {author} {\bibfnamefont {J.-Y.}\ \bibnamefont {Zhang}}, \bibinfo {author}
  {\bibfnamefont {Y.-H.}\ \bibnamefont {Lu}}, \bibinfo {author} {\bibfnamefont
  {C.-R.}\ \bibnamefont {Yi}}, \bibinfo {author} {\bibfnamefont
  {S.}~\bibnamefont {Niu}}, \bibinfo {author} {\bibfnamefont {Y.}~\bibnamefont
  {Deng}}, \bibinfo {author} {\bibfnamefont {X.-J.}\ \bibnamefont {Liu}},
  \bibinfo {author} {\bibfnamefont {S.}~\bibnamefont {Chen}}, \ and\ \bibinfo
  {author} {\bibfnamefont {J.-W.}\ \bibnamefont {Pan}},\ }\href@noop {}
  {\bibfield  {journal} {\bibinfo  {journal} {Science}\ }\textbf {\bibinfo
  {volume} {372}},\ \bibinfo {pages} {271} (\bibinfo {year}
  {2021})}\BibitemShut {NoStop}%
\bibitem [{\citenamefont {Li}\ and\ \citenamefont
  {Liu}(2021)}]{LiLiu21scibull}%
  \BibitemOpen
  \bibfield  {author} {\bibinfo {author} {\bibfnamefont {X.}~\bibnamefont
  {Li}}\ and\ \bibinfo {author} {\bibfnamefont {W.~V.}\ \bibnamefont {Liu}},\
  }\href@noop {} {\bibfield  {journal} {\bibinfo  {journal} {Science Bulletin}\
  }\textbf {\bibinfo {volume} {66}},\ \bibinfo {pages} {1253} (\bibinfo {year}
  {2021})}\BibitemShut {NoStop}%
\bibitem [{\citenamefont {Valdes-Curiel}\ \emph {et~al.}(2021)\citenamefont
  {Valdes-Curiel}, \citenamefont {Trypogeorgos}, \citenamefont {Liang},
  \citenamefont {Anderson},\ and\ \citenamefont {Spielman}}]{Spielman21nc}%
  \BibitemOpen
  \bibfield  {author} {\bibinfo {author} {\bibfnamefont {A.}~\bibnamefont
  {Valdes-Curiel}}, \bibinfo {author} {\bibfnamefont {D.}~\bibnamefont
  {Trypogeorgos}}, \bibinfo {author} {\bibfnamefont {Q.-Y.}\ \bibnamefont
  {Liang}}, \bibinfo {author} {\bibfnamefont {R.~P.}\ \bibnamefont {Anderson}},
  \ and\ \bibinfo {author} {\bibfnamefont {I.~B.}\ \bibnamefont {Spielman}},\
  }\href@noop {} {\bibfield  {journal} {\bibinfo  {journal} {Nature
  Communications}\ }\textbf {\bibinfo {volume} {12}},\ \bibinfo {pages} {593}
  (\bibinfo {year} {2021})}\BibitemShut {NoStop}%
\bibitem [{\citenamefont {Lauria}\ \emph {et~al.}(2022)\citenamefont {Lauria},
  \citenamefont {Kuo}, \citenamefont {Cooper},\ and\ \citenamefont
  {Barreiro}}]{barreiro22prl}%
  \BibitemOpen
  \bibfield  {author} {\bibinfo {author} {\bibfnamefont {P.}~\bibnamefont
  {Lauria}}, \bibinfo {author} {\bibfnamefont {W.-T.}\ \bibnamefont {Kuo}},
  \bibinfo {author} {\bibfnamefont {N.~R.}\ \bibnamefont {Cooper}}, \ and\
  \bibinfo {author} {\bibfnamefont {J.~T.}\ \bibnamefont {Barreiro}},\
  }\href@noop {} {\bibfield  {journal} {\bibinfo  {journal} {Physical Review
  Letters}\ }\textbf {\bibinfo {volume} {128}},\ \bibinfo {pages} {245301}
  (\bibinfo {year} {2022})}\BibitemShut {NoStop}%
\bibitem [{\citenamefont {Liang}\ \emph {et~al.}(2023)\citenamefont {Liang},
  \citenamefont {Wei}, \citenamefont {Zhang}, \citenamefont {Wang},
  \citenamefont {Zhang}, \citenamefont {Wang}, \citenamefont {Qi},
  \citenamefont {Liu},\ and\ \citenamefont {Zhang}}]{Liang21arxiv}%
  \BibitemOpen
  \bibfield  {author} {\bibinfo {author} {\bibfnamefont {M.-C.}\ \bibnamefont
  {Liang}}, \bibinfo {author} {\bibfnamefont {Y.-D.}\ \bibnamefont {Wei}},
  \bibinfo {author} {\bibfnamefont {L.}~\bibnamefont {Zhang}}, \bibinfo
  {author} {\bibfnamefont {X.-J.}\ \bibnamefont {Wang}}, \bibinfo {author}
  {\bibfnamefont {H.}~\bibnamefont {Zhang}}, \bibinfo {author} {\bibfnamefont
  {W.-W.}\ \bibnamefont {Wang}}, \bibinfo {author} {\bibfnamefont
  {W.}~\bibnamefont {Qi}}, \bibinfo {author} {\bibfnamefont {X.-J.}\
  \bibnamefont {Liu}}, \ and\ \bibinfo {author} {\bibfnamefont
  {X.}~\bibnamefont {Zhang}},\ }\href@noop {} {\bibfield  {journal} {\bibinfo
  {journal} {Physical Review Research}\ }\textbf {\bibinfo {volume} {5}},\
  \bibinfo {pages} {L012006} (\bibinfo {year} {2023})}\BibitemShut {NoStop}%
\bibitem [{\citenamefont {Jaynes}\ and\ \citenamefont
  {Cummings}(1963)}]{JaynesCummings63ProcIEEE}%
  \BibitemOpen
  \bibfield  {author} {\bibinfo {author} {\bibfnamefont {E.~T.}\ \bibnamefont
  {Jaynes}}\ and\ \bibinfo {author} {\bibfnamefont {F.~W.}\ \bibnamefont
  {Cummings}},\ }\href@noop {} {\bibfield  {journal} {\bibinfo  {journal}
  {Proceedings of the IEEE}\ }\textbf {\bibinfo {volume} {51}},\ \bibinfo
  {pages} {89} (\bibinfo {year} {1963})}\BibitemShut {NoStop}%
\bibitem [{\citenamefont {Shore}\ and\ \citenamefont
  {Knight}(1993)}]{Knight93JMO}%
  \BibitemOpen
  \bibfield  {author} {\bibinfo {author} {\bibfnamefont {B.~W.}\ \bibnamefont
  {Shore}}\ and\ \bibinfo {author} {\bibfnamefont {P.~L.}\ \bibnamefont
  {Knight}},\ }\href@noop {} {\bibfield  {journal} {\bibinfo  {journal}
  {Journal of Modern Optics}\ }\textbf {\bibinfo {volume} {40}},\ \bibinfo
  {pages} {1195} (\bibinfo {year} {1993})}\BibitemShut {NoStop}%
\bibitem [{\citenamefont {Allen}\ and\ \citenamefont
  {Eberly}(1975)}]{AE75twolevelbook}%
  \BibitemOpen
  \bibfield  {author} {\bibinfo {author} {\bibfnamefont {L.}~\bibnamefont
  {Allen}}\ and\ \bibinfo {author} {\bibfnamefont {J.~H.}\ \bibnamefont
  {Eberly}},\ }\href@noop {} {\emph {\bibinfo {title} {Optical resonance and
  two-level atoms}}}\ (\bibinfo  {publisher} {John Wiley \& Sons, Inc.},\
  \bibinfo {address} {New York},\ \bibinfo {year} {1975})\BibitemShut {NoStop}%
\bibitem [{\citenamefont {Foot}(2005)}]{Foot05atomicbook}%
  \BibitemOpen
  \bibfield  {author} {\bibinfo {author} {\bibfnamefont {C.~J.}\ \bibnamefont
  {Foot}},\ }\href@noop {} {\emph {\bibinfo {title} {Atomic Physics}}}\
  (\bibinfo  {publisher} {Oxford University Press},\ \bibinfo {address} {Great
  Clarendon Street, Oxford OX2 6DP},\ \bibinfo {year} {2005})\BibitemShut
  {NoStop}%
\bibitem [{\citenamefont {Zhai}(2021)}]{HuiZhai21ultracoldbook}%
  \BibitemOpen
  \bibfield  {author} {\bibinfo {author} {\bibfnamefont {H.}~\bibnamefont
  {Zhai}},\ }\href@noop {} {\emph {\bibinfo {title} {Ultracold Atomic
  Physics}}}\ (\bibinfo  {publisher} {Cambridge University Press},\ \bibinfo
  {address} {University Printing House, Cambridge CB2 8BS, United Kingdom},\
  \bibinfo {year} {2021})\BibitemShut {NoStop}%
\bibitem [{\citenamefont {Fuchs}\ \emph {et~al.}(2009)\citenamefont {Fuchs},
  \citenamefont {Dobrovitski}, \citenamefont {Toyli}, \citenamefont
  {Heremans},\ and\ \citenamefont {Awschalom}}]{Awschalom09science}%
  \BibitemOpen
  \bibfield  {author} {\bibinfo {author} {\bibfnamefont {G.~D.}\ \bibnamefont
  {Fuchs}}, \bibinfo {author} {\bibfnamefont {V.~V.}\ \bibnamefont
  {Dobrovitski}}, \bibinfo {author} {\bibfnamefont {D.~M.}\ \bibnamefont
  {Toyli}}, \bibinfo {author} {\bibfnamefont {F.~J.}\ \bibnamefont {Heremans}},
  \ and\ \bibinfo {author} {\bibfnamefont {D.~D.}\ \bibnamefont {Awschalom}},\
  }\href@noop {} {\bibfield  {journal} {\bibinfo  {journal} {Science}\ }\textbf
  {\bibinfo {volume} {326}},\ \bibinfo {pages} {1520} (\bibinfo {year}
  {2009})}\BibitemShut {NoStop}%
\bibitem [{\citenamefont {Bloch}\ and\ \citenamefont
  {Siegert}(1940)}]{BlochSiegert1940pr}%
  \BibitemOpen
  \bibfield  {author} {\bibinfo {author} {\bibfnamefont {F.}~\bibnamefont
  {Bloch}}\ and\ \bibinfo {author} {\bibfnamefont {A.}~\bibnamefont
  {Siegert}},\ }\href@noop {} {\bibfield  {journal} {\bibinfo  {journal}
  {Physical Review}\ }\textbf {\bibinfo {volume} {57}},\ \bibinfo {pages} {522}
  (\bibinfo {year} {1940})}\BibitemShut {NoStop}%
\bibitem [{\citenamefont {Milonni}\ \emph {et~al.}(1983)\citenamefont
  {Milonni}, \citenamefont {Ackerhalt},\ and\ \citenamefont
  {Galbraith}}]{Galbraith1983prl}%
  \BibitemOpen
  \bibfield  {author} {\bibinfo {author} {\bibfnamefont {P.~W.}\ \bibnamefont
  {Milonni}}, \bibinfo {author} {\bibfnamefont {J.~R.}\ \bibnamefont
  {Ackerhalt}}, \ and\ \bibinfo {author} {\bibfnamefont {H.~W.}\ \bibnamefont
  {Galbraith}},\ }\href@noop {} {\bibfield  {journal} {\bibinfo  {journal}
  {Physical Review Letters}\ }\textbf {\bibinfo {volume} {50}},\ \bibinfo
  {pages} {966} (\bibinfo {year} {1983})}\BibitemShut {NoStop}%
\bibitem [{\citenamefont {Ng}\ and\ \citenamefont
  {Burnett}(2008)}]{Burnett08njp}%
  \BibitemOpen
  \bibfield  {author} {\bibinfo {author} {\bibfnamefont {H.~T.}\ \bibnamefont
  {Ng}}\ and\ \bibinfo {author} {\bibfnamefont {K.}~\bibnamefont {Burnett}},\
  }\href@noop {} {\bibfield  {journal} {\bibinfo  {journal} {New Journal of
  Physics}\ }\textbf {\bibinfo {volume} {10}},\ \bibinfo {pages} {123014}
  (\bibinfo {year} {2008})}\BibitemShut {NoStop}%
\bibitem [{\citenamefont {Kitagawa}\ \emph {et~al.}(2012)\citenamefont
  {Kitagawa}, \citenamefont {Broome}, \citenamefont {Fedrizzi}, \citenamefont
  {Rudner}, \citenamefont {Berg}, \citenamefont {Kassal}, \citenamefont
  {Aspuru-Guzik}, \citenamefont {Demler},\ and\ \citenamefont
  {White}}]{White12nc}%
  \BibitemOpen
  \bibfield  {author} {\bibinfo {author} {\bibfnamefont {T.}~\bibnamefont
  {Kitagawa}}, \bibinfo {author} {\bibfnamefont {M.~A.}\ \bibnamefont
  {Broome}}, \bibinfo {author} {\bibfnamefont {A.}~\bibnamefont {Fedrizzi}},
  \bibinfo {author} {\bibfnamefont {M.~S.}\ \bibnamefont {Rudner}}, \bibinfo
  {author} {\bibfnamefont {E.}~\bibnamefont {Berg}}, \bibinfo {author}
  {\bibfnamefont {I.}~\bibnamefont {Kassal}}, \bibinfo {author} {\bibfnamefont
  {A.}~\bibnamefont {Aspuru-Guzik}}, \bibinfo {author} {\bibfnamefont
  {E.}~\bibnamefont {Demler}}, \ and\ \bibinfo {author} {\bibfnamefont {A.~G.}\
  \bibnamefont {White}},\ }\href@noop {} {\bibfield  {journal} {\bibinfo
  {journal} {Nature Communications}\ }\textbf {\bibinfo {volume} {3}},\
  \bibinfo {pages} {882} (\bibinfo {year} {2012})}\BibitemShut {NoStop}%
\bibitem [{\citenamefont {Hu}\ \emph {et~al.}(2015)\citenamefont {Hu},
  \citenamefont {Pillay}, \citenamefont {Wu}, \citenamefont {Pasek},
  \citenamefont {Shum},\ and\ \citenamefont {Chong}}]{Chong15prx}%
  \BibitemOpen
  \bibfield  {author} {\bibinfo {author} {\bibfnamefont {W.}~\bibnamefont
  {Hu}}, \bibinfo {author} {\bibfnamefont {J.~C.}\ \bibnamefont {Pillay}},
  \bibinfo {author} {\bibfnamefont {K.}~\bibnamefont {Wu}}, \bibinfo {author}
  {\bibfnamefont {M.}~\bibnamefont {Pasek}}, \bibinfo {author} {\bibfnamefont
  {P.~P.}\ \bibnamefont {Shum}}, \ and\ \bibinfo {author} {\bibfnamefont
  {Y.~D.}\ \bibnamefont {Chong}},\ }\href@noop {} {\bibfield  {journal}
  {\bibinfo  {journal} {Physical Review X}\ }\textbf {\bibinfo {volume} {5}},\
  \bibinfo {pages} {011012} (\bibinfo {year} {2015})}\BibitemShut {NoStop}%
\bibitem [{\citenamefont {Maczewsky}\ \emph {et~al.}(2017)\citenamefont
  {Maczewsky}, \citenamefont {Zeuner}, \citenamefont {Nolte},\ and\
  \citenamefont {Szameit}}]{Szameit17nc}%
  \BibitemOpen
  \bibfield  {author} {\bibinfo {author} {\bibfnamefont {L.~J.}\ \bibnamefont
  {Maczewsky}}, \bibinfo {author} {\bibfnamefont {J.~M.}\ \bibnamefont
  {Zeuner}}, \bibinfo {author} {\bibfnamefont {S.}~\bibnamefont {Nolte}}, \
  and\ \bibinfo {author} {\bibfnamefont {A.}~\bibnamefont {Szameit}},\
  }\href@noop {} {\bibfield  {journal} {\bibinfo  {journal} {Nature
  Communications}\ }\textbf {\bibinfo {volume} {8}},\ \bibinfo {pages} {13756}
  (\bibinfo {year} {2017})}\BibitemShut {NoStop}%
\bibitem [{\citenamefont {Mukherjee}\ \emph {et~al.}(2017)\citenamefont
  {Mukherjee}, \citenamefont {Spracklen}, \citenamefont {Valiente},
  \citenamefont {Andersson}, \citenamefont {Ohberg}, \citenamefont {Goldman},\
  and\ \citenamefont {Thomson}}]{Thomson17nc}%
  \BibitemOpen
  \bibfield  {author} {\bibinfo {author} {\bibfnamefont {S.}~\bibnamefont
  {Mukherjee}}, \bibinfo {author} {\bibfnamefont {A.}~\bibnamefont
  {Spracklen}}, \bibinfo {author} {\bibfnamefont {M.}~\bibnamefont {Valiente}},
  \bibinfo {author} {\bibfnamefont {E.}~\bibnamefont {Andersson}}, \bibinfo
  {author} {\bibfnamefont {P.}~\bibnamefont {Ohberg}}, \bibinfo {author}
  {\bibfnamefont {N.}~\bibnamefont {Goldman}}, \ and\ \bibinfo {author}
  {\bibfnamefont {R.~R.}\ \bibnamefont {Thomson}},\ }\href@noop {} {\bibfield
  {journal} {\bibinfo  {journal} {Nature Communications}\ }\textbf {\bibinfo
  {volume} {8}},\ \bibinfo {pages} {13918} (\bibinfo {year}
  {2017})}\BibitemShut {NoStop}%
\bibitem [{\citenamefont {D'Errico}\ \emph {et~al.}(2020)\citenamefont
  {D'Errico}, \citenamefont {Cardano}, \citenamefont {Maffei}, \citenamefont
  {Dauphin}, \citenamefont {Barboza}, \citenamefont {Esposito}, \citenamefont
  {Piccirillo}, \citenamefont {Lewenstein}, \citenamefont {Massignan},\ and\
  \citenamefont {Marrucci}}]{Marrucci20optica}%
  \BibitemOpen
  \bibfield  {author} {\bibinfo {author} {\bibfnamefont {A.}~\bibnamefont
  {D'Errico}}, \bibinfo {author} {\bibfnamefont {F.}~\bibnamefont {Cardano}},
  \bibinfo {author} {\bibfnamefont {M.}~\bibnamefont {Maffei}}, \bibinfo
  {author} {\bibfnamefont {A.}~\bibnamefont {Dauphin}}, \bibinfo {author}
  {\bibfnamefont {R.}~\bibnamefont {Barboza}}, \bibinfo {author} {\bibfnamefont
  {C.}~\bibnamefont {Esposito}}, \bibinfo {author} {\bibfnamefont
  {B.}~\bibnamefont {Piccirillo}}, \bibinfo {author} {\bibfnamefont
  {M.}~\bibnamefont {Lewenstein}}, \bibinfo {author} {\bibfnamefont
  {P.}~\bibnamefont {Massignan}}, \ and\ \bibinfo {author} {\bibfnamefont
  {L.}~\bibnamefont {Marrucci}},\ }\href@noop {} {\bibfield  {journal}
  {\bibinfo  {journal} {Optica}\ }\textbf {\bibinfo {volume} {7}},\ \bibinfo
  {pages} {108} (\bibinfo {year} {2020})}\BibitemShut {NoStop}%
\bibitem [{\citenamefont {Peng}\ \emph {et~al.}(2016)\citenamefont {Peng},
  \citenamefont {Qin}, \citenamefont {Zhao}, \citenamefont {Shen},
  \citenamefont {Xu}, \citenamefont {Bao}, \citenamefont {Jia},\ and\
  \citenamefont {Zhu}}]{Zhu16nc}%
  \BibitemOpen
  \bibfield  {author} {\bibinfo {author} {\bibfnamefont {Y.-G.}\ \bibnamefont
  {Peng}}, \bibinfo {author} {\bibfnamefont {C.-Z.}\ \bibnamefont {Qin}},
  \bibinfo {author} {\bibfnamefont {D.-G.}\ \bibnamefont {Zhao}}, \bibinfo
  {author} {\bibfnamefont {Y.-X.}\ \bibnamefont {Shen}}, \bibinfo {author}
  {\bibfnamefont {X.-Y.}\ \bibnamefont {Xu}}, \bibinfo {author} {\bibfnamefont
  {M.}~\bibnamefont {Bao}}, \bibinfo {author} {\bibfnamefont {H.}~\bibnamefont
  {Jia}}, \ and\ \bibinfo {author} {\bibfnamefont {X.-F.}\ \bibnamefont
  {Zhu}},\ }\href@noop {} {\bibfield  {journal} {\bibinfo  {journal} {Nature
  Communications}\ }\textbf {\bibinfo {volume} {7}},\ \bibinfo {pages} {13368}
  (\bibinfo {year} {2016})}\BibitemShut {NoStop}%
\bibitem [{\citenamefont {Wintersperger}\ \emph {et~al.}(2020)\citenamefont
  {Wintersperger}, \citenamefont {Braun}, \citenamefont {Unal}, \citenamefont
  {Eckardt}, \citenamefont {Liberto}, \citenamefont {Goldman}, \citenamefont
  {Bloch},\ and\ \citenamefont {Aidelsburger}}]{Bloch20natphysAF}%
  \BibitemOpen
  \bibfield  {author} {\bibinfo {author} {\bibfnamefont {K.}~\bibnamefont
  {Wintersperger}}, \bibinfo {author} {\bibfnamefont {C.}~\bibnamefont
  {Braun}}, \bibinfo {author} {\bibfnamefont {F.~N.}\ \bibnamefont {Unal}},
  \bibinfo {author} {\bibfnamefont {A.}~\bibnamefont {Eckardt}}, \bibinfo
  {author} {\bibfnamefont {M.~D.}\ \bibnamefont {Liberto}}, \bibinfo {author}
  {\bibfnamefont {N.}~\bibnamefont {Goldman}}, \bibinfo {author} {\bibfnamefont
  {I.}~\bibnamefont {Bloch}}, \ and\ \bibinfo {author} {\bibfnamefont
  {M.}~\bibnamefont {Aidelsburger}},\ }\href@noop {} {\bibfield  {journal}
  {\bibinfo  {journal} {Nature Physics}\ }\textbf {\bibinfo {volume} {16}},\
  \bibinfo {pages} {1058} (\bibinfo {year} {2020})}\BibitemShut {NoStop}%
\bibitem [{\citenamefont {Zhang}\ \emph {et~al.}(2023)\citenamefont {Zhang},
  \citenamefont {Yi}, \citenamefont {Zhang}, \citenamefont {Jiao},
  \citenamefont {Shi}, \citenamefont {Yuan}, \citenamefont {Zhang},
  \citenamefont {Liu}, \citenamefont {Chen},\ and\ \citenamefont
  {Pan}}]{Shuai22FloquetArxiv}%
  \BibitemOpen
  \bibfield  {author} {\bibinfo {author} {\bibfnamefont {J.-Y.}\ \bibnamefont
  {Zhang}}, \bibinfo {author} {\bibfnamefont {C.~R.}\ \bibnamefont {Yi}},
  \bibinfo {author} {\bibfnamefont {L.}~\bibnamefont {Zhang}}, \bibinfo
  {author} {\bibfnamefont {R.-H.}\ \bibnamefont {Jiao}}, \bibinfo {author}
  {\bibfnamefont {K.-Y.}\ \bibnamefont {Shi}}, \bibinfo {author} {\bibfnamefont
  {H.}~\bibnamefont {Yuan}}, \bibinfo {author} {\bibfnamefont {W.}~\bibnamefont
  {Zhang}}, \bibinfo {author} {\bibfnamefont {X.-J.}\ \bibnamefont {Liu}},
  \bibinfo {author} {\bibfnamefont {S.}~\bibnamefont {Chen}}, \ and\ \bibinfo
  {author} {\bibfnamefont {J.-W.}\ \bibnamefont {Pan}},\ }\href@noop {}
  {\bibfield  {journal} {\bibinfo  {journal} {Physical Review Letters}\
  }\textbf {\bibinfo {volume} {130}},\ \bibinfo {pages} {043201} (\bibinfo
  {year} {2023})}\BibitemShut {NoStop}%
\bibitem [{\citenamefont {Kitagawa}\ \emph {et~al.}(2010)\citenamefont
  {Kitagawa}, \citenamefont {Berg}, \citenamefont {Rudner},\ and\ \citenamefont
  {Demler}}]{Demler10prb}%
  \BibitemOpen
  \bibfield  {author} {\bibinfo {author} {\bibfnamefont {T.}~\bibnamefont
  {Kitagawa}}, \bibinfo {author} {\bibfnamefont {E.}~\bibnamefont {Berg}},
  \bibinfo {author} {\bibfnamefont {M.}~\bibnamefont {Rudner}}, \ and\ \bibinfo
  {author} {\bibfnamefont {E.}~\bibnamefont {Demler}},\ }\href@noop {}
  {\bibfield  {journal} {\bibinfo  {journal} {Physical Review B}\ }\textbf
  {\bibinfo {volume} {82}},\ \bibinfo {pages} {235114} (\bibinfo {year}
  {2010})}\BibitemShut {NoStop}%
\bibitem [{\citenamefont {Rudner}\ \emph {et~al.}(2013)\citenamefont {Rudner},
  \citenamefont {Lindner}, \citenamefont {Berg},\ and\ \citenamefont
  {Levin}}]{Levin13prx}%
  \BibitemOpen
  \bibfield  {author} {\bibinfo {author} {\bibfnamefont {M.~S.}\ \bibnamefont
  {Rudner}}, \bibinfo {author} {\bibfnamefont {N.~H.}\ \bibnamefont {Lindner}},
  \bibinfo {author} {\bibfnamefont {E.}~\bibnamefont {Berg}}, \ and\ \bibinfo
  {author} {\bibfnamefont {M.}~\bibnamefont {Levin}},\ }\href@noop {}
  {\bibfield  {journal} {\bibinfo  {journal} {Physical Review X}\ }\textbf
  {\bibinfo {volume} {3}},\ \bibinfo {pages} {031005} (\bibinfo {year}
  {2013})}\BibitemShut {NoStop}%
\bibitem [{\citenamefont {Nathan}\ and\ \citenamefont
  {Rudner}(2015)}]{Rudner15njp}%
  \BibitemOpen
  \bibfield  {author} {\bibinfo {author} {\bibfnamefont {F.}~\bibnamefont
  {Nathan}}\ and\ \bibinfo {author} {\bibfnamefont {M.~S.}\ \bibnamefont
  {Rudner}},\ }\href@noop {} {\bibfield  {journal} {\bibinfo  {journal} {New
  Journal of Physics}\ }\textbf {\bibinfo {volume} {17}},\ \bibinfo {pages}
  {125014} (\bibinfo {year} {2015})}\BibitemShut {NoStop}%
\bibitem [{\citenamefont {Roy}\ and\ \citenamefont
  {Harper}(2017)}]{Harper17prb}%
  \BibitemOpen
  \bibfield  {author} {\bibinfo {author} {\bibfnamefont {R.}~\bibnamefont
  {Roy}}\ and\ \bibinfo {author} {\bibfnamefont {F.}~\bibnamefont {Harper}},\
  }\href@noop {} {\bibfield  {journal} {\bibinfo  {journal} {Physical Review
  B}\ }\textbf {\bibinfo {volume} {96}},\ \bibinfo {pages} {155118} (\bibinfo
  {year} {2017})}\BibitemShut {NoStop}%
\bibitem [{\citenamefont {Nathan}\ \emph {et~al.}(2019)\citenamefont {Nathan},
  \citenamefont {Abanin}, \citenamefont {Berg}, \citenamefont {Lindner},\ and\
  \citenamefont {Rudner}}]{Rudner19prb}%
  \BibitemOpen
  \bibfield  {author} {\bibinfo {author} {\bibfnamefont {F.}~\bibnamefont
  {Nathan}}, \bibinfo {author} {\bibfnamefont {D.}~\bibnamefont {Abanin}},
  \bibinfo {author} {\bibfnamefont {E.}~\bibnamefont {Berg}}, \bibinfo {author}
  {\bibfnamefont {N.~H.}\ \bibnamefont {Lindner}}, \ and\ \bibinfo {author}
  {\bibfnamefont {M.~S.}\ \bibnamefont {Rudner}},\ }\href@noop {} {\bibfield
  {journal} {\bibinfo  {journal} {Physical Review B}\ }\textbf {\bibinfo
  {volume} {99}},\ \bibinfo {pages} {195133} (\bibinfo {year}
  {2019})}\BibitemShut {NoStop}%
\bibitem [{\citenamefont {Rudner}\ and\ \citenamefont
  {Lindner}(2020)}]{Rudner20nrp}%
  \BibitemOpen
  \bibfield  {author} {\bibinfo {author} {\bibfnamefont {M.~S.}\ \bibnamefont
  {Rudner}}\ and\ \bibinfo {author} {\bibfnamefont {N.~H.}\ \bibnamefont
  {Lindner}},\ }\href@noop {} {\bibfield  {journal} {\bibinfo  {journal}
  {Nature Reviews Physics}\ }\textbf {\bibinfo {volume} {2}},\ \bibinfo {pages}
  {229} (\bibinfo {year} {2020})}\BibitemShut {NoStop}%
\bibitem [{\citenamefont {Eckardt}(2017)}]{Eckardt17rmp}%
  \BibitemOpen
  \bibfield  {author} {\bibinfo {author} {\bibfnamefont {A.}~\bibnamefont
  {Eckardt}},\ }\href@noop {} {\bibfield  {journal} {\bibinfo  {journal}
  {Reviews of Modern Physics}\ }\textbf {\bibinfo {volume} {89}},\ \bibinfo
  {pages} {011004} (\bibinfo {year} {2017})}\BibitemShut {NoStop}%
\bibitem [{\citenamefont {Zhang}\ \emph {et~al.}(2020)\citenamefont {Zhang},
  \citenamefont {Zhang},\ and\ \citenamefont {Liu}}]{LongZhang20prl}%
  \BibitemOpen
  \bibfield  {author} {\bibinfo {author} {\bibfnamefont {L.}~\bibnamefont
  {Zhang}}, \bibinfo {author} {\bibfnamefont {L.}~\bibnamefont {Zhang}}, \ and\
  \bibinfo {author} {\bibfnamefont {X.-J.}\ \bibnamefont {Liu}},\ }\href@noop
  {} {\bibfield  {journal} {\bibinfo  {journal} {Physical Review Letters}\
  }\textbf {\bibinfo {volume} {125}},\ \bibinfo {pages} {183001} (\bibinfo
  {year} {2020})}\BibitemShut {NoStop}%
\bibitem [{\citenamefont {Zhang}\ and\ \citenamefont
  {Liu}(2022)}]{LongZhang22prxQ}%
  \BibitemOpen
  \bibfield  {author} {\bibinfo {author} {\bibfnamefont {L.}~\bibnamefont
  {Zhang}}\ and\ \bibinfo {author} {\bibfnamefont {X.-J.}\ \bibnamefont
  {Liu}},\ }\href@noop {} {\bibfield  {journal} {\bibinfo  {journal} {PRX
  Quantum}\ }\textbf {\bibinfo {volume} {3}},\ \bibinfo {pages} {040312}
  (\bibinfo {year} {2022})}\BibitemShut {NoStop}%
\bibitem [{\citenamefont {Floquet}(1883)}]{Floquet1883}%
  \BibitemOpen
  \bibfield  {author} {\bibinfo {author} {\bibfnamefont {M.~G.}\ \bibnamefont
  {Floquet}},\ }\href@noop {} {\bibfield  {journal} {\bibinfo  {journal}
  {Annales Scientifiques de l'Ecole Normale Superieure}\ }\textbf {\bibinfo
  {volume} {12}},\ \bibinfo {pages} {47} (\bibinfo {year} {1883})}\BibitemShut
  {NoStop}%
\bibitem [{\citenamefont {Shirley}(1965)}]{Shirley65pr}%
  \BibitemOpen
  \bibfield  {author} {\bibinfo {author} {\bibfnamefont {J.~H.}\ \bibnamefont
  {Shirley}},\ }\href@noop {} {\bibfield  {journal} {\bibinfo  {journal} {Phys.
  Rev.}\ }\textbf {\bibinfo {volume} {138}},\ \bibinfo {pages} {B979} (\bibinfo
  {year} {1965})}\BibitemShut {NoStop}%
\bibitem [{SM2()}]{SM22HanZhang}%
  \BibitemOpen
  \href@noop {} {\bibinfo  {journal} {Supporting information is available as
  Supplemental Material}\ }\BibitemShut {NoStop}%
\bibitem [{\citenamefont {Sun}\ \emph {et~al.}(2018{\natexlab{a}})\citenamefont
  {Sun}, \citenamefont {Wang}, \citenamefont {Xu}, \citenamefont {Yi},
  \citenamefont {Zhang}, \citenamefont {Wu}, \citenamefont {Deng},
  \citenamefont {Liu}, \citenamefont {Chen},\ and\ \citenamefont
  {Pan}}]{shuai18prlrobust}%
  \BibitemOpen
\bibfield  {journal} {  }\bibfield  {author} {\bibinfo {author} {\bibfnamefont
  {W.}~\bibnamefont {Sun}}, \bibinfo {author} {\bibfnamefont {B.-Z.}\
  \bibnamefont {Wang}}, \bibinfo {author} {\bibfnamefont {X.-T.}\ \bibnamefont
  {Xu}}, \bibinfo {author} {\bibfnamefont {C.-R.}\ \bibnamefont {Yi}}, \bibinfo
  {author} {\bibfnamefont {L.}~\bibnamefont {Zhang}}, \bibinfo {author}
  {\bibfnamefont {Z.}~\bibnamefont {Wu}}, \bibinfo {author} {\bibfnamefont
  {Y.}~\bibnamefont {Deng}}, \bibinfo {author} {\bibfnamefont {X.-J.}\
  \bibnamefont {Liu}}, \bibinfo {author} {\bibfnamefont {S.}~\bibnamefont
  {Chen}}, \ and\ \bibinfo {author} {\bibfnamefont {J.-W.}\ \bibnamefont
  {Pan}},\ }\href@noop {} {\bibfield  {journal} {\bibinfo  {journal} {Physical
  Review Letters}\ }\textbf {\bibinfo {volume} {121}},\ \bibinfo {pages}
  {150401} (\bibinfo {year} {2018}{\natexlab{a}})}\BibitemShut {NoStop}%
\bibitem [{\citenamefont {Zhang}\ \emph {et~al.}(2018)\citenamefont {Zhang},
  \citenamefont {Zhang}, \citenamefont {Niu},\ and\ \citenamefont
  {Liu}}]{zhanglin18scibull}%
  \BibitemOpen
  \bibfield  {author} {\bibinfo {author} {\bibfnamefont {L.}~\bibnamefont
  {Zhang}}, \bibinfo {author} {\bibfnamefont {L.}~\bibnamefont {Zhang}},
  \bibinfo {author} {\bibfnamefont {S.}~\bibnamefont {Niu}}, \ and\ \bibinfo
  {author} {\bibfnamefont {X.-J.}\ \bibnamefont {Liu}},\ }\href@noop {}
  {\bibfield  {journal} {\bibinfo  {journal} {Science Bulletin}\ }\textbf
  {\bibinfo {volume} {63}},\ \bibinfo {pages} {1385} (\bibinfo {year}
  {2018})}\BibitemShut {NoStop}%
\bibitem [{\citenamefont {Zhu}\ \emph {et~al.}(2020)\citenamefont {Zhu},
  \citenamefont {Ke}, \citenamefont {Zhong},\ and\ \citenamefont
  {Lee}}]{lee20prr}%
  \BibitemOpen
  \bibfield  {author} {\bibinfo {author} {\bibfnamefont {B.}~\bibnamefont
  {Zhu}}, \bibinfo {author} {\bibfnamefont {Y.}~\bibnamefont {Ke}}, \bibinfo
  {author} {\bibfnamefont {H.}~\bibnamefont {Zhong}}, \ and\ \bibinfo {author}
  {\bibfnamefont {C.}~\bibnamefont {Lee}},\ }\href@noop {} {\bibfield
  {journal} {\bibinfo  {journal} {Physical Review Research}\ }\textbf {\bibinfo
  {volume} {2}},\ \bibinfo {pages} {023043} (\bibinfo {year}
  {2020})}\BibitemShut {NoStop}%
\bibitem [{\citenamefont {Ye}\ and\ \citenamefont {Li}(2020)}]{ye20pra}%
  \BibitemOpen
  \bibfield  {author} {\bibinfo {author} {\bibfnamefont {J.}~\bibnamefont
  {Ye}}\ and\ \bibinfo {author} {\bibfnamefont {F.}~\bibnamefont {Li}},\
  }\href@noop {} {\bibfield  {journal} {\bibinfo  {journal} {Physical Review
  A}\ }\textbf {\bibinfo {volume} {102}},\ \bibinfo {pages} {042209} (\bibinfo
  {year} {2020})}\BibitemShut {NoStop}%
\bibitem [{\citenamefont {Li}\ \emph {et~al.}(2021)\citenamefont {Li},
  \citenamefont {Zhu},\ and\ \citenamefont {Gong}}]{gong21scibull}%
  \BibitemOpen
  \bibfield  {author} {\bibinfo {author} {\bibfnamefont {L.}~\bibnamefont
  {Li}}, \bibinfo {author} {\bibfnamefont {W.}~\bibnamefont {Zhu}}, \ and\
  \bibinfo {author} {\bibfnamefont {J.}~\bibnamefont {Gong}},\ }\href@noop {}
  {\bibfield  {journal} {\bibinfo  {journal} {Science Bulletin}\ }\textbf
  {\bibinfo {volume} {66}},\ \bibinfo {pages} {1502} (\bibinfo {year}
  {2021})}\BibitemShut {NoStop}%
\bibitem [{\citenamefont {Zhang}\ \emph {et~al.}(2022)\citenamefont {Zhang},
  \citenamefont {Jia},\ and\ \citenamefont {Liu}}]{Zhanglin2022scibull}%
  \BibitemOpen
  \bibfield  {author} {\bibinfo {author} {\bibfnamefont {L.}~\bibnamefont
  {Zhang}}, \bibinfo {author} {\bibfnamefont {W.}~\bibnamefont {Jia}}, \ and\
  \bibinfo {author} {\bibfnamefont {X.-J.}\ \bibnamefont {Liu}},\ }\href@noop
  {} {\bibfield  {journal} {\bibinfo  {journal} {Science Bulletin}\ }\textbf
  {\bibinfo {volume} {67}},\ \bibinfo {pages} {1236} (\bibinfo {year}
  {2022})}\BibitemShut {NoStop}%
\bibitem [{\citenamefont {Sun}\ \emph {et~al.}(2018{\natexlab{b}})\citenamefont
  {Sun}, \citenamefont {Yi}, \citenamefont {Wang}, \citenamefont {Zhang},
  \citenamefont {Sanders}, \citenamefont {Xu}, \citenamefont {Wang},
  \citenamefont {Schmiedmayer}, \citenamefont {Deng}, \citenamefont {Liu},
  \citenamefont {Chen},\ and\ \citenamefont {Pan}}]{shuai18prltopo}%
  \BibitemOpen
  \bibfield  {author} {\bibinfo {author} {\bibfnamefont {W.}~\bibnamefont
  {Sun}}, \bibinfo {author} {\bibfnamefont {C.-R.}\ \bibnamefont {Yi}},
  \bibinfo {author} {\bibfnamefont {B.-Z.}\ \bibnamefont {Wang}}, \bibinfo
  {author} {\bibfnamefont {W.-W.}\ \bibnamefont {Zhang}}, \bibinfo {author}
  {\bibfnamefont {B.~C.}\ \bibnamefont {Sanders}}, \bibinfo {author}
  {\bibfnamefont {X.-T.}\ \bibnamefont {Xu}}, \bibinfo {author} {\bibfnamefont
  {Z.-Y.}\ \bibnamefont {Wang}}, \bibinfo {author} {\bibfnamefont
  {J.}~\bibnamefont {Schmiedmayer}}, \bibinfo {author} {\bibfnamefont
  {Y.}~\bibnamefont {Deng}}, \bibinfo {author} {\bibfnamefont {X.-J.}\
  \bibnamefont {Liu}}, \bibinfo {author} {\bibfnamefont {S.}~\bibnamefont
  {Chen}}, \ and\ \bibinfo {author} {\bibfnamefont {J.-W.}\ \bibnamefont
  {Pan}},\ }\href@noop {} {\bibfield  {journal} {\bibinfo  {journal} {Physical
  Review Letters}\ }\textbf {\bibinfo {volume} {121}},\ \bibinfo {pages}
  {250403} (\bibinfo {year} {2018}{\natexlab{b}})}\BibitemShut {NoStop}%
\bibitem [{\citenamefont {Hu}\ and\ \citenamefont {Zhao}(2020)}]{hu20prl}%
  \BibitemOpen
  \bibfield  {author} {\bibinfo {author} {\bibfnamefont {H.}~\bibnamefont
  {Hu}}\ and\ \bibinfo {author} {\bibfnamefont {E.}~\bibnamefont {Zhao}},\
  }\href@noop {} {\bibfield  {journal} {\bibinfo  {journal} {Physical Review
  Letters}\ }\textbf {\bibinfo {volume} {124}},\ \bibinfo {pages} {160402}
  (\bibinfo {year} {2020})}\BibitemShut {NoStop}%
\bibitem [{\citenamefont {Yu}\ \emph {et~al.}(2021)\citenamefont {Yu},
  \citenamefont {Ji}, \citenamefont {Zhang}, \citenamefont {Wang},
  \citenamefont {Wu},\ and\ \citenamefont {Liu}}]{liu21prxqu}%
  \BibitemOpen
  \bibfield  {author} {\bibinfo {author} {\bibfnamefont {X.-L.}\ \bibnamefont
  {Yu}}, \bibinfo {author} {\bibfnamefont {W.}~\bibnamefont {Ji}}, \bibinfo
  {author} {\bibfnamefont {L.}~\bibnamefont {Zhang}}, \bibinfo {author}
  {\bibfnamefont {Y.}~\bibnamefont {Wang}}, \bibinfo {author} {\bibfnamefont
  {J.}~\bibnamefont {Wu}}, \ and\ \bibinfo {author} {\bibfnamefont {X.-J.}\
  \bibnamefont {Liu}},\ }\href@noop {} {\bibfield  {journal} {\bibinfo
  {journal} {PRX Quantum}\ }\textbf {\bibinfo {volume} {2}},\ \bibinfo {pages}
  {020320} (\bibinfo {year} {2021})}\BibitemShut {NoStop}%
\bibitem [{\citenamefont {Yi}\ \emph {et~al.}(2019)\citenamefont {Yi},
  \citenamefont {Zhang}, \citenamefont {Zhang}, \citenamefont {Jiao},
  \citenamefont {Cheng}, \citenamefont {Wang}, \citenamefont {Xu},
  \citenamefont {Sun}, \citenamefont {Liu}, \citenamefont {Chen},\ and\
  \citenamefont {Pan}}]{shuai19prl}%
  \BibitemOpen
  \bibfield  {author} {\bibinfo {author} {\bibfnamefont {C.-R.}\ \bibnamefont
  {Yi}}, \bibinfo {author} {\bibfnamefont {L.}~\bibnamefont {Zhang}}, \bibinfo
  {author} {\bibfnamefont {L.}~\bibnamefont {Zhang}}, \bibinfo {author}
  {\bibfnamefont {R.-H.}\ \bibnamefont {Jiao}}, \bibinfo {author}
  {\bibfnamefont {X.-C.}\ \bibnamefont {Cheng}}, \bibinfo {author}
  {\bibfnamefont {Z.-Y.}\ \bibnamefont {Wang}}, \bibinfo {author}
  {\bibfnamefont {X.-T.}\ \bibnamefont {Xu}}, \bibinfo {author} {\bibfnamefont
  {W.}~\bibnamefont {Sun}}, \bibinfo {author} {\bibfnamefont {X.-J.}\
  \bibnamefont {Liu}}, \bibinfo {author} {\bibfnamefont {S.}~\bibnamefont
  {Chen}}, \ and\ \bibinfo {author} {\bibfnamefont {J.-W.}\ \bibnamefont
  {Pan}},\ }\href@noop {} {\bibfield  {journal} {\bibinfo  {journal} {Physical
  Review Letters}\ }\textbf {\bibinfo {volume} {123}},\ \bibinfo {pages}
  {190603} (\bibinfo {year} {2019})}\BibitemShut {NoStop}%
\bibitem [{\citenamefont {Wang}\ \emph {et~al.}(2019)\citenamefont {Wang},
  \citenamefont {Ji}, \citenamefont {Chai}, \citenamefont {Guo}, \citenamefont
  {Wang}, \citenamefont {Ye}, \citenamefont {Yu}, \citenamefont {Zhang},
  \citenamefont {Qin}, \citenamefont {Wang}, \citenamefont {Shi}, \citenamefont
  {Rong}, \citenamefont {Lu}, \citenamefont {Liu},\ and\ \citenamefont
  {Du}}]{du19pra}%
  \BibitemOpen
  \bibfield  {author} {\bibinfo {author} {\bibfnamefont {Y.}~\bibnamefont
  {Wang}}, \bibinfo {author} {\bibfnamefont {W.}~\bibnamefont {Ji}}, \bibinfo
  {author} {\bibfnamefont {Z.}~\bibnamefont {Chai}}, \bibinfo {author}
  {\bibfnamefont {Y.}~\bibnamefont {Guo}}, \bibinfo {author} {\bibfnamefont
  {M.}~\bibnamefont {Wang}}, \bibinfo {author} {\bibfnamefont {X.}~\bibnamefont
  {Ye}}, \bibinfo {author} {\bibfnamefont {P.}~\bibnamefont {Yu}}, \bibinfo
  {author} {\bibfnamefont {L.}~\bibnamefont {Zhang}}, \bibinfo {author}
  {\bibfnamefont {X.}~\bibnamefont {Qin}}, \bibinfo {author} {\bibfnamefont
  {P.}~\bibnamefont {Wang}}, \bibinfo {author} {\bibfnamefont {F.}~\bibnamefont
  {Shi}}, \bibinfo {author} {\bibfnamefont {X.}~\bibnamefont {Rong}}, \bibinfo
  {author} {\bibfnamefont {D.}~\bibnamefont {Lu}}, \bibinfo {author}
  {\bibfnamefont {X.-J.}\ \bibnamefont {Liu}}, \ and\ \bibinfo {author}
  {\bibfnamefont {J.}~\bibnamefont {Du}},\ }\href@noop {} {\bibfield  {journal}
  {\bibinfo  {journal} {Physical Review A}\ }\textbf {\bibinfo {volume}
  {100}},\ \bibinfo {pages} {052328} (\bibinfo {year} {2019})}\BibitemShut
  {NoStop}%
\bibitem [{\citenamefont {Ji}\ \emph {et~al.}(2020)\citenamefont {Ji},
  \citenamefont {Zhang}, \citenamefont {Wang}, \citenamefont {Zhang},
  \citenamefont {Guo}, \citenamefont {Chai}, \citenamefont {Rong},
  \citenamefont {Shi}, \citenamefont {Liu}, \citenamefont {Wang},\ and\
  \citenamefont {Du}}]{du20prl}%
  \BibitemOpen
  \bibfield  {author} {\bibinfo {author} {\bibfnamefont {W.}~\bibnamefont
  {Ji}}, \bibinfo {author} {\bibfnamefont {L.}~\bibnamefont {Zhang}}, \bibinfo
  {author} {\bibfnamefont {M.}~\bibnamefont {Wang}}, \bibinfo {author}
  {\bibfnamefont {L.}~\bibnamefont {Zhang}}, \bibinfo {author} {\bibfnamefont
  {Y.}~\bibnamefont {Guo}}, \bibinfo {author} {\bibfnamefont {Z.}~\bibnamefont
  {Chai}}, \bibinfo {author} {\bibfnamefont {X.}~\bibnamefont {Rong}}, \bibinfo
  {author} {\bibfnamefont {F.}~\bibnamefont {Shi}}, \bibinfo {author}
  {\bibfnamefont {X.-J.}\ \bibnamefont {Liu}}, \bibinfo {author} {\bibfnamefont
  {Y.}~\bibnamefont {Wang}}, \ and\ \bibinfo {author} {\bibfnamefont
  {J.}~\bibnamefont {Du}},\ }\href@noop {} {\bibfield  {journal} {\bibinfo
  {journal} {Physical Review Letters}\ }\textbf {\bibinfo {volume} {125}},\
  \bibinfo {pages} {020504} (\bibinfo {year} {2020})}\BibitemShut {NoStop}%
\bibitem [{\citenamefont {Xin}\ \emph {et~al.}(2020)\citenamefont {Xin},
  \citenamefont {Li}, \citenamefont {ang Fan}, \citenamefont {Zhu},
  \citenamefont {Zhang}, \citenamefont {Nie}, \citenamefont {Li}, \citenamefont
  {Liu},\ and\ \citenamefont {Lu}}]{Lu20prl}%
  \BibitemOpen
  \bibfield  {author} {\bibinfo {author} {\bibfnamefont {T.}~\bibnamefont
  {Xin}}, \bibinfo {author} {\bibfnamefont {Y.}~\bibnamefont {Li}}, \bibinfo
  {author} {\bibfnamefont {Y.}~\bibnamefont {ang Fan}}, \bibinfo {author}
  {\bibfnamefont {X.}~\bibnamefont {Zhu}}, \bibinfo {author} {\bibfnamefont
  {Y.}~\bibnamefont {Zhang}}, \bibinfo {author} {\bibfnamefont
  {X.}~\bibnamefont {Nie}}, \bibinfo {author} {\bibfnamefont {J.}~\bibnamefont
  {Li}}, \bibinfo {author} {\bibfnamefont {Q.}~\bibnamefont {Liu}}, \ and\
  \bibinfo {author} {\bibfnamefont {D.}~\bibnamefont {Lu}},\ }\href@noop {}
  {\bibfield  {journal} {\bibinfo  {journal} {Physical Review Letters}\
  }\textbf {\bibinfo {volume} {125}},\ \bibinfo {pages} {090502} (\bibinfo
  {year} {2020})}\BibitemShut {NoStop}%
\bibitem [{not()}]{note2}%
  \BibitemOpen
  \href@noop {} {\bibinfo  {journal} {{We note that in Fig.~4(a), the observed
  weak connections between two BISs are contributed from the
  $|\!\!\uparrow\rangle$ atoms initially populated outside the first Brillouin
  zone~\cite{SM22HanZhang}}}\ }\BibitemShut {NoStop}%
\bibitem [{Note1()}]{Note1}%
  \BibitemOpen
\bibfield  {journal} {  }\bibinfo {note} {{\protect \color {black}The
  sub-leading-order SO coupling, which is a third-order coupling, has a
  strength estimated to be on the scale of $h \times 40$~Hz, which is not
  relevant in the current measurements; see more details in the Supplemental
  Material~\cite {SM22HanZhang}.}}\BibitemShut {Stop}%
\bibitem [{\citenamefont {Ziegler}\ \emph
  {et~al.}(2022{\natexlab{a}})\citenamefont {Ziegler}, \citenamefont {Tirrito},
  \citenamefont {Lewenstein}, \citenamefont {Hands},\ and\ \citenamefont
  {Bermudez}}]{bermudez20arxiv}%
  \BibitemOpen
  \bibfield  {author} {\bibinfo {author} {\bibfnamefont {L.}~\bibnamefont
  {Ziegler}}, \bibinfo {author} {\bibfnamefont {E.}~\bibnamefont {Tirrito}},
  \bibinfo {author} {\bibfnamefont {M.}~\bibnamefont {Lewenstein}}, \bibinfo
  {author} {\bibfnamefont {S.}~\bibnamefont {Hands}}, \ and\ \bibinfo {author}
  {\bibfnamefont {A.}~\bibnamefont {Bermudez}},\ }\href@noop {} {\bibfield
  {journal} {\bibinfo  {journal} {Physical Review Research}\ }\textbf {\bibinfo
  {volume} {4}},\ \bibinfo {pages} {L042012} (\bibinfo {year}
  {2022}{\natexlab{a}})}\BibitemShut {NoStop}%
\bibitem [{\citenamefont {Ziegler}\ \emph
  {et~al.}(2022{\natexlab{b}})\citenamefont {Ziegler}, \citenamefont {Tirrito},
  \citenamefont {Lewenstein}, \citenamefont {Hands},\ and\ \citenamefont
  {Bermudez}}]{bermudez22aop}%
  \BibitemOpen
  \bibfield  {author} {\bibinfo {author} {\bibfnamefont {L.}~\bibnamefont
  {Ziegler}}, \bibinfo {author} {\bibfnamefont {E.}~\bibnamefont {Tirrito}},
  \bibinfo {author} {\bibfnamefont {M.}~\bibnamefont {Lewenstein}}, \bibinfo
  {author} {\bibfnamefont {S.}~\bibnamefont {Hands}}, \ and\ \bibinfo {author}
  {\bibfnamefont {A.}~\bibnamefont {Bermudez}},\ }\href@noop {} {\bibfield
  {journal} {\bibinfo  {journal} {Annals of Physics}\ }\textbf {\bibinfo
  {volume} {439}},\ \bibinfo {pages} {168763} (\bibinfo {year}
  {2022}{\natexlab{b}})}\BibitemShut {NoStop}%
\bibitem [{\citenamefont {Radic}\ \emph {et~al.}(2012)\citenamefont {Radic},
  \citenamefont {Ciolo}, \citenamefont {Sun},\ and\ \citenamefont
  {Galitski}}]{Galitski12prlmagnetic}%
  \BibitemOpen
  \bibfield  {author} {\bibinfo {author} {\bibfnamefont {J.}~\bibnamefont
  {Radic}}, \bibinfo {author} {\bibfnamefont {A.~D.}\ \bibnamefont {Ciolo}},
  \bibinfo {author} {\bibfnamefont {K.}~\bibnamefont {Sun}}, \ and\ \bibinfo
  {author} {\bibfnamefont {V.}~\bibnamefont {Galitski}},\ }\href@noop {}
  {\bibfield  {journal} {\bibinfo  {journal} {Physical Review Letters}\
  }\textbf {\bibinfo {volume} {109}},\ \bibinfo {pages} {085303} (\bibinfo
  {year} {2012})}\BibitemShut {NoStop}%
\bibitem [{\citenamefont {Reuther}\ \emph {et~al.}(2012)\citenamefont
  {Reuther}, \citenamefont {Thomale},\ and\ \citenamefont
  {Rachel}}]{Rachel12prb}%
  \BibitemOpen
  \bibfield  {author} {\bibinfo {author} {\bibfnamefont {J.}~\bibnamefont
  {Reuther}}, \bibinfo {author} {\bibfnamefont {R.}~\bibnamefont {Thomale}}, \
  and\ \bibinfo {author} {\bibfnamefont {S.}~\bibnamefont {Rachel}},\
  }\href@noop {} {\bibfield  {journal} {\bibinfo  {journal} {Physical Review
  B}\ }\textbf {\bibinfo {volume} {86}},\ \bibinfo {pages} {155127} (\bibinfo
  {year} {2012})}\BibitemShut {NoStop}%
\bibitem [{\citenamefont {Barnett}\ \emph {et~al.}(2012)\citenamefont
  {Barnett}, \citenamefont {Boyd},\ and\ \citenamefont
  {Galitski}}]{Galitski12prlsu3}%
  \BibitemOpen
  \bibfield  {author} {\bibinfo {author} {\bibfnamefont {R.}~\bibnamefont
  {Barnett}}, \bibinfo {author} {\bibfnamefont {G.~R.}\ \bibnamefont {Boyd}}, \
  and\ \bibinfo {author} {\bibfnamefont {V.}~\bibnamefont {Galitski}},\
  }\href@noop {} {\bibfield  {journal} {\bibinfo  {journal} {Physical Review
  Letters}\ }\textbf {\bibinfo {volume} {109}},\ \bibinfo {pages} {235308}
  (\bibinfo {year} {2012})}\BibitemShut {NoStop}%
\bibitem [{\citenamefont {Jimenez-Garcia}\ \emph {et~al.}(2015)\citenamefont
  {Jimenez-Garcia}, \citenamefont {LeBlanc}, \citenamefont {Williams},
  \citenamefont {Beeler}, \citenamefont {Qu}, \citenamefont {Gong},
  \citenamefont {Zhang},\ and\ \citenamefont {Spielman}}]{Spielman15prl}%
  \BibitemOpen
  \bibfield  {author} {\bibinfo {author} {\bibfnamefont {K.}~\bibnamefont
  {Jimenez-Garcia}}, \bibinfo {author} {\bibfnamefont {L.~J.}\ \bibnamefont
  {LeBlanc}}, \bibinfo {author} {\bibfnamefont {R.~A.}\ \bibnamefont
  {Williams}}, \bibinfo {author} {\bibfnamefont {M.~C.}\ \bibnamefont
  {Beeler}}, \bibinfo {author} {\bibfnamefont {C.}~\bibnamefont {Qu}}, \bibinfo
  {author} {\bibfnamefont {M.}~\bibnamefont {Gong}}, \bibinfo {author}
  {\bibfnamefont {C.}~\bibnamefont {Zhang}}, \ and\ \bibinfo {author}
  {\bibfnamefont {I.~B.}\ \bibnamefont {Spielman}},\ }\href@noop {} {\bibfield
  {journal} {\bibinfo  {journal} {Physical Review Letters}\ }\textbf {\bibinfo
  {volume} {114}},\ \bibinfo {pages} {125301} (\bibinfo {year}
  {2015})}\BibitemShut {NoStop}%
\end{thebibliography}%


\newcommand{\pzcS}{\mathpzc{S}}
\newcommand{\pzci}{\mathpzc{i}}

\renewcommand{\theequation}{S\arabic{equation}}
\renewcommand{\thefigure}{S\arabic{figure}}
\renewcommand{\thetable}{S\arabic{table}}

\newcommand{\f}{|f\rangle}
\newcommand{\e}{|e\rangle}
\newcommand{\fth}{|f;\,+\frac{3}{2}\rangle}
\renewcommand{\eth}{|e;\,+\frac{3}{2}\rangle}
\renewcommand{\theequation}{S\arabic{equation}}
\renewcommand{\thefigure}{S\arabic{figure}}

\newcommand{\boldrho}{\mbox{\boldmath$\rho$}}
\newcommand{\bg}[1]{\mbox{\boldmath$#1$}}
\definecolor{red}{rgb}{0.7,0,0}
\definecolor{green}{rgb}{0.,0.35,0.}
\definecolor{blue}{rgb}{0.2,0.2,0.7}
\definecolor{black}{rgb}{0.15,0.15,.15}
\newcommand{\com}[1]{{\color{blue}\small\   #1 }}
\newcommand{\ad}{a^{\dagger}}
\newcommand{\al}{\alpha^{\dagger}}
\newcommand{\be}{\beta^{\dagger}}
\newcommand{\ga}{\gamma^{\dagger}}
\newcommand{\de}{\delta^{\dagger}}
\newcommand{\fd}{f^{\dagger}}
\newcommand{\an}{\mathcal N}
\newcommand{\bfn}{\mathbf n}
\newcommand{\one}{\mbox{$1 \hspace{-1.0mm}  {\bf l}$}}
\newcommand{\spec}{\mbox{spec}}
\newcommand{\bracket}{\rangle \langle}
\newcommand{\vac}{|0\rangle }
\newcommand{\ket}[1]{\left|#1\right\rangle}
\newcommand{\bra}[1]{\left\langle#1\right|}
\newcommand{\rem}[1]{\textbf{\textcolor{red}{[#1]}}}

\newcommand{\ree}{\rho_{ee}^{mm}}
\newcommand{\rgg}{\rho_{gg}^{mm}}
\newcommand{\reg}{\rho_{eg}^{mm}}
\newcommand{\rge}{\rho_{ge}^{mm}}

\setcounter{figure}{0}
\setcounter{equation}{0}
\clearpage

\section*{SUPPLEMENTAL MATERIAL}

\section{Experimental setup and methods}
\subsection{Preparation and detection of Fermi gases} 
The preparation of strontium ($^{87}$Sr) ultracold Fermi gases and the detection of two spin states are performed in a way similar to Ref.~\citenum{Liang21arxiv}. The two relevant spin ground states are $\ket{\uparrow}=\mathrm{5s}^2\,{}^1\mathrm{S}_0|F=\frac{9}{2}, m_F=-\frac{9}{2}\rangle$ and $\ket{\downarrow}=|9/2, -7/2\rangle$. 
Fermions of ${}^{87}$Sr are first laser-cooled and then evaporatively cooled to a temperature below 200~nK.  
The corresponding ultracold Fermi gas has about $4\times10^4$ atoms in a far-detuned crossed dipole trap, with more than $85\%$ of these atoms initially populated in the $\ket{\uparrow}$ state. 

After the Fermi gas is loaded into the optical Raman lattice and the experiment is performed, we shut off all lasers within 1~$\mu$s and perform spin-resolved time-of-flight measurements to extract the distribution of the $|\!\!\uparrow\rangle$ and $|\!\!\downarrow\rangle$ atoms~\cite{Liang21arxiv}.

\subsection{Two-dimensional optical Raman lattice}
\begin{figure}
	\centering
	\includegraphics[scale=0.45]{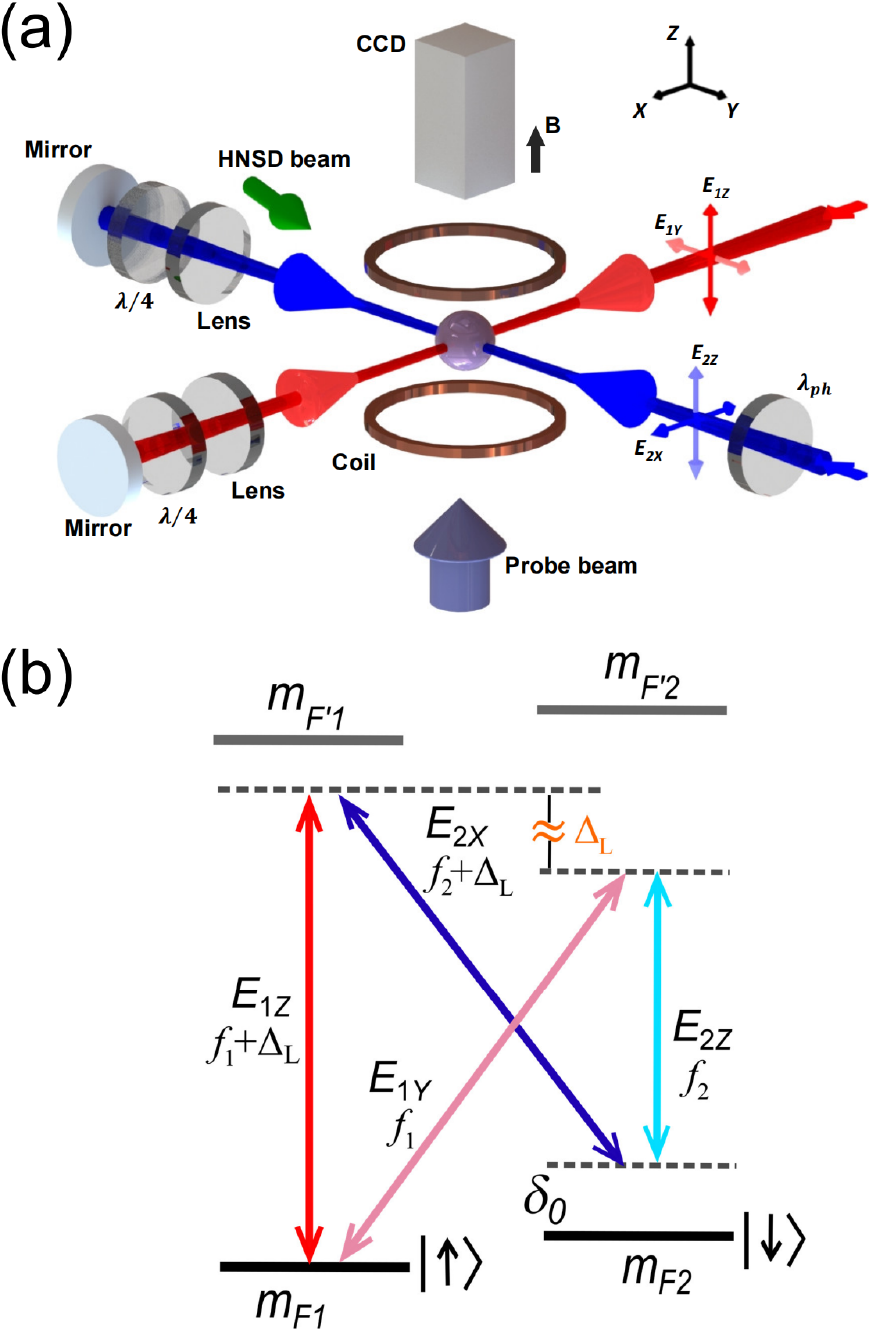}
	\caption{{\color{black}Setup and diagram for two-dimensional optical Raman lattice. (a) Schematic of the experimental setup. The two incident Raman coupling beams propagate along the $\hat{X}$ and $-\hat{Y}$ directions. In the optical path along the $-\hat{Y}$ direction, $\lambda_{ph}$ denotes a composite wave plate used for controlling the relative phase $\delta\varphi$~\cite{Liang21arxiv}. (b) Energy level diagram and  frequencies of the four polarization components.  Among the four linearly-polarized electric fields ($E_{1Z}$, $E_{1Y}$, $E_{2Z}$, and $E_{2X}$), the field $E_{1Z}$ ($E_{2X}$)  is additionally frequency-shifted with respect to $E_{1Y}$ ($E_{2Z}$) by a large amount of $\Delta_{\mathrm{L}}$, as also illustrated in Fig.~2(a) of the main text.}
	}
	\label{fig:sm_setup}
\end{figure}

The experimental setup for the two-dimensional optical Raman lattice is illustrated in Fig.~\ref{fig:sm_setup}(a). Two Raman coupling beams propagate along the $\hat{X}$ and $-\hat{Y}$ horizontal directions, intersect at the atoms, and are each phase-shifted and retro-reflected to form two-dimensional (2D) optical lattices for the $|\!\!\uparrow\rangle$ and $|\!\!\downarrow\rangle$ states. Here, $\hat{X}$, $\hat{Y}$, and $\hat{Z}$ denote a set of orthogonal spatial axes, with $-\hat{Z}$ being the direction of the gravity. 

The Raman coupling beams of wavelength $\lambda_0 \approx 689.4$~nm are set to a frequency of about -0.8~GHz with respect to the ${}^1$S${}_0 (F=\frac{9}{2}) \rightarrow {}^3$P${}_1$($F'=\frac{11}{2}$) intercombination transition for the ${}^{87}$Sr atoms, where the transition linewidth is as narrow as 7.5~kHz. A total of four two-photon Raman transitions between the  $|\!\!\uparrow\rangle$ and $|\!\!\downarrow\rangle$ states can be induced by pairs of electric field components: $(E_{1Z}, E^{*}_{2X})$, $(E_{1Y},E^{*}_{2Z})$, $(E_{2Z},E^{*}_{1Y})$, and $(E_{2X},E^{*}_{1Z})$. Here, in the combination $(E_a, E^{*}_b)$, $E_a$ corresponds to the absorption of one photon from the $E_a$ beam, and $E_b^{*}$ corresponds to the emission of one photon into the $E_b$ beam. As shown in Fig.~1(a) of the main text, the first two combinations, $(E_{1Z}, E^{*}_{2X})$ and $(E_{1Y},E^{*}_{2Z})$, correspond to Raman couplings with a two-photon detuning $\delta_0$. By contrast, the rest two combinations, $(E_{2Z},E^{*}_{1Y})$ and $(E_{2X},E^{*}_{1Z})$, correspond to Raman couplings with a distinct two-photon detuning $\delta'_0 = \delta_0 - \omega$, where $\omega/2\pi$ equals to twice the frequency difference between two Raman coupling beams (say $E_{1Z}$ and $E_{2X}$); see the main text below Eq.~2.

\textit{Realization of essentially static 2D optical lattice.---} 
It has been observed that the interference between similarly polarized electric field components (such as $E_{1Z}$ and $E_{2Z}$) can cause detrimental moving lattice potentials that heat up the atoms~\cite{Liang21arxiv}. Such detrimental effect can be suppressed by implementing a large frequency difference between two pairs of electric fields, namely the vertically polarized pair $(E_{1Z},E_{2Z})$ and the horizontally polarized pair $(E_{1Y}, E_{2X})$. 

{\color{black} As shown in Fig.~\ref{fig:sm_setup}(b),} we introduce large frequency differences (close to $\Delta_{\mathrm{L}} = 4$~MHz) between these pairs of similarly polarized electric fields; see also Fig.~2(a) and Fig.~1(a) in the main text. Here, $\Delta_{\mathrm{L}}$ is on the megahertz scale and is three orders of magnitude larger than two physical quantities: (a) the kilohertz-scale level splitting $\Delta_0$ between the  $|\!\!\uparrow\rangle$ and  $|\!\!\downarrow\rangle$ states and (b) typical two-photon detuning values that are also on the kilohertz scale. In the experimental implementation, for each Raman coupling beam, the vertical and horizontal linear polarization components are two single-frequency laser beams of distinct frequencies, with a large frequency difference $\Delta_{\mathrm{L}}$ between the two. \textit{We note that the large $\Delta_{\mathrm{L}}$ and the two-photon detuning ($\delta_0,\delta'_0$) for Raman couplings are fully independent parameters that are precisely controlled in the experiment.} The detailed structure of this optical implementation is not required for the understanding of this work, and will be presented elsewhere in a more specialized journal. 

The choice of $\Delta_{\mathrm{L}}$ is based on two considerations. First, in our experiment, we have verified that as long as $\Delta_{\mathrm{L}}$ reaches a few MHz, the heating effect due to moving lattice potentials can be sufficiently suppressed. Second, $\Delta_{\mathrm{L}}$ can not be too large, either. This is because, under the same physical distance between atoms and the retro-reflection mirror, the frequency difference between two polarization components will cause a mismatch between the lattice potential minima for these two polarization components. Under $\Delta_{\mathrm{L}} = 4$~MHz and a distance of $L_{\mathrm{a,m}}\approx 0.5$~m between atoms and the retro-reflection mirror, the mismatch is about $1.3\times10^{-2}a$, where $a = \lambda_0/2\approx 344.7$~nm is the lattice spacing for both polarization components. This mismatch is sufficiently small and thus neglected in our analysis.

\subsection{The highly nonlinear spin-discriminating method}
We design and implement a highly nonlinear spin-discriminating (HNSD) method  to isolate an effective two-spin manifold out of the ten nuclear spin ground states of ${}^{87}$Sr. In our case of two-photon Raman transitions, we need to well separate the two nearby spin ground states (${}^1\mathrm{S}_0|F=\frac{9}{2}, m_F=-\frac{5}{2}\rangle$ and $|\frac{9}{2},-\frac{3}{2}\rangle$) from the two-spin manifold of the $|\!\!\uparrow\rangle$ and  $|\!\!\downarrow\rangle$ states. To achieve this goal, we first magnetically separate the energy levels of ${}^3$P${}_1$ excited states, and then apply HNSD beams that are (1) near-resonance for the $\pi$-transitions of the $m_F = -\frac{5}{2}$ or $-\frac{3}{2}$ state and (2) relatively farther-detuned from transitions of the $m_F = -\frac{9}{2}$ and $-\frac{7}{2}$ states.

In our experiment, a relatively large magnetic field of 35~G is applied along the $+\hat{Z}$ direction, which both determines the quantization axis for the atomic states and induces a large level splitting of $\Delta_e \approx 13$~MHz between adjacent spin excited states in the  $\mathrm{5s5p}\,{}^3\mathrm{P}_1|F'=\frac{11}{2},m_{F'}\rangle$ excited manifold. By contrast, adjacent spin ground states are separated by a much smaller amount of about 6~kHz under the same field. Under such magnetic Zeeman splittings, we apply to atoms a $\pi$-polarized  HNSD beam that propagate in the horizontal plane along a direction 20 degree with respect to $\hat{X}$. This HNSD beam, with 80~$\mu$W power and a $1/e^2$ beam radius of 200~$\mu$m, contains two frequency modes in its spectrum. Each frequency mode has a narrow linewidth below 1~kHz and is about 100~kHz blue-detuned relative to the resonance of the $\pi$-transition ${}^1$S${}_0|F=\frac{9}{2},m_F\rangle \rightarrow {}^3$P${}_1|F'=\frac{11}{2},m_F\rangle$ at $m_F = -\frac{5}{2}$ or $-\frac{3}{2}$. As a result, the HNSD beam induces large upward level shifts (more than 100~kHz) for the $|\frac{9}{2},-\frac{5}{2}\rangle$ and $|\frac{9}{2},-\frac{3}{2}\rangle$ spin ground states. By contrast, this beam only induces kilohertz-scale small level shift for the $|\!\!\uparrow\rangle$ and  $|\!\!\downarrow\rangle$ states, which has a high stability on the 10-hertz scale based on laser intensity stabilization technique. Thus, compared to the Zeeman shift that is linear to the lowest order (and at small magnetic field), the HNSD method engineers highly nonlinear energy level shifts among the spin states, which both enhances the lifetime of SO-coupled fermions and holds promise for creating an isolated manifold with an arbitrary number of spin states.

\subsection{Lifetime measurement and analysis}
The lifetime $\tau_0$ of SO-coupled fermions is determined as follows. We hold the atoms in the optical Raman lattice for various length of time,  measure the total number of atoms in the $|\!\!\uparrow\rangle$ and  $|\!\!\downarrow\rangle$ states as a function of holding time, and fit the measure with a first-order exponential decay function, extracting a $1/e$ decay lifetime.

{\color{black}As described in the main text, we enhance the lifetime of 2D-SO-coupled fermions by an order of magnitude. Such enhancement builds on two technical improvements that are also described in the main text, namely the HNSD method and the upgraded method for generating the optical Raman lattice. These two  improvements for the experimental realization are both indispensable to the lifetime enhancement. In fact, during our experimental period, we first implemented the optical Raman lattice upgrade, and observed the lifetime doubled, which was encouraging but still insufficient. Later we developed and implemented the HNSD method, and achieved another enhancement factor of about 5 for the lifetime. These two improvements together enable the one-order-of-magnitude lifetime enhancement. }

In addition, we perform a similar measurement when the SO couplings are far-off-resonance, and extract a reference lifetime $\tau_{\mathrm{Ref}} \approx 265$~ms under the same lattice depths and Raman coupling strengths as those in Fig.~2(b) in the main text. Comparing $\tau_0 \approx 73$~ms and $\tau_{\mathrm{Ref}}$, we see that the {\color{black}near-resonant} SO-couplings still play an important role in affecting the lifetime. We further extract a SO-coupling-constrained characteristic $1/e$ decay lifetime, $t_{\mathrm{SOC}}$, based on an empirical form: $1/\tau_0 \equiv 1/\tau_{\mathrm{SOC}} + 1/\tau_{\mathrm{Ref}}$. This analysis yields $\tau_{\mathrm{SOC}} \approx 101$~ms, which is shorter than $\tau_{\mathrm{Ref}}$. Therefore, compared to other factors (revealed by $\tau_{\mathrm{Ref}}$), the SO couplings currently do impose a primary constraint on the lifetime of SO-coupled fermions, {\color{black}which can be further improved in future experiments; see next paragraph}. At the same time, such a SO-coupling constrained lifetime already reaches the 100~ms scale and can enable a number of precise measurements for various topological phases.

{\color{black} The lifetime of 2D-SO-coupled fermions in our system is currently limited by several factors that can be further improved in future experiments. First, from the fundamental aspect, when the SO coupling is close to the two-photon resonance, the fluctuation of the two-photon detuning can induce heating effects, which is similar to the magnetic-field-induced heating effect in the 2D-SO-coupled bosonic systems~\cite{shuai16science}. Thus, the lifetime can be further enhanced by improving the stability of the relative energy splitting between the $|\!\!\uparrow\rangle$ and $|\!\!\downarrow\rangle$ states. Second, we can further reduce the photon scattering in our setup. For example, applying a larger magnetic field can reduce the residual scattering rate due to the HNSD beams for the $|\!\!\uparrow\rangle$ and $|\!\!\downarrow\rangle$ states. The photon scattering due to the Raman beams can also be reduced by implementing a Ti:Sapphire laser that provides a cleaner spectrum, higher optical power and larger single-photon detunings for the optical Raman lattices.  Third, the heating due to residual moving lattice potentials can be further suppressed by improving the laser polarization purity with better polarizing optics. The systematic technical improvements can further remove the major technical impediments to enhancing the lifetime and allow us to focus on improving the fundamental, SO-coupling-related factors for the Fermi gases. Overall, our system of 2D-SO-coupled Fermi gas holds the promise to reach an even longer lifetime that approaches the performance in Ref.~\citenum{Lev16prx} or even that in the bosonic systems~\cite{shuai16science,shuai18prlrobust}.
	
	We also note that, when the lattice depth is increased to even higher values, the aforementioned several improvements will similarly alleviate the issues of increased scattering rate and residual moving lattice potentials. This is beneficial for reaching a sufficiently long lifetime under the corresponding experimental condition and for future studies of novel correlated topological physics in the interacting regimes.

}

\subsection{Optimization of PPQM pulse length}
The long lifetime enables us to sharpen the detection resolution of the PPQM measurement. Under a given PPQM pulse area, the Fourier-limited width of atomic distribution for the observed $|\!\!\downarrow\rangle$-state atoms decreases as the pulse becomes longer. The key in this optimization is to search for a balance between the Fourier broadening and other factors. 

In Fig.~3(c) of the main text, we employ a fixed pulse area of $\sqrt{\Omega_{01}^2+\Omega_{02}^2}t_{\mathrm{p}}/2\hbar \approx 0.6\pi$, the measured momentum-width of pumped $|\!\!\downarrow\rangle$-state atoms indeed decreases as the pulse length $t_{\mathrm{p}}$ increases from a small value (about 100~$\mu$s), which is consistent with a reduction of the Fourier limit of a finite pulse ($\sim 1/t_{\mathrm{p}}$); see the blue dashed line in Fig.~3(c). For $t_{\mathrm{p}}$ exceeding 800~$\mu$s, the atomic width stops improving, which is likely due to other factors such as the finite imaging resolution. An optimum value of $t_{\mathrm{p}} \approx 700$~$\mu$s is found, providing substantial better resolving power than the 200~$\mu$s used in Ref.~\citenum{Liang21arxiv}.

\section{Key characteristic feature of SO couplings beyond RWA}

The two Raman potentials, $\Omega$ and $\Omega'$, are generated from the same electric fields in two different but intrinsically related ways. Below we shall show that the sum of the two relative phases for SO couplings driven by Raman potentials $\Omega$ and $\Omega'$ equals to a deterministic value of $\pi$, which is dictated by the phase relation between dipole oscillations with respect to different spatial axes.

For the Raman potential $\Omega$, the relative phase between the two Raman processes 
[$\Omega_{01}$ and  $\Omega_{02}$ in Figs.~1(a) and 2(a) in the main text] is determined by 
\begin{equation}\label{Sa1}
	\delta\phi_1=(-\theta_{79\hat{X}})-\theta_{97\hat{Y}}+\delta\psi,
\end{equation}
where $\delta\psi$ is the relative phase shift induced by the laser and optics in the setup~\cite{Liang21arxiv}, and $\theta_{ij\hat{e}}$  denotes the phase of the dipole matrix element  $\langle m_{F^\prime}=-\frac{j}{2}|\vec{d}\cdot\hat{e} |m_{F}=-\frac{i}{2}\rangle$ for the transition from the ground state ${}^1$S${}_0|F=\frac{9}{2},m_F = -\frac{i}{2}\rangle$ to an excited state ${}^3$P${}_1|F',m_{F'} = -\frac{j}{2}\rangle$, with $i,j\in\{9,7\}$, where the transition is driven by the dipole operator $\vec{d}$ and an unit vector $\hat{e}$ along the relevant electric field polarization, with $\hat{e}\in\{ \hat{X},\hat{Y},\hat{Z} \}$ in our experimental implementation. In Eq.~(\ref{Sa1}), the minus sign in $-\theta_{79\hat{X}}$ marks the ``emission'' of a photon. Here for simplicity of expression, we only show the terms for $\sigma$ transitions driven by horizontally polarized electric fields.
The phase terms of the $\pi$ transitions driven by the vertically polarized electric fields are neglected because  they do not influence the main conclusion in this section [Eq.~(\ref{eq:phasesum1and2})].

Likewise, in the Raman potential $\Omega'$, the relative phase between Raman processes $\Omega^{\prime}_{01}$ and  $\Omega^{\prime}_{02}$ is given by 
\begin{equation}\label{Sa2}
	\delta\phi_2=\theta_{97\hat{X}}-(-\theta_{79\hat{Y}})-\delta\psi,
\end{equation}
where the minus sign in $-\theta_{79\hat{Y}}$ again marks the ``emission'' of a photon. Here in Eq.~(\ref{Sa2}), the contribution from lasers and optics ($-\delta\psi$) reverses sign as compared to that in Eq.~(\ref{Sa1}), which is because each pair of electric fields in two-photon Raman transitions reverse their roles of absorption and emission.

Finally, it is straightforward to verify the relation that $\theta_{79\hat{Y}}-\theta_{79\hat{X}}=\theta_{97\hat{X}}-\theta_{97\hat{Y}}$, and that these two phase differences both equal to $\frac{\pi}{2}$~\cite{liu18pra}. We thus derive the key characteristic feature of SO couplings beyond RWA --- that the two relative phases $\delta\phi_1$ and $\delta\phi_2$ must have a deterministic relation: 
\begin{eqnarray}\label{eq:phasesum1and2}
	\delta\phi_1+\delta\phi_2 & = & \pi,
\end{eqnarray}
which corresponds to the expression $\delta\varphi + (\pi-\delta\varphi) = \pi$ in the ``Verification of SO couplings beyond RWA'' section of the main text, with the correspondence $\delta\phi_1 = \delta\varphi$.

\section{Numerical simulations}
\subsection{Numerical method and the Floquet topological phase diagram}
\textit{Numerical method.---}The 2D-SO-coupled fermions beyond the rotating wave approximation are described by the Hamiltonian in  Eqs.~(1) to (3) in the main text. We rewrite the Bloch Hamiltonian in Eq.~(3) as follows (neglecting the overall energy shift $U_0(\textbf{q})$):
\begin{eqnarray}\label{eq:SM_Hamiltonian}
	\hat{H}(\textbf{q},t) & = &
	\left (
	\begin{matrix}
		h_\uparrow (\textbf{q})&\Omega(\textbf{q})+\Omega'(\textbf{q}) e^{i \omega t}\\
		\Omega^*(\textbf{q})+\Omega'^*(\textbf{q}) e^{-i \omega t}&h_\downarrow(\textbf{q})
	\end{matrix}
	\right ),\nonumber\\
\end{eqnarray}
where $h_{\uparrow}(\textbf{q}) = -h_{\downarrow}(\textbf{q}) = h^z(\textbf{q})$, $\Omega(\textbf{q})$ and $\Omega'(\textbf{q})$ are each given by $h_0^x(\textbf{q}) - i h_0^y(\textbf{q})$, with $h^z(\textbf{q})$ and $h_0^{x,y}(\textbf{q})$ defined in the main text following Eq.~(3). This Hamiltonian can be treated as a Floquet system where a time-dependent term $\hat{V}(\textbf{q},t)=\Omega'(\textbf{q})  e^{i \omega t} \ket{\uparrow} \bra{\downarrow}+\mathrm{h.c.}$ periodically drives the static part $\hat{H}_{\mathrm{s}}(\textbf{q})=\hat{H}(\textbf{q},t)-\hat{V}(\textbf{q},t)$ with a driving period of $t_{\mathrm{F}} = 2\pi/\omega$. Here h.c. denotes the Hermitian conjugate. We then employ the Floquet theory~\cite{Floquet1883, Shirley65pr, Eckardt17rmp, LongZhang20prl,LongZhang22prxQ} to compute the band topology of this Hamiltonian.

Based on the Floquet-Magnus expansion scheme, we construct a matrix $\hat{H}_{\mathrm{sF}}$ of the static Floquet Hamiltonian for describing the system evolution  governed by $\hat{H}(\textbf{q},t)$, namely,   
\begin{widetext}\label{eq:HsF}
	\begin{equation}
		\hat{H}_{\mathrm{sF}}(\textbf{q})=
		\left (
		\begin{matrix}
			\ddots &\\
			& h_\uparrow(\textbf{q})-\omega&\Omega(\textbf{q})\\
			&\Omega^*(\textbf{q})& h_\downarrow(\textbf{q})-\omega&\Omega'^*(\textbf{q})\\
			&&\Omega'(\textbf{q})& h_\uparrow(\textbf{q})&\Omega(\textbf{q})\\
			&&&\Omega^*(\textbf{q})& h_\downarrow(\textbf{q})&\Omega'^*(\textbf{q})\\
			&&&&\Omega'(\textbf{q})& h_\uparrow(\textbf{q})+\omega&\Omega(\textbf{q})\\
			&&&&&\Omega^*(\textbf{q})& h_\downarrow(\textbf{q})+\omega\\
			&&&&&&&\ddots
		\end{matrix}
		\right ),
	\end{equation}
\end{widetext}
where the unshown matrix elements are zero. \textit{Under finite lattice depths in the experiment, the form of the Hamiltonian matrix requires more elaboration and can be accurately determined based on expansions under a plane-wave basis~\cite{Liang21arxiv}}.

{\color{black}We note that under proper conditions, in the derivation of a simplified time-independent Floquet Hamiltonian under the $|\!\!\uparrow\rangle$ and $|\!\!\downarrow\rangle$ basis,   one can retain the form of the Bloch Hamiltonian by introducing renormalized SO coupling strength and other parameters~\cite{Spielman17njp,Spielman15prl, LongZhang22prxQ, LongZhang20prl}. If the SO coupling strength is relatively small compared to the modulation frequency, the corresponding 0th-order Bessel function approximately takes the value of unity, and the effective Hamiltonian can be further simplified. }

The Hamiltonian matrix can be numerically diagonalized~\cite{Liang21arxiv}, revealing the properties of this Floquet topological system, such as quasi-energies as functions of quasi-momentum. We have numerically confirmed that this exact diagonalization method yields results in agreement with results based on an evolution operator scheme. In the latter method, an effective Floquet Hamiltonian can be deduced by  $\hat{H}_{\mathrm{eF}}= i\,\log\, \hat{U}(t_{\mathrm{F}})/t_{\mathrm{F}}$,  where $\hat{U}(t)=\hat{T}{\rm exp}[-i \int^t_0 \hat{H}(\tau) d\tau]$ with $\hat{T}$ denoting the time-ordering operator.

\textit{Topological invariants and Floquet topological phase diagram.---}
As shown in the phase diagram in Fig.~1(b) of the main text, 
the topology of our model system is characterized by two winding numbers $W_0$ and $W_{\pi}$, with the Chern number of the Floquet bands given by $\mathrm{Ch}_1=W_0-W_{\pi}$~\cite{Levin13prx,Rudner15njp}.
Here, the Chern number can be also numerically computed as the integral of Berry curvature $\mathcal{F}$ over the first Brillouin zone, namely,  
\begin{equation}\label{chen}
	\begin{split}
		\text{Ch}_1=\frac{1}{2\pi}\int_{-\frac{\pi}{a}}^{\frac{\pi}{a}}\int_{-\frac{\pi}{a}}^{\frac{\pi}{a}} \mathcal{F} d q_x d q_y,
	\end{split}
\end{equation}
where the Berry curvature is given by $ \mathcal{F}=i(\frac{\partial}{\partial q_x} \langle \psi(\textbf{q})| \frac{\partial}{\partial q_y}|\psi(\textbf{q})\rangle-\frac{\partial}{\partial q_y} \langle \psi(\textbf{q})| \frac{\partial}{\partial q_x}|\psi(\textbf{q})\rangle)$, with $\psi(\textbf{q})$ denoting a state of the ground quasi-energy band~\cite{Kane10rmp,sczhang11rmp,asboth2016short}.

We use the topological characterization theory based on band-inversion surfaces (BISs)~\cite{zhanglin18scibull,LongZhang20prl,LongZhang22prxQ} to determine the invariants $W_{0,\pi}$. 
Here a BIS denotes the momentum subspace where the spin bands are inverted, and is determined by $h^z(\textbf{q})-n\omega/2=0$ ($n=0,\pm1,\cdots$). 
We denote the gap around the quasienergy 0 ($\frac{\pi}{T}$) as the 0-gap ($\pi$-gap) and the BIS living in this gap as the 0-BIS ($\pi$-BIS)~\cite{LongZhang20prl,LongZhang22prxQ}.
According to the BIS characterization, the winding number $W_{0}$ ($W_{\pi}$) is contributed by all the 0-BISs ($\pi$-BISs)~\cite{LongZhang20prl,LongZhang22prxQ}:
\begin{align}
	W_{0/\pi}=\sum_j v_j^{(0/\pi)},
\end{align}
where $v_j^{(0)}$ ( $v_j^{(\pi)}$) denotes the topological invariant associated with the $j$th 0-BIS ($\pi$-BIS), characterizing the winding of spin-orbit coupling on this BIS. Under the experimental conditions with finite lattice depths, the positions of the BISs can be numerically computed under a plane-wave basis~\cite{Liang21arxiv}.

\textit{Analytical forms.---}{\color{black}Under SO coupling strengths that are weaker compared to $\hbar\omega$, the topological phase boundaries can be computed by performing BIS analysis and considering the coupling between pairs of quasi-energy bands for $\hat{H}_{\mathrm{sF}}$~\cite{LongZhang20prl,LongZhang22prxQ,zhanglin18scibull}.} We provide the characteristic phase boundary equations $f_1[n]\approx0$ and $f_2[n]\approx0$ (marked by solid lines in Fig.~1(b) of the main text), where in the tight-binding regime, $f_1$ and $f_2$ take the following analytical forms:
\begin{equation}\label{eq:f1f2}
	\begin{split}
		f_1[n]&=8+\text{sgn}[n](2m_z+\frac{1-n}{2}\omega),\\
		f_2[n]&=m_z+\frac{1-n}{4}\omega.
	\end{split}
\end{equation}
Here the sign function $\textrm{sgn}[x]$ takes values of 1, 0, -1 under the conditions $x>0$, $x=0$, $x<0$, respectively. Here, $n$ takes odd integer values ($n=\pm1,\pm3,\cdots$) and  $|n|$ denotes the order number of Raman couplings that induces a certain band inversion. By considering the contributions from all BISs, we find an analytical form for the Chern number:
\begin{eqnarray}\label{ch1}
	\text{Ch}_1& =&\sum_{n}(-1)^{\frac{|n-1|}{2}}\times \frac{\text{sgn}[f_1[n]]+1}{2}
	\times\text{sgn}[f_2[n]] \nonumber\\
	& = & \sum_{l} \frac{\text{sgn}[f_1[l]]+1}{2}
	\times\text{sgn}[f_2[l]] \nonumber\\
	& - & \sum_{m} \frac{\text{sgn}[f_1[m]]+1}{2}
	\times\text{sgn}[f_2[m]] 
\end{eqnarray}
where $l$ and $m$ are odd integers given by $l = \cdots, -7, -3, 1, 5, \cdots$ and $m = \cdots, -5, -1, 3, 7, \cdots$. The analytical form in Eq.~(\ref{ch1}) agrees with our numerical computation based on Eq.~(\ref{chen}).

\begin{widetext}
	
	\begin{figure}
		\includegraphics[scale=0.4]{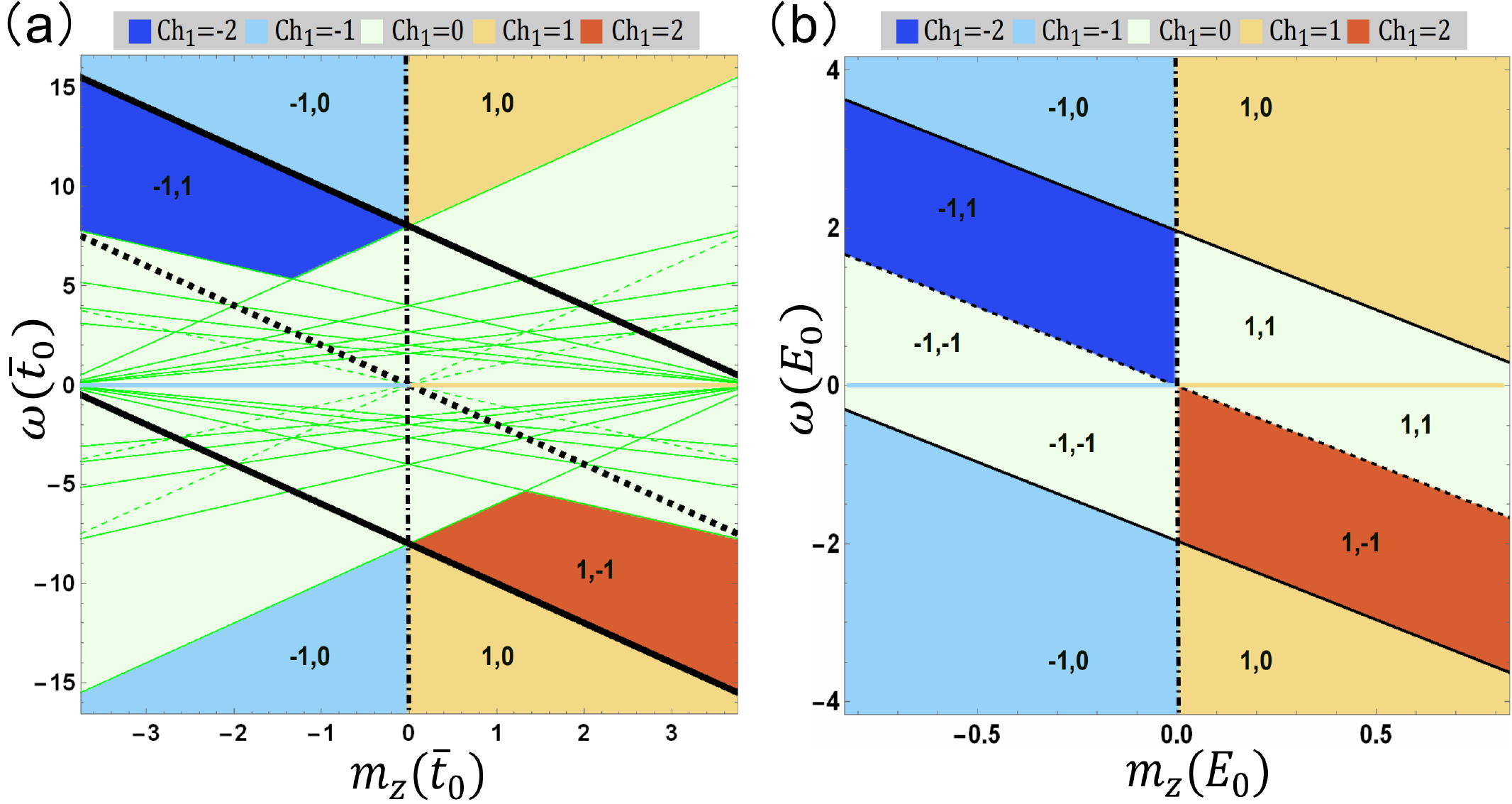}
		\caption{{\color{black}Simplification of the topological phase diagram when only the leading-order SO couplings are considered. (a) A full topological phase diagram is plotted for both the $\omega>0$ part (Fig.~1(b) of the main text) and $\omega <0$ part. Furthermore, those phase boundary lines marked by green color are due to higher-order SO couplings and are neglected when only the leading-order SO couplings are considered. Thus, in a simplified phase diagram, the only remaining phase boundary lines consist of the three negative-slope lines and the horizontal and vertical axes.  (b) For convenience of comparison,  Fig.~4(d) of the main text is provided here on the right side with the same aspect ratio, with $E_0\approx 4 \bar{t}_0$ under the experimental condition of Fig.~4.}
		}
		\label{fig:sm_phasediagramcomparison}
	\end{figure}  
	
\end{widetext}

\subsection{Negligibility of higher-order SO couplings and the simplified phase diagram}
In our present setup, Raman couplings and optical lattices are relatively weak, such that the relevant topology is dominated by leading-order (i.e., $|n|=1$) SO couplings. Specifically, the observed double-ring BISs reveal the lowest-order Raman transition processes allowed by our model. Here we further note that the higher-order processes correspond to energy gaps on the scale of tens of hertz or smaller (namely $10^{-2}E_0$ or smaller) , which are much weaker than $\Omega_{01,02}$ or $\Omega'_{01,02}$. {\color{black}For example, under the experimental parameters listed in the Fig.~4 caption of the main text, the leading-order SO coupling strength $\sqrt{\Omega_{01}^2+\Omega_{02}^2}$ is about $0.18E_0\approx h\times 0.9$kHz. The largest higher-order coupling is a third-order coupling, and can be estimated as $ (\sqrt{\Omega_{01}^2+\Omega_{02}^2})^3/(\hbar\omega)^2$. We note that the value of $\omega$ cannot be exactly zero, otherwise all BISs will overlap, which prevents the identification of a certain BIS. Thus, using a characteristic value of $\omega/4\pi \sim 2$kHz (namely $\hbar \omega \sim h \times 4$kHz), we derive a higher-order strength on the scale of $h \times 40$Hz. Such strength for the higher-order coupling is weak and rather marginal for the detection of the corresponding higher-order SO effect based on the present setup.}

Thus, even with our current long lifetime of SO-coupled fermions, the corresponding higher-order couplings can hardly cause any band inversion or detectable spin flipping in a realistic experimental sequence. {\color{black}Thus, when only the leading-order SO couplings are considered, a full topological phase diagram (Fig.~1(b) in the main text) can be simplified into Fig.~4(d), as further illustrated and explained by Fig.~\ref{fig:sm_phasediagramcomparison}.} The topological invariant can be described by Eq.~(\ref{ch1}) with  summation over $n=1$ and $-1$, which leads to the following expression
\begin{eqnarray}\label{eq:analyticPhaseBoundary}
	\text{Ch}_1 & = &  \frac{\text{sgn}[f_1[1]]+1}{2}
	\times\text{sgn}[f_2[1]] \nonumber\\
	&  - & \frac{\text{sgn}[f_1[-1]]+1}{2}
	\times\text{sgn}[f_2[-1]],
\end{eqnarray}
for the Chern number as a function of $m_z$ and $\omega$ (see Eq.~(\ref{eq:f1f2}) for example).
Equation~(\ref{eq:analyticPhaseBoundary}) provides a useful expression for depicting and understanding the simplified phase diagram that is numerically computed based on leading-order SO couplings only (Fig.~4(d) in the main text), {\color{black}where $n=1$ and $n=-1$ correspond to the rotating and counter-rotating leading-order terms for SO coupling beyond RWA}. 

{\color{black}In the future, if we improve the setup (including implementing a laser with larger optical power) and increase the leading-order SO coupling strength by a factor of two, the third-order coupling strength can in principle increase by almost an order of magnitude to $h\times$ several hundred hertz, which will provide a much more favorable condition for the detection of higher-order SO physics.}

\begin{figure}
	\includegraphics[scale=0.25]{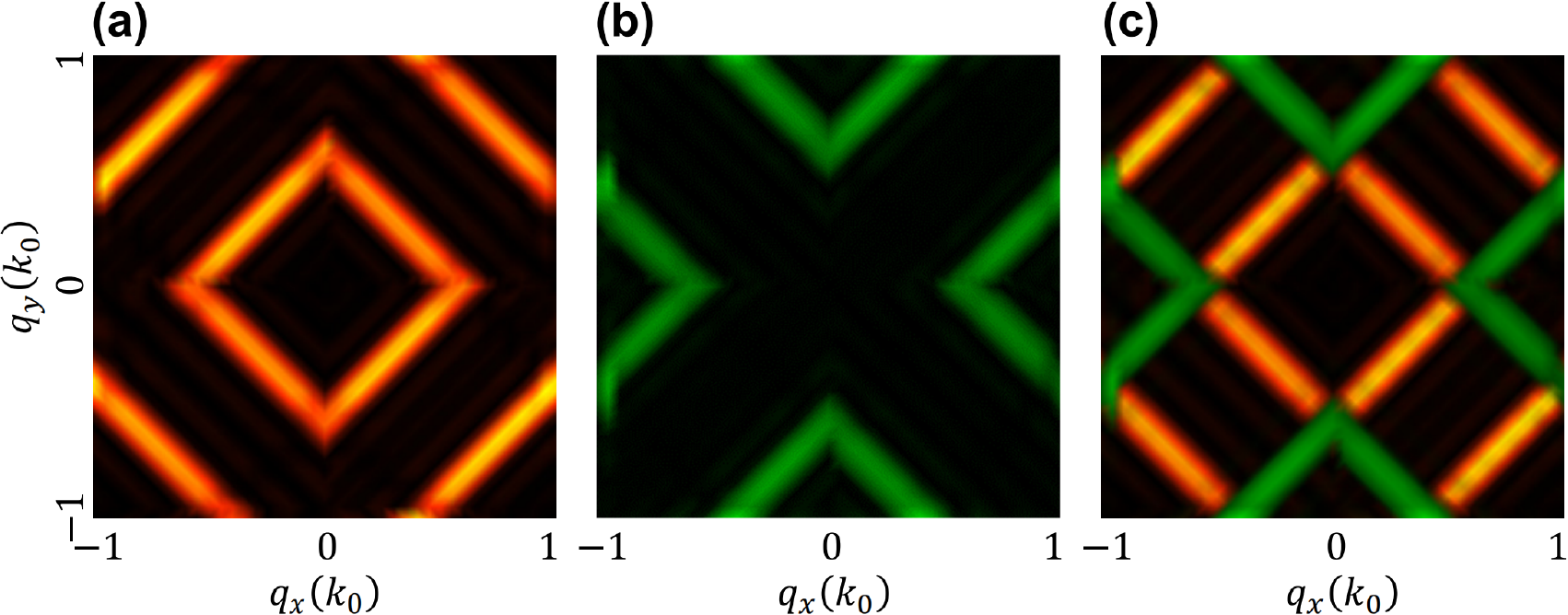}
	\caption{Effect of $|\!\!\!\uparrow\rangle$ atoms initially populated outside the first Brillouin zone. (a) PPQM signal contributed by $|\!\!\uparrow\rangle$ atoms initially populated in the first Brillouin zone. (b) PPQM signal contributed by atoms initially populated in the second Brillouin zone. (c) Sum of signal in (a) and (b). Orange and green colors are used to facilitate distinguishing between contributions from different atomic group. Here we set $\Omega_{01} = \Omega'_{02} = 0.17 E_0$, $\Omega_{02} = \Omega'_{01} = 0.07 E_0$, $V_{0X\uparrow} = V_{0Y\uparrow} = 0.13E_0$, $V_{0X\downarrow} = V_{0Y\downarrow} = 0.02E_0$, $m_z = 0.42E_0$, $\omega/4\pi = 0.83E_0$.}
	\label{fig:sm_sbz}
\end{figure}

\subsection{Effect of $|\!\!\uparrow\rangle$ atoms initially populated outside the first Brillouin zone}
This subsection provides supporting numerical evidence for the main-text note that ``in Fig.~4(a), the observed weak connections between two BISs are contributed from the $|\!\!\uparrow\rangle$ atoms initially populated outside the first Brillouin zone''. As shown in Fig.~4(a) in the main text, in addition to the two-ring band-inversion surface, we notice some weak connections between the two rings. Here, in Fig.~\ref{fig:sm_sbz}, we simulate the contributions to the PPQM result based on (a) $|\!\!\uparrow\rangle$ atoms initially populated in the first Brillouin zone, (b) $|\!\!\uparrow\rangle$ atoms initially populated in the second Brillouin zone, and (c) $|\!\!\uparrow\rangle$ atoms initially populated in the first and second Brillouin zones. Indeed, in Fig.~\ref{fig:sm_sbz}(a) we observe the two-ring BIS configuration. In Fig.~\ref{fig:sm_sbz}(b) and (c), we see that the atoms initially populated in the second Brillouin zone contributes to the ``connections'' (in green) between the two BIS rings (in orange) that reflect the ground band topology.

{\color{black}
	\subsection{On the future experimental exploration of the 2D Floquet topological phase diagram}
	At present, Fig.~4 in the main text explores a line in the 2D Floquet topological phase diagram because the modulation angular frequency $\omega$ and the effective Zeeman term $m_z$ are linearly related. In future experiments, our system holds promise for exploring more phase regions in the 2D Floquet phase diagram. This can be realized via independent control of $\omega$ and the energy splitting between the $|\!\!\uparrow\rangle$ and $|\!\!\downarrow\rangle$ states. For example, by tuning the magnetic field independent of $\omega$, the magnetic-field-induced energy splitting between $|\!\!\uparrow\rangle$ and $|\!\!\downarrow\rangle$ will be tunable independent of $\omega$. Thus $m_z$ will be similarly tunable, too. Based on independent control of two parameters like $\omega$ and the magnetic field, our system will be capable of exploring the 2D Floquet topological phase diagram in the future.

	In addition, the energy splitting between $|\!\!\uparrow\rangle$ and $|\!\!\downarrow\rangle$ can also be changed by choosing a different pair of spin states to act as $|\!\!\uparrow\rangle$ and $|\!\!\downarrow\rangle$. For example, instead of the current choice of $|m_{_F} = -\frac{9}{2}\rangle$ and $| -\frac{7}{2}\rangle$, if we choose $|\!\!\uparrow'\rangle = |+\frac{9}{2}\rangle$ and $|\!\!\downarrow'\rangle = |+\frac{7}{2}\rangle$, the magnetic-field-induced energy splitting between the two states will reverse its sign, which again makes $m_z$ change independent of $\omega$. This method further enlarges the phase region that can be explored in the 2D Floquet phase diagram.
	
}

\end{document}